\documentclass[12pt,a4paper]{article}

\usepackage[font=small,labelfont=bf,width=0.85\textwidth]{caption}
\usepackage{mathtools} 
\usepackage{amssymb}
\usepackage{bm}
\usepackage{accents}
\usepackage{color}
\usepackage{hyperref}
\usepackage{graphicx}
\usepackage[nosort]{cite}
\usepackage[bulletsep]{collref}
\usepackage{tensor}
\usepackage[british]{babel}
\usepackage[utf8]{inputenc}

\usepackage[a4paper,text={173mm,216mm},centering]{geometry}

\allowdisplaybreaks[3]

\numberwithin{equation}{section}

\newcommand{\nn}{\nonumber}

\let\oldbfseries=\bfseries
\renewcommand{\bfseries}{\oldbfseries\boldmath}

\newcommand{\eqn}[1]{(\ref{#1})}

\newcommand{\sfrac}[2]{{\textstyle\frac{#1}{#2}}}
\newcommand{\half}{\sfrac{1}{2}}
\newcommand{\ihalf}{\sfrac{i}{2}}
\newcommand{\third}{\sfrac{1}{3}}
\newcommand{\quarter}{\sfrac{1}{4}}

\DeclareMathOperator{\tr}{tr}
\DeclareMathOperator{\diag}{diag}
\newcommand*{\diff}{{\mathrm d}}

\newcommand{\Tr}{\mathop{\mathrm{Tr}}}
\newcommand{\str}{\mathop{\mathrm{str}}}

\renewcommand{\a}{\alpha}
\renewcommand{\b}{\beta}
\newcommand{\g}{\gamma}

\newcommand{\da}{{\dot{\alpha}}}
\newcommand{\db}{{\dot{\beta}}}
\newcommand{\dg}{{\dot{\gamma}}}
\newcommand{\la}{\lambda}
\newcommand{\bla}{\bar{\lambda}}
\newcommand{\dbla}{\dot{\bar{\lambda}}}

\newcommand{\eps}{\epsilon}
\newcommand{\dx}{{\dot x}}
\newcommand{\s}{\sigma}
\newcommand{\bs}{\bar{\sigma}}
\newcommand{\btheta}{\bar\theta}
\newcommand{\vart}{\vartheta}
\newcommand{\bvart}{\bar\vartheta}

\newcommand{\Qb}{{\bar{Q}}}
\newcommand{\Sb}{{\bar{S}}}

\begin{document}
\thispagestyle{empty}

\begingroup\raggedleft\footnotesize\ttfamily
HU-EP-15/13\
\vspace{15mm}
\endgroup

\begin{center}
{\Large\bfseries Minimal Surfaces of the \texorpdfstring{$AdS_5\times S^5$ }{AdS5 x S5} Superstring and the Symmetries of Super Wilson Loops at Strong Coupling\par}
\vspace{25mm}

\begingroup\scshape\large 
Hagen M\"unkler,  Jonas Pollok
\endgroup
\vspace{5mm}

\textit{Institut f\"ur Physik and IRIS Adlershof, Humboldt-Universit\"at zu Berlin, \phantom{$^\S$}\\
Zum Gro{\ss}en Windkanal 6, D-12489 Berlin, Germany} \\[0.1cm]
\texttt{\small\{muenkler,pollok\}@physik.hu-berlin.de\phantom{\ldots}} \\ \vspace{5mm}

\vspace{18mm}

\textbf{Abstract}\vspace{5mm}\par
\begin{minipage}{14.7cm}
Based on an extension of the holographic principle to superspace, we provide a strong-coupling description of smooth super Wilson loops in $\mathcal{N}=4$ super Yang-Mills theory in terms of minimal surfaces of the $AdS_5 \times S^5$ superstring. We employ the classical integrability of the Green-Schwarz superstring on $AdS_5 \times S^5$ to derive the superconformal and Yangian $Y[\mathfrak{psu}(2,2|4)]$ Ward identities for the super Wilson loop, thus extending the strong coupling results obtained for the Maldacena-Wilson loop. In the course of the derivation, we determine the minimal surface solution up to third order in an expansion close to the conformal boundary.

\end{minipage}\par
\end{center}
\newpage

\setcounter{tocdepth}{2}
\hrule height 0.75pt
\setcounter{tocdepth}{3}
\tableofcontents
\vspace{0.8cm}
\hrule height 0.75pt
\vspace{1cm}

\setcounter{tocdepth}{2}

\hyphenation{ei-gen-spa-ces}

\newpage

\section{Introduction and Summary}

The holographic AdS/CFT correspondence~\cite{Maldacena:1997re, Gubser:1998bc,
Witten:1998qj} has proven to be very successful in studying quantum
field theories as it allows to investigate the strong coupling regime which is
inaccessible to perturbation theory. The most famous and best
studied example of this correspondence is the $\mathcal{N} = 4$ supersymmetric
Yang-Mils (SYM) theory in the planar limit which corresponds to type
IIB superstring theory on an $AdS_5 \times S^5$ background. The string theory 
is known to be classically integrable \cite{Bena:2003wd} and both theories appear
to be integrable also at the quantum level. 

Integrable structures, which are often related to an infinite dimensional extension
of the underlying superconformal symmetry into a Yangian symmetry algebra, have 
by now been detected for a wide range of observables. 
The earliest examples were found while investigating two-point functions 
in $\mathcal{N}=4$ SYM. These could be related to integrable
spin chain models and generalized Bethe ans\"atze were formulated as a solution
to the spectral problem, see~\cite{Beisert:2010jr} for an overview. Other
observables showed hints of integrability in a different form. A duality
between certain scattering amplitudes and null-polygon Wilson loops was first
discovered at strong coupling~\cite{Alday:2007hr} and later also observed
at weak coupling \cite{Drummond:2007aua, Brandhuber:2007yx, Drummond:2007cf, 
Bern:2008ap, Drummond:2008aq}. Subsequently, the duality was extended to relate
any amplitude to polygonal super Wilson loops~\cite{CaronHuot:2010ek, Mason:2010yk}.
On the weak coupling side the duality between Wilson loops and amplitudes inspired
the discovery of a dual superconformal symmetry~\cite{Drummond:2008vq} in
addition to the ordinary superconformal symmetry. This symmetry has been
studied thoroughly since and was shown to combine with the usual superconformal 
symmetry into a Yangian symmetry~\cite{Drummond:2009fd}. It can most naturally be 
understood as the invariance of the $AdS_5 \times S^5$ superstring under a generalized T-duality~\cite{Berkovits:2008ic, Beisert:2009cs}. 
Due to infrared divergences, the symmetry of the amplitude is deformed at loop
level~\cite{Beisert:2010gn, Sever:2009aa}, but holds for the loop integrand~\cite{ArkaniHamed:2010kv}. 
The symmetries of null-polygonal Wilson loops suffer from the
same divergences as the amplitudes, which due to dual coordinates appear as UV
divergences in the Wilson loop picture~\cite{Belitsky:2012nu, CaronHuot:2011ky,Beisert:2012xx}.
A recent constructive application of integrability to cusped Wilson loops at any coupling uses their decomposition 
into so-called pentagon transitions which can be fixed from integrability \cite{Basso:2013vsa,Basso:2013aha,Basso:2014koa,Basso:2014nra,Basso:2014hfa}.
The attempt to make the Yangian invariance of amplitudes/Wilson loops manifest
led to the investigation of the positive Grassmanian 
and its generalization the Amplituhedron \cite{ArkaniHamed:2012nw, Arkani-Hamed:2013jha,
Arkani-Hamed:2013kca, Bai:2014cna}.
The developments in this area certainly point towards the existence
of rich integrable structures for both amplitudes and Wilson loops, the uncovering of
which is impeded by the breakdown or deformation of the symmetries due to the appearance 
of divergences.

In this paper we turn to a class of finite observables: Smooth super Wilson loops, which are a generalization of the Maldacena-Wilson loop~\cite{Maldacena:1998im,Rey:1998ik} and have already been considered in the early days of the AdS/CFT correspondence \cite{Ooguri:2000ps}. The Maldacena-Wilson loop is a generalization of the Wilson loop, which is specific for $\mathcal{N}=4$ SYM as it also includes the scalar fields $\Phi_I$,
\begin{align}
W (C) = \frac{1}{N} \, \mathcal{P} \exp{ \left( i \int \diff s \left( A_\mu \dx ^\mu + \Phi_I \lvert \dx \rvert n^I \right) \right)} \,. \label{MWLOOP}
\end{align}
Here, $n^I$ describes a point on $S^5$ as $n^2 = 1$ and may also depend on the loop parameter $s$. The expectation value of the Maldacena-Wilson loop is finite for smooth contours, which is related to the local $1/2$ BPS symmetry of the loop operator. At strong coupling, the Maldacena-Wilson loop is described by the renormalized area of a minimal surface in anti de Sitter space ending on the conformal boundary on the loop contour $C$,
\begin{align}
\left \langle W (C) \right \rangle \overset{\la \gg 1}{=} e^{-\frac{\sqrt{\la}}{2 \pi} A_\mathrm{ren}(C)} \,. \label{StrongMWL} 
\end{align}
The classical integrability of the string action which describes the area can be applied to derive the invariance of $\left \langle W (C) \right \rangle$ under the Yangian symmetry $Y[\mathfrak{so}(2,4)]$ over the conformal algebra \cite{Muller:2013rta}, for which one finds the level-1 generators
\begin{align}
J_a ^{(1)} = f \indices{^{cb} _a} \, \int \diff s_1 \diff s_2 \, \varepsilon(s_1-s_2) \, \xi^\mu _b (x_1) \, \xi^\nu _c (x_2)  \dfrac{\delta^2}{\delta x^\mu _1 \delta x^\nu _2} + \frac{\la}{2 \pi ^2} \int \limits _0 ^L \diff s \, \xi^\mu _a (x) \left( \dx_\mu \, \ddot{x}^2 + \dddot{x}_\mu \right) \,. \label{Level1Old}
\end{align}
Here, $\xi ^\mu _a (x)$ are conformal Killing vectors and $f \indices{^{cb} _a}$ denote the dual structure constants of the conformal algebra. Apart from the typical bi-local part of level-1 Yangian generators $J_a ^{(1)}$ also involves a coupling dependent local piece, which is for simplicity written in an arc-length parametrization, $\lvert \dx \rvert = 1$.

Of course, the derivation is only valid in the strong coupling regime $\la \gg 1$ and so it is natural to ask whether the observed Yangian invariance holds for any value of $\la$. This question was addressed in \cite{Muller:2013rta} by considering $\left \langle W (C) \right \rangle$ for small $\la$ in perturbation theory and it was shown that a Yangian symmetry of the Maldacena Wilson loop is not present at weak coupling. However, it was shown that a supersymmetric extension $\mathcal{W}(C)$ of the Maldacena Wilson loop, in which the fermionic fields of $\mathcal{N}=4$ SYM couple to the coordinates of a non-chiral superspace, exhibits signs of a Yangian symmetry over the superconformal algebra $\mathfrak{psu}(2,2 \vert 4)$. Specifically it was demonstrated that -- to first order in perturbation theory and to lowest order in an expansion in the anticommuting superspace coordinates $\theta$ and $\btheta$ -- the expectation value $\left \langle \mathcal{W}(C) \right \rangle$ is annihilated by the level-1 Yangian generator
\begin{align}
P^{(1) \, \mu} = f \indices{^{cb} _{P^\mu}} \, \int \diff s_1 \diff s_2 \, \varepsilon(s_1-s_2) j_b (s_1) \, j_c(s_2) \, + \frac{7 \la}{96 \pi ^2} \int \limits _0 ^L \diff s \,   \dx^\mu \, \ddot{x}^2 \label{P1weak}
\end{align} 
Here, the $j_a(s_1)$ form a representation of the superconformal algebra $\mathfrak{psu}(2,2 \vert 4)$ in terms of differential operators and it is understood that the derivatives in $j_b(s_1)$ do not act on $j_c(s_2)$. The super Maldacena Wilson loop on which this generator acts may be viewed as the smooth counterpart of the lightlike polygonal non-chiral super Wilson loops constructed in \cite{Beisert:2012xx,CaronHuot:2011ky}, although providing an explicit relation between them is obstructed by the incomplete knowledge of $\mathcal{W}(C)$ that follows from the order-by-order construction performed in \cite{Muller:2013rta}. The field theory description of smooth super Maldacena-Wilson loops is being worked out in parallel \cite{Beisert:2015jxa,Beisert:2015uda}, whereas we focus on the strong coupling description in this paper.

Interestingly, the contour dependence of the local piece in \eqn{P1weak} agrees with that of the Yangian generator derived for the bosonic Maldacena Wilson loop at strong coupling and it was suspected that this structural agreement of weak and strong coupling symmetries would continue to hold true for the full super Wilson loop and that even the exact $\la$-dependence could coincide upon including fermionic contributions at strong coupling. The latter supposition can be disproved in this paper, a comparison of the contour dependence of the local term is postponed until the results of \cite{Beisert:2015uda} are available.

In this paper we turn to the strong coupling description of the super Maldacena-Wilson loop $\mathcal{W}(C)$, for which we use the natural generalization of \eqn{StrongMWL} that is given by
\begin{align}
\left \langle \mathcal{W} (C) \right \rangle \overset{\la \gg 1}{=} e^{-\frac{\sqrt{\la}}{2 \pi} \mathcal{A}_\mathrm{ren}(C)} \,.
\end{align}
Here, the minimal area $\mathcal{A}_\mathrm{ren}(C)$ is computed from the Green-Schwarz superstring action \cite{Metsaev:1998it} in the supercoset space $PSU(2,2\vert4)/(SO(4,1) \times SO(5))$. The appropriate boundary conditions follow from the construction of the superconformal boundary of this space \cite{Ooguri:2000ps}, which also provides the appropriate superspace for the super Maldacena-Wilson loop. This space includes the spherical coordinates appearing in \eqn{MWLOOP}, whose inclusion in the superspace was subject to speculation in \cite{Muller:2013rta}. We provide a renormalization procedure for the minimal area, which is given by
\begin{align}
	\mathcal{A}_\mathrm{ren}(C) = \lim \limits_{\varepsilon \to 0} \left \lbrace
		\mathcal{A}_{\mathrm{min}}(C) \Big \vert _{y \geq \varepsilon}
		- \frac{\mathcal{L}(C)}{\varepsilon}
	\right \rbrace \, ,
	&& \mathcal{L}(C) = \int \diff s \, \lvert \pi(s) \rvert \, ,
\end{align}
where
$\pi ^\mu = \dx ^\mu + i \big(
	\dot{\bar{\la}} \s ^\mu \la - \bar{\la} \s ^\mu \dot{\la}
\big)$ is the supermomentum of a superparticle moving along the contour $C$. This generalizes the construction applied in $AdS_5$. Moreover, we derive the first few orders of the parametrization of the minimal surface in an expansion away from the boundary, thereby generalizing the results obtained by Polyakov and Rychkov for minimal surfaces in $AdS_5$ \cite{Polyakov:2000ti,Polyakov:2000jg}. These insights are then applied to show that the super Maldacena-Wilson loop is Yangian invariant at strong coupling by explicitly constructing all level zero and level one  generators of the Yangian algebra $Y[\mathfrak{psu}(2,2 \vert 4)]$, which we provide in equations \eqn{Ja0} and \eqn{Ja1}. The derivation relies on the classical integrability \cite{Bena:2003wd} of the superstring.

Let us explain briefly how this paper is structured. The simplicity of the construction we apply to derive the Yangian symmetries of the super Wilson loop is obscured by the technical difficulties that arise in dealing with the superstring action on the supercoset space $PSU(2,2 \vert 4)/(SO(4,1) \times SO(5))$. We therefore reconsider the purely bosonic situation of a minimal surface in $AdS_5$, where the structure of the derivation is more transparent. In contrast to \cite{Muller:2013rta} we provide the derivation in the language of a coset construction that can be generalized to the full supercoset. This account forms the most part of section~\ref{sec:MinimalSurfaces}, which also contains a discussion of minimal surfaces in $S^5$, again employing a coset construction that generalizes to the full supercoset.

In section~\ref{sec:SuperWL} we introduce the strong coupling description of the super Maldacena-Wilson loop based on a minimal surface in the supercoset space $PSU(2,2 \vert 4)/(SO(4,1) \times SO(5))$. Following the review article \cite{Arutyunov:2009ga} we introduce the action \cite{Metsaev:1998it} of the Green-Schwarz superstring in $AdS_5 \times S^5$ and briefly discuss the properties needed in the remainder of this paper. We then go on to discuss the boundary conditions for the minimal surface which follow from the description of the conformal boundary of the $AdS_5 \times S^5$ superspace given by Ooguri et al. \cite{Ooguri:2000ps}. 

Section~\ref{sec:symm} comprises the new results obtained in this paper. We relate the first orders of the parametrization of the minimal surface to the boundary data by iteratively solving the equations of motion and Virasoro constraints. The coefficients that are not fixed in this way can be related to variational derivatives of the minimal area. The evaluation of the conserved charges obtained from the integrability of the string model then leads to the desired superconformal and Yangian Ward identities for the super Maldacena Wilson loop. 

Wherever possible we try not to burden the exposition with too much technical detail. The details of certain calculations as well as a collection of our conventions are provided in the appendices~\ref{app:spinorconv}~-~\ref{app:kappa}.

\section{Minimal surfaces in \texorpdfstring{$AdS_5$}{AdS5} and \texorpdfstring{$S^5$}{S5}}
\label{sec:MinimalSurfaces}
\subsection{The Maldacena-Wilson Loop at strong Coupling}
\label{sec:bosAdS}
We rederive the Yangian symmetry of the Maldacena-Wilson loop at strong coupling, which was discovered in \cite{Muller:2013rta}. Here, we employ a coset description of $AdS_5 \simeq SO(4,2) / SO(4,1)$ to prepare ourselves for the discussion of the super Wilson loop, which will also be based on a coset description.  

At strong coupling, the expectation value of the Maldacena-Wilson loop is given by \cite{Maldacena:1998im}
\begin{align}
\left \langle W(C) \right \rangle \overset{\la \gg 1}{=} \exp{\left(- \sfrac{\sqrt{\la}}{2 \pi} A_\mathrm{ren}(C) \right) } \label{strongMWL} \,. 
\end{align}
Here, $A_\mathrm{ren}(C)$ is the renormalized area of a minimal surface that ends on the curve $C$ on the conformal boundary of $AdS_5$. Let us point out here, that the use of the renormalized area is a consequence of the AdS/CFT prescription for computing $\langle W(C) \rangle$ at strong coupling, see e.g. \cite{Drukker:1999zq}. It does not correspond to a renormalization of the Maldacena-Wilson loop, which is finite for smooth loops.

The minimal surface is most naturally described in Poincar{\'{e}} coordinates, which may be obtained from coset constructions in different ways. We follow \cite{Ooguri:2000ps} and use a construction which may be generalized to the super Wilson loop. The description of the coset space $SO(4,2) / SO(4,1)$ is based on the $\mathbb{Z}_2$ decomposition of the conformal algebra $\mathfrak{so}(4,2)$, which we discuss in appendix~\ref{app:su224},
\begin{alignat}{2}
\mathfrak{so}(4,2) &= \left( \mathfrak{g}^{(0)} \simeq \mathfrak{so}(4,1) \right) \oplus \mathfrak{g}^{(2)} \, ,& \qquad  &\left[ \mathfrak{g}^{(k)} , \mathfrak{g}^{(l)} \right] \subset \mathfrak{g}^{(k+l \, \mathrm{mod} \, 4)} \, , \\
\mathfrak{g}^{(0)} &= \mathrm{span} \left \lbrace P_\mu - K_\mu , M_{\mu \nu} \right \rbrace \; \: , & \qquad  &\mathfrak{g}^{(2)} = \mathrm{span} \left \lbrace P_\mu + K_\mu , D \right \rbrace \,. \label{basis1}
\end{alignat}
The coset representatives are given by
\begin{align}
g(X,y) = e^{X \cdot P} \, y^D    \quad \Rightarrow \quad A = - g^{-1} \, \diff g = - \frac{\diff X^\mu}{y} \, P_\mu - \frac{\diff y}{y} \, D \,, 
\end{align}
and we note the projections
\begin{align}
A^{(0)} = - \frac{\diff X^\mu}{2y} \left( P_\mu - K_\mu \right) \, ,  \qquad A^{(2)} = - \frac{\diff X^\mu}{2y} \left( P_\mu + K_\mu \right) - \frac{\diff y}{y} \, D \, .
\end{align}
The metric of the coset space is obtained from the group metric introduced in appendix~\ref{app:su224} and the Cartan form $A$ as 
\begin{align}
\diff s^2 = \left \langle A^{(2)} \, , \, A^{(2)} \right \rangle = \frac{\diff y ^2 + \eta^{\mu \nu}  \diff X_\mu \diff X_\nu }{y^2} \, , \qquad \eta = \mathrm{diag}(- , + , + , + ) \,.
\end{align}
showing that the parametrization $g(X,y)$ reproduces the well-known Poincar{\'{e}} coordinates for $AdS_5$. Correspondingly, we may describe the area functional in these coordinates by the sigma-model action
\begin{align}
A[X,\gamma] =  \frac{1}{2} \int \diff s \, \diff \tau \, \gamma^{i j} \left \langle A_i^{(2)} \, , \, A_j^{(2)} \right \rangle \,. \label{bosact} 
\end{align}
Here, $\gamma ^{i j}= \sqrt{h} \,  h^{ij}$ denotes the Weyl-invariant combination formed from the world-sheet metric and its determinant. In \cite{Muller:2013rta}, the authors considered Euclidean $AdS_5$ where the boundary space is Euclidean. Here, we restrict ourselves to boundary curves, for which all tangent vectors are spacelike. Then the world-sheet metric is Euclidean and in conformal gauge we have $\gamma^{ij} = \delta^{ij}$.  

The minimal surface is subject to the boundary conditions
\begin{align}
y(\tau = 0 , s) = 0 \, , \qquad X^\mu(\tau=0 , s) = x^\mu (s) \,.
\end{align}
Here, $x^\mu(s)$ denotes a parametrization of the boundary curve $C$. The minimization of \eqn{bosact} leads to the equations of motion and Virasoro constraints (in conformal gauge)
\begin{align}
\delta^{ij} \left( \partial _i \, A_j^{(2)} - \left[ A_i ^{(0)} , A_j ^{(2)} \right] \right) = 0  \, ,  \qquad \left \langle A_i^{(2)} \, , \, A_j^{(2)} \right \rangle - \half \delta_{ij} \, \delta^{kl} \left \langle A_k^{(2)} \, , \, A_l^{(2)} \right \rangle = 0 \,.
\end{align}
Due to the divergence of the metric on the conformal boundary $y=0$, one can fix the first coefficients in the $\tau$-expansion of $X$ and $y$ from the equations of motion. Introducing the notation
\begin{align}
X^\mu (\tau , s ) = X^\mu _{(0)}(s) +  \sum \limits _{n=1} ^\infty   X^\mu _{(n)}(s) \, \frac{\tau^n}{n} \, , \qquad y(\tau , s ) = y _{(0)}(s) + \sum \limits _{n=1} ^\infty   y _{(n)}(s) \, \frac{\tau^n}{n} \, , \label{notationdef}
\end{align} 
one obtains that \cite{Polyakov:2000ti,Polyakov:2000jg}
\begin{align}
y_{(1)} = \lvert \dx \rvert  , \qquad y_{(2)} = 0 \, , \qquad  X^\mu_{(1)} = 0 \, , \qquad  X^\mu_{(2)} = \dx ^2 \partial _s \left( \frac{\dx^\mu}{\dx^2} \right)  \,. \label{PR1}
\end{align}
This shows that that the minimal surface moves away from the boundary perpendicularly as one would expect from the divergence of the metric on the conformal boundary. The divergence of the minimal area is correspondingly given by
\begin{align}
A(C) \Big \vert _{y \geq \varepsilon} = \frac{L(C)}{\varepsilon} + \mathcal{O}\left( \varepsilon^0 \right) \, ,
\end{align}
where $L(C)$ denotes the length of the curve. The renormalized area appearing in \eqn{strongMWL} is defined by
\begin{align}
A_\mathrm{ren}(C) := \lim \limits _{\varepsilon \to 0} \left \lbrace A(C) \Big \vert_{y \geq \varepsilon} - \frac{L(C)}{\varepsilon} \right \rbrace \,.
\end{align}
An interpretation of this specific renormalization procedure for the renormalized area has been discussed in \cite{Drukker:1999zq}. 

Computing the variation of $A_\mathrm{ren}(C)$ with respect to a variation of the boundary data $x^\mu(s)$ one may identify another coefficient in the $\tau$-expansion \cite{Polyakov:2000ti,Polyakov:2000jg},
\begin{align}
X_{(3)} ^\mu (s) = - \dx^2 \, \dfrac{\delta A_\mathrm{ren}(C) }{\delta x_\mu (s) } \,. \label{funcder}
\end{align}
In order to derive this result, consider a variation $\delta x^\mu(s)$ of the boundary curve. The variation of the boundary curve induces a variation $\left(\delta X^\mu , \delta y \right)$ of the parametrization of the minimal surface. From the solution \eqn{PR1} of the equations of motion we know that $\delta X^\mu = \delta x^\mu + \mathcal{O}(\tau^2)$. Let us then compute the variation of the minimal area, which is regulated by demanding $y \geq \varepsilon$, or equivalently $\tau \geq \tau_0(s)$, where $\tau_0(s)$ is defined by $y(\tau_0(s),s) =\varepsilon$. Since we are varying around a minimal surface solution, we may employ that $(X^\mu,y)$ satisfy the equations of motion and hence the variation is given by a boundary term,
\begin{align*}
\delta A \big\vert _{y \geq \varepsilon} = \int \diff s \int \limits _{\tau_0(s)} ^a \diff \tau \, \partial_i \, \frac{\gamma ^{ij}\left( \partial_j X^\mu \, \delta X_\mu + \partial_j y \, \delta y \right) }{y^2}  = \frac{1}{\varepsilon^2}  \int \diff s \left \lbrace \tau_0^\prime (s) \,  \partial_s X^\mu \delta X_\mu - \,  \partial_\tau X^\mu \delta X_\mu \right \rbrace .
\end{align*}
Here, we used that $\delta y(\tau_0(s),s)=0$ due to the definition of $\tau_0$. Inserting the results \eqn{PR1} one finds
\begin{align*}
\delta A \big\vert _{y \geq \varepsilon} = \frac{\delta L(C)}{\varepsilon} - \int \diff s \frac{X_{(3)}^\mu}{\dx^2} \, \delta x_\mu \, ,
\end{align*}
from which one can read off the result \eqn{funcder}. To fix the higher coefficients in the $\tau$-expansion, it is convenient to restrict the parametrization of the boundary curve to satisfy $\dx ^2 \equiv 1$. The residual reparametrization invariance in conformal gauge is sufficient to do so. We have refrained from fixing the parametrization until now, since in the derivation of \eqn{funcder} one has to be careful about restricting the parametrization of the boundary curve, since $\delta \lvert \dx \rvert \neq 0$, which one tends to overlook after setting $\lvert \dx \rvert \equiv 1$. 

For the higher-order coefficients one may derive
\begin{align}
y_{(3)} = - \ddot{x} ^2 \, , \qquad X_{(4)}^\mu = \half \, \ddddot{x}^{\,\mu}+ \sfrac{4}{3} \left(\ddot{x} ^\mu \, \ddot{x}^2 + \dx^\mu \, \ddot{x} \cdot \dddot{x} \right) 
\end{align}
from the Virasoro constraints and equations of motion respectively \cite{Muller:2013rta}. Given these findings one may employ the classical integrability of the bosonic string theory on $AdS_5$ to derive Ward identities for the Wilson loop \eqn{strongMWL} at strong coupling. The Noether current $ J_i = g A_i ^{(2)} g^{-1}$ of the model is both flat and conserved,
\begin{align}
\partial _i \, J^i = 0 \, , \qquad \qquad \partial _i \, J_j - \partial _j \, J_i + 2 \left[ J_i \, , \, J_j \right] = 0 \,.
\end{align}
Using these properties one can construct non-local conserved charges
\begin{align}
\mathcal{Q}^{(0)} = \int \diff s \, J_\tau \, , \qquad 
\mathcal{Q}^{(1)} = \frac{1}{2} \int \diff s_1 \, \diff s_2 \, \varepsilon(s_1-s_2) \, \left[ J_\tau (s_1) \, , \, J_\tau (s_2) \right] - \int \diff s \, J_s \,. \label{chargedef}
\end{align}
\begin{figure}[t]
\begin{center}
\includegraphics[height=6cm]{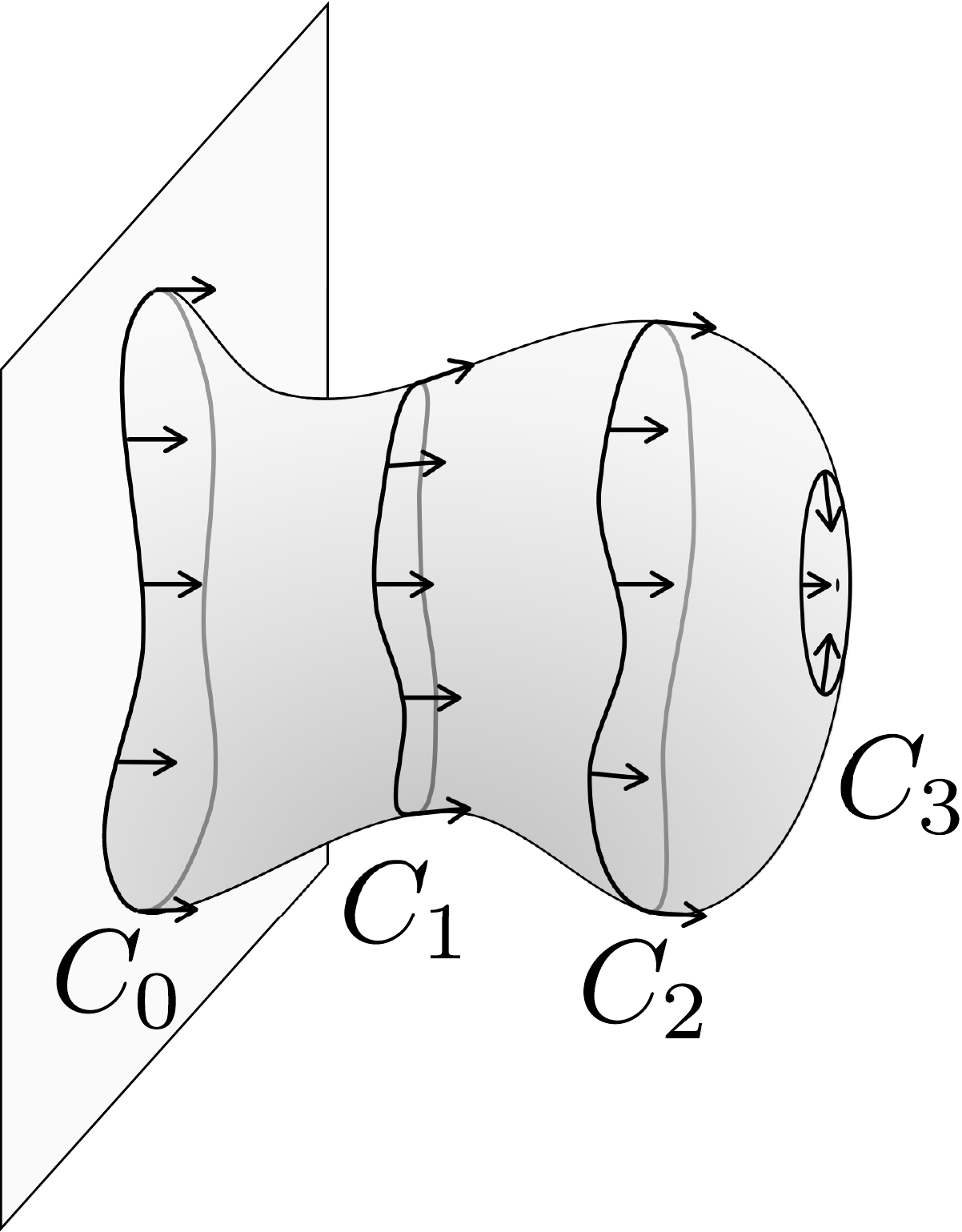}
\caption{As any cycle $C$ on the minimal surface worldsheet is contractible and the charges $\mathcal{Q}^{(0)}$ and $\mathcal{Q}^{(1)}$ do not change under cycle deformations, the initial cycle $C_{0}$  at the boundary may be shrunk to zero at the tip of the surface. This forces the charges $\mathcal{Q}^{(0)}$ and $\mathcal{Q}^{(1)}$ to vanish.}
\label{fig:one}
\end{center}
\end{figure}
Since any curve on the minimal surface is contractible, these charges are not only conserved, but also vanishing, $\mathcal{Q}^{(0)} = \mathcal{Q}^{(1)} = 0$, see figure \ref{fig:one}. The condition of vanishing charges is a global property of the minimal surface and it provides the necessary global information to derive the symmetries of the minimal area. Our analysis is otherwise restricted to the local expansion of the minimal surface around the conformal boundary. Another way to view the condition of vanishing charges is explained in \cite{Kruczenski:2014bla}. The minimal surface is not determined as an initial value problem, since the coefficients $X_{(3)}^\mu$ are not determined by the boundary data. If one provides an arbitrary function for the coefficient $X_{(3)}^\mu$, the solution of the initial value problem develops a singularity and the conserved charges are non-vanishing. The condition of vanishing charges may thus in turn be used to determine the coefficients $X_{(3)}^\mu$, the higher-order terms can then be fixed from the equations of motion.  

From the vanishing of the charges, we infer that in a Laurent expansion in $\tau$ all coefficients of the charges vanish. For the $\tau$-dependent term this follows trivially from the equations of motion, the vanishing of the $\tau^0$-coefficients leads to the desired Ward identities for the strongly coupled Maldacena-Wilson loop. The evaluation of the level-zero charge $\mathcal{Q}^{(0)}$ shows that 
\begin{align}
\mathcal{Q}^{(0)} = \mathcal{Q}^{(0)} \big \vert _{\tau^0}  =  \frac{1}{2}  \int \diff s \, e^{x\cdot P} \left( \dfrac{\delta A_\mathrm{ren}(C)}{\delta x^\mu (s) } \, K^\mu \right) e^{-x\cdot P}  = 0 \,. \label{Qzero}
\end{align}
Here, we have already calculated the conjugation with $y^D$ and extracted the $\tau^0$-term. The vanishing of $\mathcal{Q}^{(0)}$ encodes the conformal invariance of the minimal area, as the conjugation of $K_\mu$ with $e^{x\cdot P}$ gives the conformal Killing vectors $\xi _a ^\mu$, 
\begin{align}
 e^{x\cdot P} \left( \half \, K^\mu \right) e^{-x\cdot P} &= \xi ^\mu _a \, \hat{T}^a = \hat{P}^\mu + x_\nu \, \hat{M}^{\nu \mu} + x^\mu \hat{D} + \left( x^2 \delta ^\mu _\nu - 2 x^\mu x_\nu \right) \hat{K}^\nu \, , \label{cosetkilling} \\
 \left \lbrace \xi_a , \xi_b \right \rbrace ^\mu &=  \xi^\rho _a  \,  \partial _\rho \, \xi^\mu _b  - \xi^\rho _b  \,  \partial _\rho \, \xi^\mu _a  = f _{a b} {} ^c  \xi^\mu _c \,.
\end{align}
Here, $\hat{T}^a = G^{a b} T_b$ denotes the dual basis to the basis used in \eqn{basis1}. We can rewrite \eqn{Qzero} as the conformal invariance of the Maldacena-Wilson loop at strong coupling,
\begin{align}
J_a ^{(0)} \left \langle W(C) \right \rangle = \int \diff s \, \xi ^\mu _a(x) \, \dfrac{\delta}{\delta x^\mu (s) } \, \left \langle W(C) \right \rangle = 0 \,.
\end{align}
The evaluation of the level-one charge gives
\begin{align}
\mathcal{Q}^{(1)} \big \vert _{\tau^0} &= \half \, f \indices{^{cb} _a} \, \hat{T}^a \int \diff s_1 \, \diff s_2 \, \varepsilon(s_1-s_2)\,  \xi ^\mu _b(x_1) \, \xi ^\nu _c(x_2) \, \dfrac{\delta A_\mathrm{ren}}{\delta x_1^\mu  } \, \dfrac{\delta A_\mathrm{ren}}{\delta x_2^\nu } \nn \\
& \quad + \int \limits _0 ^L \diff s \left \lbrace e^{X \cdot P} \left( \frac{\dx_\mu}{\tau^2} \, K^\mu + \half \left( \dx_\mu \, \ddot{x}^2 + \dddot{x}_\mu \right) K^\mu - \dx_\mu \, \ddot{x}_\nu \,  M^{\mu \nu} \right) e^{-X \cdot P} \right \rbrace _{(0)} \, ,
\end{align}
where we have abbreviated $x_i=x(s_i)$. The local term also receives contributions from boundary terms of the bi-local term in \eqn{chargedef}\footnote{More details of the derivation are provided in the discussion around \eqn{Q1step}.}. The notation $\lbrace \ldots \rbrace_{(0)}$ denotes the $\tau^0$-coefficient of the term inside the brackets as in \eqn{notationdef}. Making use of \eqn{PR1} one easily shows that
\begin{align}
\left \lbrace e^{X \cdot P} \left( \frac{\dx_\mu}{\tau^2} \, K^\mu   - \dx_\mu \, \ddot{x}_\nu \,  M^{\mu \nu} \right) e^{-X \cdot P} \right \rbrace _{(0)} = 0 \, , \label{bosoniczero}
\end{align}
and thus, by virtue of \eqn{cosetkilling}, we have 
\begin{align}
\mathcal{Q}^{(1)} \big \vert _{\tau^0} & = \half \, f \indices{^{cb} _a} \, \hat{T}^a \int \diff s_1 \, \diff s_2 \, \varepsilon(s_1-s_2)\,  \xi ^\mu _b(x_1) \, \xi ^\nu _c(x_2) \, \dfrac{\delta A_\mathrm{ren}}{\delta x_1^\mu  } \, \dfrac{\delta A_\mathrm{ren}}{\delta x_2^\nu } \nn \\ 
& \qquad + \int \limits _0 ^L \diff s \, \xi ^\mu _a(x) \left( \dx_\mu \, \ddot{x}^2 + \dddot{x}_\mu \right) \, \hat{T}^a \,.
\end{align}
The vanishing of $\mathcal{Q}^{(1)}$ can thus be written as the invariance of the Maldacena-Wilson loop under the level-1 Yangian generator\footnote{The second-derivative term appearing upon application of this generator to \eqn{strongMWL} is subleading and can be neglected for $\la \gg 1$.} 
\begin{align}
J_a ^{(1)} = f \indices{^{cb} _a} \, \int \diff s_1 \diff s_2 \, \varepsilon(s_1-s_2) \, \xi^\mu _b (x_1) \, \xi^\nu _c (x_2)  \dfrac{\delta^2}{\delta x^\mu _1 \delta x^\nu _2} + \frac{\la}{2 \pi ^2} \int \limits _0 ^L \diff s \, \xi^\mu _a (x) \left( \dx_\mu \, \ddot{x}^2 + \dddot{x}_\mu \right) \,.
\end{align}
The bilocal part of this generator shows the typical structure of a level-one Yangian symmetry generator as it is known from 2d integrable field theories or scattering amplitudes, see e.g.\ \cite{MacKay:2004tc}, \cite{Drummond:2009fd}. In \cite{Dolan:2003uh,Dolan:2004ps} it was shown that generators of this form satisfy the commutation relations of the Yangian algebra,
\begin{align}
\left[ J^{(0)} _a \, , \,  J^{(1)} _b \right] = f \indices{_{ab}^c} \, J^{(1)} _c , 
\end{align}
as well as the Serre relations, a generalized Jacobi-like identity. In appendix~\ref{app:transf}, we show that also the local term obeys the above commutation relation. There is thus strong evidence that the generators $J^{(0)} _a$ and $J^{(1)} _a$ satisfy the commutation relations of the Yangian algebra $Y(\mathfrak{so}(2,4))$.

The level-1 generators $J_a^{(1)}$ depend on the choice of a starting point along the curve $C$ due to the path-ordering in the bi-local term. Consider a curve $C$ parametrized by $x:[0,L] \to \mathbb{R}^{(1,3)}$. Instead of $x(0)$ we could equally well choose a different starting point $x(\Delta)$ and obtain a different level-1 generator $\tilde{J}_a^{(1)}$. As our above line of arguing does not distinguish a specific starting point, both of these generators give symmetries of the Maldacena-Wilson loop in the limit of large $\la$. One would then expect that the difference between the two generators also gives a symmetry. A simple calculation shows that it is given by 
\begin{align}
J_a^{(1)} - \tilde{J}_a^{(1)} = f\indices{^{cb}_a} \left( j_b ^{\Delta} \, J_c ^{(0)} - J_b^{(0)} j_c^{\Delta} \right) = f\indices{^{cb}_a} \big[j_b ^{\Delta} , J_c ^{(0)}    \big \rbrace = f\indices{^{cb}_a} \, f\indices{_{bc}^d} \,  j_d ^{\Delta} \, . \label{level1diff}
\end{align}
Here, we defined
\begin{align*}
j_a ^\Delta = \int \limits _0 ^\Delta \xi^\mu _a (x) \, \dfrac{\delta}{\delta x^\mu(s)} \, .
\end{align*}
Note however, that acting with this generator on $\langle W(C) \rangle$ leads to a term that is of order $\sqrt{\la} \langle W(C) \rangle$ and hence subleading in $\la$ compared to the action of the level-1 generators $J_a^{(1)}$ and $\tilde{J}_a^{(1)}$. We have thus not shown that the difference \eqn{level1diff} between two level-1 generators defined with respect to different starting points annihilates the Maldacena-Wilson loop for large $\la$. Indeed, this seems to be rather unlikely. For $\mathfrak{so}(2,4)$ we have $f\indices{^{cb}_a} f\indices{_{bc}^d} = N \delta ^d _a$ with $N$ a non-vanishing numerical constant, which can be extracted from computing the Killing form. In particular, since $\Delta$ is arbitrary, the functional derivative $\delta/\delta x^\mu(s)$ at any point on the loop would have to annihilate the result, which can clearly not be the case. We thus see that the Yangian invariance of the Maldacena-Wilson loop which we showed for asymptotically large $\la$ cannot extend in an expansion in $1/\sqrt{\la}$. This matches well with the finding that the Yangian over $\mathfrak{so}(2,4)$ does not provide a symmetry of the Maldacena-Wilson loop at weak coupling \cite{Muller:2013rta}. For the super-Wilson loop the situation is different as the contraction $f\indices{^{cb}_a} f\indices{_{bc}^d}$ vanishes over $\mathfrak{psu}(2,2\vert4)$.

\subsection{A Coset Description of \texorpdfstring{$S^5$}{S5}}
\label{sec:S5}
We consider a minimal surface in $S^5$ and show that the area is invariant under $SO(6)$ rotations of the sphere. This is quite a trivial exercise if one considers the sphere in embedding coordinates. We will, however, employ a coset construction based on $SU(4)$ matrices to introduce coordinates on the sphere. This parametrization is also employed to describe the $SU(4)$-part of the supercoset $SU(2,2\vert4)/\left(SO(4,1)\times SO(5) \right)$ and the results of this section will be useful in our later discussion. The main purpose of this section is thus to familiarize ourselves with the coset description of $S^5$ in terms of $SU(4)$ matrices. We note the following $\mathbb{Z}_2$ decomposition of $\mathfrak{su}(4)$:
\begin{equation}
\begin{aligned}
\mathfrak{su}(4) ^{(0)} &= \mathrm{span} \left \lbrace \gamma^{a b} = \quarter \, \left[ \gamma ^a \, , \, \gamma ^b \right]  \, , \gamma ^{a 6} = \quarter \, \left[ \gamma ^a \, , \, \gamma ^5 \right] \, \vert a,b \in   \left \lbrace 1 \, , \ldots , 4 \right \rbrace \right \rbrace \simeq \mathfrak{so}(5) \, , \\
\mathfrak{su}(4) ^{(2)} &= \mathrm{span} \left \lbrace \gamma^{a 5} = \ihalf \gamma^a \, , \gamma ^{5 6} = - \ihalf \gamma^5 \, \vert a \in   \left \lbrace 1 \, , \ldots , 4 \right \rbrace \right \rbrace \,.
\end{aligned}
\end{equation}
Here, the matrices $\lbrace \gamma^{IJ} = - \gamma^{JI} \, , I,J \in \lbrace 1, \ldots , 6 \rbrace \rbrace$ are constructed from the gamma matrices $\lbrace \gamma^1 , \ldots , \gamma^5 \rbrace$, which satisfy the $SO(5)$ Clifford algebra, see appendix~\ref{app:su4} for details. Following \cite{Arutyunov:2009ga}, we choose the following coset representatives in $SU(4)$:
\begin{align}
u(\phi , z ) = \exp{\left( \ihalf \phi \gamma^5 \right)} \, \left(1 + z^2 \right)^{-1/2} \bigg( \mathbb{I}_4 + \sum _{a=1} ^4 i\, z^a \gamma^a \bigg)  \,. \label{eqn:su4coset}
\end{align} 
The Cartan form is given by
\begin{align*}
a = -u^{-1} \diff u = - i \, \frac{(1-z^2)\, \diff \phi }{2(1+z^2)}\, \gamma ^5 - i \, \frac{\diff z^a}{1+z^2} \, \gamma ^a - \frac{z^a \diff z^b}{2(1+z^2)} \left[ \gamma ^a \, , \, \gamma ^b \right] - \frac{z^a \diff \phi}{2(1+z^2)} \left[ \gamma ^a \, , \, \gamma ^5 \right] \,.
\end{align*}
Under the $\mathbb{Z}_2$ grading of $\mathfrak{su}(4)$ this decomposes into
\begin{align*}
a^{(0)} = - \frac{z^a \diff z^b}{2(1+z^2)} \left[ \gamma ^a \, , \, \gamma ^b \right] - \frac{z^a \diff \phi}{2(1+z^2)} \left[ \gamma ^a \, , \, \gamma ^5 \right] \, , \qquad a^{(2)} = - i \, \frac{(1-z^2)\, \diff \phi }{2(1+z^2)}\, \gamma ^5 - i \, \frac{\diff z^a}{1+z^2} \, \gamma ^a \,.
\end{align*}
The metric of the coset space is thus given by
\begin{align}
\diff s^2 = \left \langle a^{(2)}  , \, a^{(2)} \right \rangle = \frac{4 \, \diff z ^2}{(1+z^2)^2} + \left( \frac{1-z^2}{1+z^2} \right)^2 \, \diff \phi ^2  = \diff N^I \, \diff N^I
\end{align}
Here, the group metric $\left \langle b , c  \right \rangle = - \tr \left(b c \right)$ is inherited from the group metric given in appendix~\ref{app:su224}. The coordinates $\left( \phi , z^i \right)$ are related\footnote{We require $z^2\leq 1$ to get a one-to-one map between the coordinates.} to the embedding coordinates of the sphere by
\begin{align}
N^a = \frac{2\, z^a}{1+z^2} \, , \qquad N^5 + i N^6 = \frac{1-z^2}{1+z^2} \, e^{i \phi} \,.
\end{align}
Consider now the following boundary conditions for the minimal surface in $S^5$:
\begin{align}
N^I (\tau = 0 , s ) = n^I(s)\quad \Rightarrow \quad  N^I (\tau , s )= n^I(s)+ \tau N^I_{(1)}(s) + \ldots \,.
\end{align}
The minimal surface minimizes the area functional
\begin{align}
A[N,h] = \frac{1}{2} \int \diff \tau \diff s  \, \gamma^{i j}   \left \langle a_i^{(2)} , \, a_j^{(2)} \right \rangle \,.
\label{S5action}
\end{align}
The Noether current of the coset model is given by $J^{S^5} _i = u A_i ^{(2)} u^{-1}$ and may be computed to be
\begin{align}
J^{S^5} _i = N^I \left( \partial_i N^J \right) \, \gamma ^{I J} \,.  \label{S5Noether}
\end{align}
The corresponding Noether charge is conserved and due to the possibility to contract the boundary curve to a point we have
\begin{align}
\mathcal{Q} = \int \diff s J^{S^5} _{\tau \, (0)}   = \int \diff s \,  n^I  N_{(1)}^J \, \gamma ^{IJ} = 0 \,. \label{S5charge}
\end{align}
In order to determine the coefficient $N_{(1)}^I$, consider a variation $\delta n^I$ of the boundary curve. This induces a variation $\delta N^I$ of the parametrization of the minimal area. Using that the parametrization of the minimal area satisfies the equations of motion, one only picks up a boundary term in computing the variation of the area and thence (we use conformal gauge)
\begin{align}
\delta A_{\mathrm{min}} =  - \int \diff s \, N_{1} ^I \delta n^I \, .
\end{align}
Due to the use of embedding coordinates $N^I$ we have $n^I \delta n^I = 0$ and we conclude that
\begin{align*}
\dfrac{\delta A_{\mathrm{min}} }{\delta n^I (s)} = - N_{1} ^I (s) + \alpha(s) n^I(s)\,.
\end{align*}
The coefficient $\alpha(s)$ is determined from the condition $n^I N_{(1)}^I = 0$ and we find
\begin{align}
N_{(1)}^I (s) = - \dfrac{\delta A_{\mathrm{min}} }{\delta n^I (s)} + \left( n^J(s) \, \dfrac{\delta A_{\mathrm{min}} }{\delta n^J (s)} \right) n^I(s) \,.
\end{align} 
Equation \eqn{S5charge} thus encodes the $SO(6)$ invariance of the minimal area,
\begin{align}
\int \diff s \left( n^I(s) \, \dfrac{\delta A_{\mathrm{min}} }{\delta n^J (s)} - n^J(s) \, \dfrac{\delta A_{\mathrm{min}} }{\delta n^I (s)} \right) = 0 \,. 
\end{align}

\section{The super Wilson Loop at strong Coupling}
\label{sec:SuperWL}
In this section we describe the supersymmetric generalization of the strong coupling description of the Maldacena-Wilson loop which is given by replacing the renormalized minimal area in $AdS_5$ by a minimal area in the supercoset space $PSU(2,2\vert4) / \left(SO(4,1) \times SO(5)\right)$, which we also renormalize appropriately. We thus describe the expectation value of the super Maldacena-Wilson loop at strong coupling by
\begin{align}
\left \langle \mathcal{W}(C) \right \rangle = e^{ -\frac{\sqrt{\la}}{2 \pi} \mathcal{A}_\mathrm{ren}(C) } \, , \label{strongsuperWL}
\end{align}
where the area functional $\mathcal{A}$ is based on the superstring action \cite{Metsaev:1998it}, which we review in section~\ref{superstring}. The boundary conditions follow from the generalized Poincar{\'e} coordinates introduced in \cite{Ooguri:2000ps}, which we discuss in section~\ref{sec:boundcond}. In section~\ref{sec:symm} we show explicitly that the divergence of the minimal area for these boundary conditions is proportional to the super-length of the curve, 
\begin{align}
\mathcal{A}_\mathrm{min}(C) \Big \vert _{y \geq \varepsilon} = \frac{\mathcal{L}(C)}{\varepsilon} + \mathcal{A}_\mathrm{ren}(C) \, , \qquad \mathcal{L}(C) = \int \diff s \lvert \pi (s) \rvert \,.
\end{align}
Here $\pi ^\mu = \dx ^\mu + i \big( \dot{\bar{\la}} \s ^\mu \la - \bar{\la} \s ^\mu \dot{\la} \big)$ describes the supermomentum of a superparticle moving along the respective contour in the boundary superspace and we have regulated the minimal area by imposing a cut-off $\varepsilon$ in the coordinate $y$ of $AdS_5$ \cite{Drukker:1999zq}.

\subsection{The Area Functional}
\label{superstring}

Following the review article \cite{Arutyunov:2009ga}, we discuss those aspects of classical type IIB superstring theory in $AdS_5 \times S^5$ that will be needed in the remainder of this paper. 
The theory can be described by a sigma model type action with target space
\begin{align*}
\frac{PSU(2,2\vert 4)}{SO(4,1) \times SO(5)} \,.
\end{align*}
For a function $g(\tau, s) \in SU(2,2\vert4)$ of the world-sheet coordinates, the Cartan form $A_i = - g^{-1} \partial_i g$ provides a flat connection, 
\begin{align}
\eps^{i j} \left( 2 \partial_i \, A_j - \left[A_i \, , \, A_j \right] \right) = 0 \, ,
\end{align}
taking values in the Lie superalgebra $\mathfrak{su}(2,2\vert4)$. This algebra may be endowed with a $\mathbb{Z}_4$-grading: 
\begin{align}
\mathfrak{su}(2,2\vert4) &= \mathfrak{g}^{(0)} \oplus \mathfrak{g}^{(2)} \oplus \mathfrak{g}^{(1)} \oplus \mathfrak{g}^{(3)} \, , \qquad \qquad  \left[ \mathfrak{g}^{(k)} \, , \, \mathfrak{g}^{(l)} \right] \subset \mathfrak{g}^{(k+l) \, \mathrm{mod} 4 }  \,.
\end{align}
Here, $\mathfrak{g}^{(0)} \oplus \mathfrak{g}^{(2)}$ is the bosonic subalgebra of $\mathfrak{su}(2,2\vert4)$ and $\mathfrak{g}^{(1)}\oplus \mathfrak{g}^{(3)}$ comprises the fermionic generators. Based on the projection operators $P^{(k)}: \mathfrak{su}(2,2\vert4) \to \mathfrak{g}^{(k)}$ onto these graded components, we introduce the short-hand notation
\begin{align}
B^{(k)} = P^{(k)} \left( B \right) \, , \qquad B^{(1)\pm (3)} = B^{(1)} \pm B^{(3)} \,.
\end{align}
More details on the $\mathbb{Z}_4$ decomposition of $\mathfrak{su}(2,2\vert4)$ can be found in appendix~\ref{app:su224}. There we also introduce a metric $\langle \cdot , \cdot \rangle$ on the algebra based on the supertrace in the fundamental representation. This metric is non-degenerate on $\mathfrak{psu}(2,2\vert4)$, which is obtained from $\mathfrak{su}(2,2\vert4)$ by projecting out the central element $C$. The area functional\footnote{The area functional differs from the superstring action by a factor of $\frac{\sqrt{\la}}{2 \pi}$ which appears explicitly in \eqn{strongsuperWL}} can then be written as
\begin{align}
\mathcal{A} = \frac{1}{2}  \int \diff \tau \, \diff s \left \lbrace \gamma ^{i j} \,  \left \langle A_i ^{(2)} , A_j ^{(2)} \right \rangle + i \, \tilde{\kappa} \, \eps^{ij}  \left \langle A_i ^ {(1)} ,  A_j ^ {(3)} \right \rangle \right \rbrace \,. \label{action}
\end{align}
Here, $\tilde{\kappa}$ is a numerical constant, $\gamma^{i j} = \sqrt{\det(h_{ij})} \, h^{ij}$ is manifestly Weyl-invariant and we fix the convention $\eps^{\tau s} = 1$. Note also, that we work with a Euclidean world-sheet metric -- resulting in the factor of $i$ in front of the fermionic term -- since the induced metric on the world-sheet is Euclidean for the boundary conditions we consider. The variations with respect to $g$ and $\gamma$ give the equations of motion\footnote{The equations of motion are written over $\mathfrak{psu}(2,2\vert 4)$. Over $\mathfrak{su}(2,2\vert 4)$ one would have $\partial_i \, \Lambda ^i - \left[ A_i \, , \, \Lambda ^i \right] = \rho \, C $.} and the Virasoro constraints, 
\begin{equation}
\begin{aligned}
0&= \partial_i \, \Lambda ^i - \left[ A_i \, , \, \Lambda ^i \right] \, , \qquad \qquad  \left \langle A_i ^{(2)} , A_j ^{(2)} \right \rangle - \half \gamma _{i j} \, \gamma ^{k l}  \left \langle A_k ^{(2)} , A_l ^{(2)} \right \rangle = 0 \, , \\
\Lambda ^i &=  \gamma ^{i j} \, A_j ^ {(2)} - \sfrac{i}{2} \, \tilde{\kappa} \, \eps^{i j} A_j^{(1)-(3)} \,. \label{defLambda}
\end{aligned}
\end{equation}
The equations of motion may be split into purely bosonic or fermionic equations, which are given by
\begin{align}
\partial _i \left( \gamma^{ij} A_j ^{(2)} \right) - \gamma ^{i j} \left[ A_i^{(0)} \, , \, A_j^{(2)} \right] + \sfrac{i}{2} \, \tilde{\kappa} \, \eps^{ij} \left[ A_i^{(1)-(3)} \, , \, A_j^{(1)+(3)} \right] &= 0 \, , \label{boseom}\\
\gamma ^{ij} \left[ A_i ^{(2)} \, , A_j^{(1)+(3)} \right] + i \, \tilde{\kappa} \, \eps^{i j} \left[ A_i ^{(2)} \, , A_j^{(1)-(3)} \right] &= 0 \, , \label{fermeom}
\end{align}
For $\tilde{\kappa} = \pm 1$, the action \eqn{action} exhibits kappa symmetry and we will set $\tilde{\kappa} =1$ from now on. The respective transformations are given by
\begin{align}
g \mapsto g^\prime \sim g \cdot \exp{\kappa} \, , \qquad \delta_\kappa \, A = -\mathrm{d}\kappa + \left[A \, , \, \kappa\right] + h^{(0)} \,. 
\end{align}
The kappa symmetry transformations also include a variation of the world-sheet metric, which we do not specify as we will fix conformal gauge for the world-sheet metric. The transformations constitute a local gauge symmetry of the action provided that the supermatrix $\kappa = \kappa^{(1)+(3)}$ is given by
\begin{align}
\kappa^{(1)} &= A_{i , -} ^{(2)} \, \mathcal{K} _+ ^{(1),i} + \mathcal{K} _+ ^{(1),i} \, A_{i , -} ^{(2)} \, ,  \qquad \qquad
\kappa^{(3)} = A_{i , +} ^{(2)} \, \mathcal{K}_- ^{(3),i} + \mathcal{K} _- ^{(3),i} \, A_{i , +} ^{(2)} \, , \quad \text{where} \label{kappatransform} \\
V_{\pm} ^i &= P_{\pm} ^{i j} \, V_j = \half \left( \gamma^{ij} \pm i \, \eps^{i j} \right) V_j \,. \nn
\end{align}
Here $\mathcal{K}$ denotes an arbitrary fermionic one form taking values in $\mathfrak{su}(2,2\vert4)$ and $\mathcal{K}^{(1)}$, $\mathcal{K}^{(3)}$ are its projections in the algebra. The string model is classically integrable\cite{Bena:2003wd} and a Lax connection can be parametrized\cite{Zarembo:2010yz} as
\begin{align}
L_i = A_i ^{(0)} + \frac{{\tt{x}}^2+1}{{\tt{x}}^2-1} \, A_i^{(2)} - \frac{2i \, {\tt{x}}}{{\tt{x}}^2 -1} \, \gamma_{ij} \, \eps^{jk} \, A_k ^{(2)} + \frac{\sqrt{{\tt{x}}+1}}{\sqrt{{\tt{x}}-1}} \, A_i ^{(1)} + \frac{\sqrt{{\tt{x}}-1}}{\sqrt{{\tt{x}}+1}} \, A_i ^{(3)} \,.
\end{align}
Here, $\tt{x} \in \mathbb{C}$ denotes the spectral parameter. The flatness of the Lax connection, 
\begin{align}
\eps^{ij} \left(2 \partial_i L_j - \left[L_i \, , \, L_j \right] \right)= 0 \, ,
\end{align}
is equivalent to the equations of motion. We consider the gauge transformed Lax connection
\begin{align}
l_i &= g L_i g^{-1} + \left(\partial_i g \right) g^{-1} = g \left( L_i - A_i \right) g^{-1} \notag \\
&= \left( \frac{{\tt{x}}^2+1}{{\tt{x}}^2-1}  -1 \right) a_i^{(2)} - \frac{2i \, {\tt{x}}}{{\tt{x}}^2 -1} \, \gamma_{ij} \, \eps^{jk} \, a_k ^{(2)} + \left( \frac{\sqrt{ {\tt{x}}+1}}{\sqrt{{\tt{x}}-1}} -1 \right) a_i ^{(1)} + \left( \frac{\sqrt{{\tt{x}}-1}}{\sqrt{{\tt{x}}+1}} -1 \right) a_i ^{(3)} \, ,
\end{align}
where we defined $a_i ^{(k)} = g \, A_i ^{(k)} \, g^{-1}$. A tower of multi-local conserved charges can be extracted from expanding the monodromy matrix \cite{Beisert:2005bm} associated to the gauge transformed Lax connection around ${\tt{x}}=\infty$. Concretely, in conformal gauge and for ${\tt{z}}= {\tt{x}}^{-1}$ we have
\begin{align*}
l_s =  2i{\tt{z}} \, J_\tau + 2 {\tt{z}}^2 \left( a_s ^{(2)} + \quarter a_s^{(1)+(3)} \right) + \mathcal{O}({\tt{z}}^3) \, .
\end{align*} 
Here, $J_i = g \Lambda _i g^{-1}$ is the Noether current corresponding to the global $PSU(2,2\vert4)$ symmetry of the action \eqn{action}. The expansion of $T[{\tt{z}}]= \mathcal{P}\exp{\left(\int \diff s \,  l_s \right)}$ around ${\tt{z}}=0$ leads to the conserved charges 
\begin{align}
\mathcal{Q}^{(0)} &= \oint \diff s \, J_\tau \, , \\
\mathcal{Q}^{(1)} &=  \frac{1}{2}  \int \diff s_1 \diff s_2 \, \varepsilon(s_1-s_2) \, \left[ J_\tau(s_1) \, , \, J_\tau (s_2) \right] - \oint \diff s \left( a_s ^{(2)} + \quarter a_s ^{(1)+(3)} \right) \,.
\end{align}
Here we have subtracted a term proportional to $(\mathcal{Q}^{(0)})^2$ from $\mathcal{Q}^{(1)}$ in order to reach a more convenient form. Note that the local term of the charge $\mathcal{Q}^{(1)}$ takes a different form than in section~\ref{sec:bosAdS}, since the Noether current for the full superstring is not flat. 

These charges are conserved by construction, but it is still instructive to show their conservation explicitly. The conservation of the Noether charge $\mathcal{Q}^{(0)}$ follows easily from the conservation of the Noether current and the periodicity of the boundary curve. Moreover, since the curve is contractible on the minimal surface, it follows that 
\begin{align}
\mathcal{Q}^{(0)}=0 \, , \label{Q00}
\end{align}
which is crucial for the conservation of $\mathcal{Q}^{(1)}$:
\begin{align}
\partial_\tau \, \mathcal{Q}^{(1)} =  \left[ \mathcal{Q}^{(0)} \, , \, J_s(L) + J_s(0) \right] = 0 \,.
\end{align}
Here, $0$ and $L$ are the end-points of the parametrization and we used that
\begin{align}
\partial _\tau \left( a_s ^{(2)} + \quarter a_s^{(1)+(3)} \right) = \partial _s \left( a_\tau ^{(2)} + \quarter a_\tau^{(1)+(3)} \right) - 2 \left[ J_\tau \, , \, J_s \right] \, ,
\end{align}
which one may obtain from the expanding the flatness condition for $l_i$ around ${\tt{z}}=0$. Using once more that the curve can be contracted on the minimal surface, we find that
\begin{align}
\mathcal{Q}^{(1)} = 0 \,.
\end{align}
This condition as well as the condition \eqn{Q00} allow to derive the local and non-local symmetries of the super Wilson loop.

\subsection{The Boundary Conditions}
\label{sec:boundcond}
In the case of the Maldacena-Wilson loop the minimal surface is required to end on the respective curve on the conformal boundary of $AdS_5$. The notion of the conformal boundary is easiest to picture in Poincar{\'e} coordinates, where it is simply given by\footnote{Since Poincar{\'e} coordinates do not cover $AdS_5$ completely, the topological properties of the conformal boundary cannot be inferred from this picture, but this is not our concern here.} the Minkowski space at $y=0$. The boundary conditions for the super Wilson loop follow from a generalization of Poincar{\'e} coordinates to the coset superspace $PSU(2,2\vert4)/\left(SO(4,1) \times SO(5) \right)$, which has been introduced in \cite{Ooguri:2000ps}. There are two important aspects that need to be considered in the construction of the conformal boundary of our superspace. The geometric relation between bulk and boundary space requires that the super-isometries of the bulk space should reduce to superconformal transformations on the conformal boundary space when taking the boundary limit. Moreover, we should impose the right number of boundary conditions on the bulk coordinates in order to determine a minimal surface. For the bosonic coordinates, the equations of motion are second order differential equations and we impose a boundary condition for all bosonic coordinates. For the fermionic coordinates, the equations of motion are first order differential equations and we will thus only impose boundary conditions on half of the fermionic coordinates.

Let us first consider the $AdS_5$ part once more. As we have seen in section~\ref{sec:bosAdS}, the coset parametrization 
\begin{align}
g_1(X,y) = e^{X \cdot P} \, y^D
\end{align}
provides Poincar{\'e} coordinates on $AdS_5 \simeq SO(4,2) / SO(4,1)$. In this case, we split the algebra $\mathfrak{so}(4,2)$ according to
\begin{align*}
\mathfrak{so}(4,2) = \left( \mathfrak{g}^{(0)} \simeq \mathfrak{so}(4,1) \right) \oplus \mathfrak{g}^{(2)} \, , \quad \mathfrak{g}^{(0)} &= \mathrm{span} \left \lbrace P_\mu - K_\mu , M_{\mu \nu} \right \rbrace \, , \quad  \mathfrak{g}^{(2)} = \mathrm{span} \left \lbrace P_\mu + K_\mu , D \right \rbrace \,.
\end{align*}
The boundary Minkowski space can be thought of as the coset space $SO(4,2)/H$, where $H$ is the subgroup of $SO(4,2)$ generated by the subalgebra $\mathfrak{h} = \mathrm{span}\lbrace K_\mu , M_{\mu \nu}, D \rbrace$. The algebra $\mathfrak{so}(4,2)$ is hence split according to
\begin{align*}
\mathfrak{so}(4,2) = \mathrm{span}\lbrace K_\mu , M_{\mu \nu}, D \rbrace \oplus \mathrm{span} \lbrace P_\mu \rbrace \, ,
\end{align*}
and we parametrize the coset space by 
\begin{align}
g_2(x) = e^{x \cdot P} \,.
\end{align}
An isometry or conformal transformation on the coset spaces is obtained from left-multiplication with a generic group element $t = e^{\bm{t}}$,
\begin{align*}
t \cdot g_1(X,y) = g_1(X^\prime , y ^\prime) \cdot h(X,y) \, , \qquad   t \cdot g_2(x) = g_2(x^\prime) \cdot h(x) \, ,
\end{align*} 
where $h = e^{ \bm{h}}$ are compensating gauge transformations. Infinitesimally, we then have
\begin{equation}
\begin{aligned}
\delta g_1(X,y) &= \bm{t} \cdot g_1 - g_1 \cdot \bm{h} (X,y) = \partial_\mu g_1 \, \delta_{\bm{t}} X^\mu + \partial_y g_1 \, \delta_{\bm{t}} y \, , \\
\delta g_2(x) &= \bm{t} \cdot g_2 - g_2 \cdot \bm{h} (x) = \partial_\mu g_2 \, \delta_{\bm{t}} x^\mu  \,.
\end{aligned}
\end{equation}
Consider now for example a transformation parametrized by $\bm{t}= \varepsilon^\mu K_\mu$. One computes easily that the coordinates of the two coset spaces transform as
\begin{equation}
\begin{aligned}
\delta _{\varepsilon \cdot K } X^\mu &= X^2 \, \varepsilon^\mu - 2 \left( \varepsilon \cdot X \right) X^\mu + y^2 \, \varepsilon^\mu \, , \qquad \delta _{\varepsilon \cdot K }  y = -2y \left( \varepsilon \cdot X \right) \, ,  \\
\delta _{\varepsilon \cdot K }  x^\mu &= x^2 \, \varepsilon^\mu - 2 \left( \varepsilon \cdot x \right) x^\mu \,.
\end{aligned}
\end{equation}
We recognize that the transformations indeed agree in the boundary limit $y \to 0$ if we identify the $X$ coordinates. Before turning to the superspace, we reformulate the criterion, that the isometries of the bulk space reduce to conformal transformations on the boundary space, for a more general situation.

Assume we have two cosets $C_1 = G / H_1$ and $C_2 = G / H_2$ with the same group $G$ but different stability groups $H_1$ and $H_2$. Correspondingly we have two decompositions of the Lie algebra $\mathfrak{g}$,
\begin{align}
\mathfrak{g} = \mathfrak{h}_1 \oplus \mathfrak{f}_1 \, , \qquad \mathfrak{g} = \mathfrak{h}_2 \oplus \mathfrak{f}_2 \,.
\end{align}
Let the coset space $C_1$ represent the bulk space and $C_2$ the boundary space. Then we have $\mathrm{dim}(H_1) < \mathrm{dim}(H_2)$, but not necessarily that $H_1$ is a subset of $H_2$. Furthermore, we assume that the cosets can be parametrized by $g_1(x^m, y^i) \in C_1$ and $g_2(x^m) \in C_2$ in such a way that they realize
\begin{align}
	\label{eqn:ReprRelation}
	g_1(x, y) = g_2(x) \, h_2(x, y) \; ,
\end{align}
where $h_2 \in H_2$. In particular, the Cartan forms for the two parametrizations are related by
\begin{align}
A^1 = - g_1^{-1} \diff g_1 = h_2 A^2 h_2 ^{-1} - h_2 ^{-1} \diff h_2 \,.  
\end{align}
Under the left action of some group element $t = e^{\bm{t}} \in G$ a coset representative
transforms as
\begin{align}
	g(Z) \mapsto g(Z') = t g(Z) h(Z) \, , \qquad \delta g(Z) = \bm{t} g(Z) - g(Z) \bm{h}(Z) \,.
	\label{gtransf}
\end{align}
This transformation can be related to the Cartan form,
\begin{align}
	\label{eqn:mcCosetTrans}
	\delta Z^M A_M = - \delta Z^M g(Z)^{-1} \partial_M g(Z) = - g(Z)^{-1} \left (
		\bm{t} g(Z) - g(Z) \bm{h}(Z)
	\right) \;.
\end{align}
Splitting the left hand side of this formula into the coordinates on the first coset $Z^M = (x^m, y^i)$ yields
\begin{align}
\delta_1 Z^M A_M^1 =
	\delta_1 x^m h_2^{-1} A_m^2 h_2
	- \delta_1 x^m h_2^{-1} \partial_m h_2
	- \delta_1 y^i h_2^{-1} \partial_i h_2 \, ,
\end{align}
which can be resolved to
\begin{align}
\delta_1 x^m A_m^2 =
	\delta_1 Z^M h_2 A_M^1 h_2^{-1}
	+ \delta_1 x^m \left ( \partial_m h_2 \right ) h_2^{-1}
	+ \delta_1 y^i \left ( \partial_i h_2 \right ) h_2^{-1} \;.
\end{align}
Plugging in (\ref{eqn:mcCosetTrans}) leads to
\begin{align}
\delta_1 x^m A_m^2 &=
	- h_2 \, g_1^{-1} \left (
		\bm{t} g_1 - g_1 \bm{h}_1
	\right)h_2^{-1}
	+ \delta_1 x^m \left ( \partial_m h_2 \right ) h_2^{-1}
	+ \delta_1 y^i \left ( \partial_i h_2 \right ) h_2^{-1} \nn \\
&= - g_2^{-1} \bm{t} g_2 + h_2 \bm{h}_1 h_2^{-1}
	+ \delta_1 x^m \left ( \partial_m h_2 \right ) h_2^{-1}
	+ \delta_1 y^i \left ( \partial_i h_2 \right ) h_2^{-1} \;.
\end{align}
Evaluating (\ref{eqn:mcCosetTrans}) again for the second coset gives
\begin{align}
\delta_2 x^m A^2_m = -g_2^{-1} \bm{t} g_2 + \bm{h}_2
\end{align}
and we find the difference between the variations of the coordinates:
\begin{align}
\Delta \left (\delta x^m \right ) A^2_m
&= \left( \delta_1 x^m - \delta_2 x^m \right) A^2_m \nn \\
&= h_2 \bm{h}_1 h_2^{-1} - \bm{h}_2
	+ \delta_1 x^m \left ( \partial_m h_2 \right ) h_2^{-1}
	+ \delta_1 y^i \left ( \partial_i h_2 \right ) h_2^{-1} \;.
\end{align}
The interesting part of the above equation is contained in the projection on the subalgebra $\mathfrak{f}_2$:
\begin{align}
\Delta \left (\delta x^m \right ) A^2_m \Big \vert _{\mathfrak{f}_2} =  \Delta \left (\delta x^m \right ) E^{2} {} _m {} ^a \, T_a = h_2 \bm{h}_1 h_2^{-1} \Big \vert _{\mathfrak{f}_2}
\end{align}
Here, $T_a$ denotes a basis of $\mathfrak{f}_2$ and  $E^{2} {} _m {} ^a$ are the vielbein components associated to $A^2$, which form an invertible matrix. We thus conclude that 
\begin{align}
\Delta \left (\delta x^m \right ) \xrightarrow{\scriptscriptstyle y \to 0} 0 \quad \Leftrightarrow \quad h_2 \bm{h}_1 h_2^{-1} \Big \vert _{\mathfrak{f}_2} \xrightarrow{\scriptscriptstyle y \to 0}  0 \,. \label{crit1}
\end{align}
To understand this formulation of the boundary criterion better, let us consider $AdS_5$ once more. The Cartan form for the bulk is given by 
\begin{align}
A^1 = - g_1 ^{-1} \diff g_1 = - \frac{\diff X^\mu}{y} \, P_\mu - \frac{\diff y}{y} \, D \,.
\end{align}
Thus evaluating \eqn{eqn:mcCosetTrans} for $\bm{t} = \varepsilon \cdot K$ yields
\begin{align}
\delta _{\varepsilon \cdot K} X^M \, A^1 _M &= - \frac{1}{y} \left( \delta_{\varepsilon \cdot K}  X^\mu \, P_\mu + \delta_{\varepsilon \cdot K} y \, D \right) = - y^{-D} e^{- X \cdot P} \left ( \varepsilon \cdot K \right ) e^{X \cdot P} y^D + \bm{h}_{1} \nn \\
&= \frac{1}{y} \left( X^2 \, \varepsilon^\mu - 2 \left( \varepsilon \cdot X \right) X^\mu \right) P_\mu - 2 \left( \varepsilon \cdot X \right) D - 2 \, \varepsilon^\mu x^\nu M_{\mu \nu} + y\,  \varepsilon^\mu K_\mu + \bm{h}_{1}
\end{align}
We read off that
\begin{align}
\bm{h}_{1} = 2 \, \varepsilon^\mu x^\nu M_{\mu \nu} - y \, \varepsilon^\mu \left(K_\mu -P_\mu \right)
\end{align}
in order to cancel the contributions proportional to $K_\mu$ and $M_{\mu \nu}$. Noting that in our case $\mathfrak{f}_2 = \mathrm{span}\lbrace P_\mu \rbrace$ we find
\begin{align}
h_2 \bm{h}_1 h_2^{-1} \Big \vert _{\mathfrak{f}_2} = y^D \left(2 \, \varepsilon^\mu x^\nu M_{\mu \nu} - y \, \varepsilon^\mu \left(K_\mu -P_\mu \right) \right) y^{-D} \Big \vert _{P} = y^2 \varepsilon^\mu P_\mu \, ,
\end{align}
which vanishes for $y \to 0$ as it should be.

We now turn to the discussion of the full supercoset $PSU(2,2\vert 4)/\left(SO(4,1) \times SO(5) \right)$. The authors of \cite{Ooguri:2000ps} suggest the following coset parametrization:
\begin{align}
g(X,N,y,\theta, \vartheta) = e^{ X \cdot P } \, e^{\theta _\a {} ^A \, Q_A {}^\a + \btheta_{A \da} \, \Qb^{\da A} } \, e^{\vart _A {} ^\a \, S_\a {} ^A + \bvart^{\da A} \, \Sb_{A \da} } \, U(N) \, y^D \,. \label{cosetrep}
\end{align}
Here $(X,y)$ and $N$ are bosonic coordinates parametrizing the $AdS_5$ and the $S^5$ part, respectively. The 32 fermionic degrees of freedom are parametrized by the Grassmann odd coordinates $\theta, \btheta, \vart$ and $\bvart$. The coset representative for the spherical coordinates is given in the same way as in (\ref{eqn:su4coset}) but written as a $\left(4 \vert 4 \right)$ supermatrix, 
\begin{align}
U(N) = \begin{pmatrix}
\mathbb{I}_4 & 0 \\ 0 & u(N) 
\end{pmatrix} \,.
\end{align}
This specific choice of coset parametrization is not at all arbitrary. A crucial aspect is that all exponents have definite weight and that they are ordered by these weights. Moreover, the $y$-coordinate which vanishes on the conformal boundary is associated to the dilatation generator and is put to the right of the coset representative. We shall see below, why these are important aspects in the choice of the coset representative \eqn{cosetrep}. 

\subsubsection{The Cartan Form}
\label{sec:Cartanform}
The Cartan form for the coset representative \eqn{cosetrep} has been partially derived in \cite{Ooguri:2000ps}, as this provides some insight into the geometry of the space, which is also helpful in the discussion of the conformal boundary space. Since we will need the complete Cartan form in order to determine the expansion of the minimal surface into the bulk space, we provide a full derivation. The calculation is lengthy but straightforward. Abbreviating
\begin{align}
\Omega := \theta _\a {} ^A \, Q_A {}^\a + \btheta_{A \da} \, \Qb^{\da A}  \, , \qquad \eta := \vart _A {} ^\a \, S_\a {} ^A + \bvart^{\da A} \, \Sb_{A \da}
\end{align}  
we need to compute
\begin{align*}
g^{-1} \diff g = y^{-D} \, U^{-1} \, e^{-\eta} \left( \diff X \cdot P + e^{-\Omega} \diff e^ {\Omega} \right) e^\eta \, U \, y^D + y^{-D} \, U^{-1} \left( e^{-\eta}  \diff e^\eta \right) U \, y^D + U^{-1} \diff U + \frac{\diff y}{y} D \,.
\end{align*}
Noting that $e^{-\Omega} \diff e^ {\Omega} = \diff \Omega + i \left(\theta _\a {} ^A \, \diff \btheta _{A \da} - \diff \theta _\a {} ^A \, \btheta _{A \da} \right) P^{\da \a} $ everything that is left to do are conjugations. Most of them can be done via the formula
\begin{align*}
e^{A} \, B \, e^{-A} = \sum \limits _{n=0} ^\infty \frac{1}{n!} \, \left[ A  \, ,  B \right]_{(n)} \, , \qquad \left[ A \, ,  B \right]_{(n)} = \big[A \, ,  \left[ A \, ,  B \right]_{(n-1)} \big] \, , \qquad \left[ A \, ,  B \right]_{(0)} = B \,.
\end{align*}
Here it is crucial that the exponents $X\cdot P$, $\Omega$ and $\eta$ have a definite non-zero weight. Since the weights are additive,
\begin{align*}
\big[ D , A \big] = \Delta _A   A \, , \qquad \big[ D , B \big] = \Delta _B   B \quad \Rightarrow \quad \big[ D , \big[ A , B \big] \big] = \left( \Delta _A + \Delta_B \right) \big[ A , B \big] 
\end{align*}
the expansion above breaks off after at most four orders, if the generator $A$ has a definite non-zero weight. In order to do the conjugations with $U(N)$, one can consider the supermatrices explicitly using the definitions given in appendix~\ref{app:su224}, e.g.:
\begin{align}
U^{-1} \, \left( \theta  _\a {}^A \, Q _A {} ^\a \right) U &=   \begin{pmatrix} \mathbb{I}_2 & 0 & 0 \\ 0 & \mathbb{I}_2 & 0 \\ 0 & 0 & u^{-1} \end{pmatrix}  \begin{pmatrix} 0 & 0 & 2 \, \theta \\ 0 & 0 & 0 \\ 0 & 0 & 0 \end{pmatrix} \begin{pmatrix} \mathbb{I}_2 & 0 & 0 \\ 0 & \mathbb{I}_2 & 0 \\ 0 & 0 & u \end{pmatrix} \nn \\
&=  \begin{pmatrix} 0 & 0 & 2 \, \theta u \\ 0 & 0 & 0 \\ 0 & 0 & 0 \end{pmatrix} = \left( \theta  _\a {}^B u \indices{_B ^A} \right) Q _A {} ^\a = \left( \theta u \right)_\a {}^A  \,  Q _A {} ^\a 
\end{align}
Similarly, one finds:
\begin{equation}
\begin{alignedat}{2}
U^{-1}  \left( \bvart^{\da A} \, \Sb_{A \da} \right) U & = \left(\bvart u  \right)^{\da A}  \Sb_{A \da} \, , & \qquad U^{-1} \left( \btheta_{A \da} \, \Qb^{\da A}  \right) U &= \left( u^{-1}  \btheta \right)_{A \da}  \Qb^{A \da} \, ,  \\
U^{-1} \left( \vart _A {} ^\a \, S_\a {} ^A  \right) U &= \left( u^{-1}  \vart \right) _A {} ^\a  S_\a {} ^A \, , & \qquad
U^{-1} \left(  \Lambda \indices{_A ^B} R \indices{ ^A _B} \right) U &= \left( u^{-1} \Lambda  u \right)\indices{_A ^B}  R \indices{ ^A _B} \,.
\end{alignedat}
\end{equation}
The conjugations with $y^D$ follow from the weights of the generators,
\begin{align*}
\left[ D , T ^\Delta \right] = \Delta \, T^\Delta \quad \Rightarrow y^{-D} \, T^\Delta \, y^D = y^{-\Delta} \, T^\Delta \,.
\end{align*}
These findings allow to do all of the conjugations and we arrive at the following result:
\begin{align}
A = - g^{-1} \, \diff g &= \frac{r_{\a \da} }{2 y} \, P^{\da \a} - \frac{\sigma}{y} D - \frac{1}{\sqrt{y}} \left( \varepsilon _\a {}^A \, Q_A {} ^\a + \bar{\varepsilon}_{A \da} \, \Qb^{\da A} \right) - \la \indices{_\a ^\b} \, M \indices{_\b ^\a} - \bar{\la}  \indices{^\db _\da} \, \bar{M} \indices{ ^\da _\db}  \nn \\
& \qquad - \Lambda \indices{_A ^B} R \indices{^A _B} - \gamma \, C - \sqrt{y} \left( \chi_A {}^\a \, S_\a {} ^A + \bar{\chi}^{\da A} \, \Sb_{A \da} \right) + \frac{y}{2}\, \kappa ^{\da \a}   \, K_{\a \da} - U^{-1} \diff U \label{eqn:MaurerCartanForm}
\end{align} 
Here, we defined:
\begin{equation}
\begin{alignedat}{2}
r_{\a \da} &=  \diff X_{\a \da} + 2 i \left( \diff \theta _\a {} ^A \, \btheta _{A \da} - \theta _\a {}^A \, \diff \btheta _{A \da}      \right) \, & \qquad r_\mu &=  \diff X_{\mu} - i \, \tr \left( \diff \btheta \s_\mu \theta  -   \btheta \s_\mu \diff  \theta \right) \\
\zeta _\a {} ^A &= i \, r_{\a \da} \, \bar{\vartheta}^{\da A} \, & \qquad \bar{\zeta}_{A \da} &= - i \, \vartheta _A {} ^\a \, r_{\a \da} \\
\varepsilon _\a {}^A &= \left( \diff \theta + \zeta \right) _\a {} ^B \, u(N) \indices{_B  ^A} \, & \qquad \bar{\varepsilon}_{A \da} &= u^{-1}(N) \indices{_A ^ B} \left( \diff \btheta + \bar{\zeta} \right)_{B \da} \\
\la \indices{_\a ^\b} &= - i \, \left(2 \diff \theta + \zeta \right) _\a {} ^A \, \vart _A {} ^\b \, & \qquad \bar{\la} \indices{^\db _\da} &= i \, \bvart ^{\db A} \left(2 \diff \btheta +  \bar{\zeta} \right)_{A \da} \\ 
\sigma &= \diff y + 2 y\,  \tr \left( \diff \btheta \bvart + \vart \diff \theta \right) \, & \qquad \gamma &=  \tr \left( \vart \left(2 \diff\theta + \zeta \right) - \left(2 \diff\btheta + \bar{\zeta} \right) \bvart \right) ,
\end{alignedat} 
\end{equation}
The remaining terms are given by:
\begin{equation}
\begin{aligned}
\Lambda _A {} ^B &=
	\left[ u^{-1} \left(
		\vart \left(\diff\theta + \half \zeta\right)
		- \left(\diff\btheta + \half \bar{\zeta} \right) \bvart
	\right) u \right]_A {} ^B  \\
\chi _A {}^\a &=
	\left[ u^{-1} \left[
		\left(
			4 \vart \left(\diff\theta + \third \zeta \right)
			- 2 \left(\diff\btheta + \third \bar{\zeta} \right) \bvart
		\right) \vart
		+ \diff\vart
	\right] \right] _A {}^\a \\
\bar{\chi} ^{\da A} &=
	\left[ \left[
		\diff\bvart
		+ \bvart \left(
			4 \left(\diff\btheta + \third \bar{\zeta} \right) \bvart
			- 2 \vart \left(\diff\theta + \third \zeta \right)
		\right)
	\right] u \right] ^{\da A} \\
\kappa ^{\da \a} &=
	\left[ - 8 i \, \bvart \left[
		\left( \diff\btheta + \quarter \bar{\zeta} \right) \bvart
		- \vart \left( \diff\theta + \quarter \zeta \right)
	\right] \vart
	- 2 i \left( \diff\bvart \, \vart - \bvart \, \diff\vart \right) \right] ^{\da \a}
\end{aligned}
\end{equation}
In the above expression the coordinates are viewed as matrices of different size and the relation between matrix multiplication and the contraction of indices can be understood from the fixed index positions, see also appendix~\ref{app:spinorconv}. For our further calculations we note the $\mathbb{Z}_4$-decomposition of $A= - g^{-1} \diff g$:
\begin{align}
A^{(0)} &= \frac{r_{\a \da} - y^2 \, \kappa_{\a \da}}{4y}   \left( P^{\da \a} - K^{\da \a} \right) - \la \indices{_\a ^\b} \, M \indices{_\b ^\a} - \bar{\la}  \indices{^\db _\da} \, \bar{M} \indices{ ^\da _\db} -  \left(\Lambda \indices{_A ^B} R \indices{^A _B}  + U^{-1} \diff U   \right)^{(0)}   \\
A^{(2)} &=  \frac{r_{\a \da} + y^2 \,  \kappa_{\a \da}}{4y} \left( P^{\da \a} + K^{\da \a} \right)  - \frac{\s}{y} D  - \gamma \, C  -  \left(\Lambda \indices{_A ^B} R \indices{^A _B}  + U^{-1} \diff U   \right)^{(2)}  \label{A2} \\ 
A^{(1)+(3)} &= - \frac{1}{\sqrt{y}} \left( \varepsilon _\a {}^A \, Q_A {} ^\a + \bar{\varepsilon}_{A \da} \,  \Qb^{\da A}  + y \left(  \chi_A {}^\a \, S_\a {} ^A + \bar{\chi}^{\da A} \, \Sb_{A \da} \right) \right) \\
A^{(1)-(3)} &= - \frac{i}{\sqrt{y}} \left( \varepsilon_\a {} ^A  \, K_{A B}  \, S^{\a B} -  \bar{\varepsilon}_{A \da} \,  K^{A B} \, \Sb^{\, \da} {} _B   +y \left(  \chi _A {} ^\a \, K^{A B} \, Q _{B \a} -  \bar{\chi}^{A \da} \, K_{A B} \, \Qb^B {} _\da \,  \right) \right)
\end{align}
The matrix $K = \big( K^{A B} \big) = \big( K_{A B} \big)$ appears explicitly in the $\mathbb{Z}_4$ decomposition introduced in appendix~\ref{app:su224} and is given by
\begin{align}
K = \begin{pmatrix}
0 & -1 & 0 & 0 \\ 1 & 0 & 0 & 0 \\ 0 & 0 & 0 & -1 \\ 0 & 0 & 1 & 0
\end{pmatrix} \,.
\end{align}

\subsubsection{The Conformal Boundary}
\label{sec:ConfBound}

We now turn to the conformal boundary space, which is given by the supercoset $PSU(2,2 \vert 4) / H_2$, where $H_2$ is the subgroup generated by the subalgebra
\begin{align}
\mathfrak{h}_2 = \mathrm{span} \left \lbrace M_{\mu \nu}, D ,  K_\mu , S_\a {} ^A , \Sb_{A \da} \right \rbrace \oplus \mathfrak{so}(5)
\end{align} 
A suitable coset representative \cite{Ooguri:2000ps} is given by
\begin{align}
g_2(x,\theta,N) = e^{ X \cdot P } \, e^{\theta _\a {} ^A \, Q_A {}^\a + \btheta_{A \da} \, \Qb^{\da A} } \, U(N) \,.
\label{boundarycoset}
\end{align}
The boundary superspace has only half as many fermionic degrees of freedom as the bulk space, which is due to the fermionic part of the superstring equations of motion being first order differential equations. In section~\ref{sec:symm}, we will see explicitly that the respective boundary conditions determine the minimal surface in the bulk space. 

We now apply the criterion \eqn{crit1} to show that the bulk isometries reduce to superconformal transformations on the boundary space. The discussion follows \cite{Ooguri:2000ps}. The relation between the coset representatives of bulk and boundary space is given by $g=g_2 h_2$, where
\begin{align}
h_2 = U(N)^{-1} \, e^{\vart _A {} ^\a \, S_\a {} ^A + \bvart^{\da A} \, \Sb_{A \da} } \, U(N) \, y^D =  e^{ (u^{-1} \vart ) _A {} ^\a \, S_\a {} ^A + ( \bvart u ) ^{\da A} \, \Sb_{A \da} } \, y^D \,.
\end{align}
Consider now an isometry of the bulk space parametrized by $\bm{t} \in \mathfrak{psu}(2,2 \vert 4)$. The coset representative transforms according to \eqn{gtransf},
\begin{align}
g^{-1} \, \delta g = g^{-1} \bm{t}\,  g - \bm{h}_1 \, , \qquad \bm{h}_1 \in \mathfrak{h}_1 = \big(\mathrm{span} \lbrace M_{\mu \nu} , P_\mu - K_\mu \rbrace \simeq \mathfrak{so}(4,1)  \big) \oplus \mathfrak{so}(5) \, ,
\end{align}
and we want to show that $h_2 \bm{h}_1 h_2^{-1} \Big \vert _{\mathfrak{f}_2} \xrightarrow{\scriptscriptstyle y \to 0}  0$. For this purpose it is not necessary to actually compute $\bm{h}_1$. Rather, decompose $\bm{h}_1$ as
\begin{align}
\bm{h}_1 = \bm{h}_1 \Big \vert _{\mathfrak{h}_2} + \bm{h}_1 ^\prime \, \quad \Rightarrow \quad h_2 \, \bm{h}_1 \, h_2^{-1} \Big \vert _{\mathfrak{f}_2} = h_2 \, \bm{h}_1 ^\prime \, h_2^{-1} \Big \vert _{\mathfrak{f}_2} \,.
\end{align}
In our case, this means that we only need to find the $y$-dependence of $\bm{h}_1 ^\prime$,
\begin{align}
\bm{h}_1 ^\prime = y^\omega c^\mu P_\mu \, \quad \Rightarrow \quad  h_2 \, \bm{h}_1 ^\prime \, h_2^{-1} = \mathcal{O}\left( y^{\omega +1}  \right) \,.
\end{align}
It is thus sufficient to show that for $\bm{h}_1 = y^\omega c^\mu \left( P_\mu - K_\mu \right)+ \ldots $ we have $\omega = 1$. In order to see this, we note that 
\begin{align}
g = g_+ \, g_- \, , \qquad g_+ = e^{ X \cdot P } \, e^{\theta _\a {} ^A \, Q_A {}^\a + \btheta_{A \da} \, \Qb^{\da A} } \, , \qquad g_- = \, e^{\vart _A {} ^\a \, S_\a {} ^A + \bvart^{\da A} \, \Sb_{A \da} } \, U(N) \, y^D \, ,
\end{align}
where $g_+$ only contains generators with positive weights, while $g_-$ only contains generators with weights $\leq 0$. The group elements $g_+$ can be considered as a set of coset representatives for a superspace obtained by factoring out the subgroup $H_+$ with Lie algebra
\begin{align}
\mathfrak{h}_+ = \mathrm{span} \left \lbrace M_{\mu \nu}, D ,  K_\mu , S_\a {} ^A , \Sb_{A \da} \right \rbrace \oplus \mathfrak{su}(4) \,.
\end{align}
In particular, $\mathfrak{h}_+$ only contains generators of weights $\leq 0$. The Cartan form for this coset representative can be obtained from \eqn{eqn:MaurerCartanForm} by setting $\vart = 0$, $N$ constant and $y=1$. Let us first consider the transformation $g_+ ^{-1} \bm{t} g_+ = g_+ ^{-1} \delta g_+ + \bm{h}_+$, where $\bm{h}_+ \in \mathfrak{h}_+$. We thus have:
\begin{align}
- \delta Z^M \, A_M = g^{-1} \, \delta g = g_- ^{-1} \left( g_+ ^{-1} \delta g_+ + \bm{h}_+ \right) g_- - \bm{h}_1
\end{align}
In the above formula $\bm{h}_1$ compensates for those terms of $g_- ^{-1} \left( g_+ ^{-1} \delta g_+ + \bm{h}_+ \right) g_-$ which may not be put into the form $\delta Z^M \, A_M$. In particular, there is no need to compensate a term proportional to $P_\mu$ as these terms are the same in $g^{-1} \, \delta g$ and $g_- ^{-1} \left( g_+ ^{-1} \delta g_+  \right) g_-$, which may be seen easily from our calculation of the Cartan form in section~\ref{sec:Cartanform}. It is thus clear that the term in $\bm{h}_1$ proportional to $(P_\mu - K_\mu)$ compensates for a term in $g_- ^{-1} \left( g_+ ^{-1} \delta g_+ + \bm{h}_+ \right) g_-$, which is proportional to $K_\mu$. This term, however, is of order $\mathcal{O}(y)$, since $K_\mu$ has weight $-1$ and we are doing the conjugation with $y^D$ last. We thus see that $\bm{h}_1 = y \, c^\mu \left( P_\mu - K_\mu \right)+ \ldots $, as we have claimed above. 

This shows that the space parametrized by \eqn{boundarycoset} may indeed be viewed as the superconformal boundary of the space $PSU(2,2\vert4) / \left( SO(4,1) \times SO(5) \right)$. Correspondingly, we impose the following set of boundary conditions on the minimal surface that describes the super Wilson loop at strong coupling:
\begin{equation}
\begin{alignedat}{3}
X^\mu (\tau = 0 , s) &= x^\mu (s) \, & \qquad \quad 
y(0, s) &= 0 \, & \qquad \quad 
N^I(0,s) &= n^I(s)  \\
\theta _\a {} ^A (0, s ) &= \la_\a {} ^A (s) \, & \qquad \quad
\btheta_{A \da} (0, s ) &= \bar \la _{A \da} (s)
\end{alignedat}
\end{equation}

\section{Symmetries of the Super Wilson Loop}
\label{sec:symm}

In this section we derive the superconformal and level-1 Yangian symmetries of the super Wilson loop at strong coupling. Just like in section~\ref{sec:bosAdS}, this can be achieved by evaluating the conserved charges derived from the classical integrability of the string model. On the way, we determine how the minimal surface behaves close to the conformal boundary of the superspace. 

\subsection{The Bulk Expansion}

We determine the coordinates of the minimal surface in an expansion away from the conformal boundary $y=0$. As for the bosonic case discussed in section~\ref{sec:bosAdS} the first few coefficients in the $\tau$-expansion can be given in terms of the boundary curve and variational derivatives of the minimal area by iteratively solving the equations of motion and the Virasoro constraints as well as computing the variation of the renormalized minimal area $\mathcal{A}_\mathrm{ren}$ under a variation of the boundary curve. 

The superstring equations of motion only have a unique solution if one fixes a kappa symmetry gauge. This is often done by setting half of the coordinates of the coset representative to zero. In our case it is however more efficient to set half of the coefficients of the fermionic part of $A=-g^{-1} \diff g$ to zero. Working to linear order in Gra{\ss}mann variables, the Cartan form transforms as follows under a kappa symmetry transformation
\begin{align}
A \mapsto A^\prime = A -\diff \kappa + \left[ A \, , \, \kappa \right] + h^{(0)}\,.
\end{align}
Here, the kappa symmetry parameter $\kappa$ is given by \eqn{kappatransform},
\begin{align*}
\kappa^{(1)} &= A_{i , -} ^{(2)} \, \mathcal{K} _+ ^{(1),i} + \mathcal{K} _+ ^{(1),i} \, A_{i , -} ^{(2)} \, ,  \qquad \qquad
\kappa^{(3)} = A_{i , +} ^{(2)} \, \mathcal{K}_- ^{(3),i} + \mathcal{K} _- ^{(3),i} \, A_{i , +} ^{(2)} \, , \quad \text{where}\\
V_{\pm} ^i &= P_{\pm} ^{i j} \, V_j = \half \left( \gamma^{ij} \pm i \, \eps^{i j} \right) V_j \,.
\end{align*}
We use the kappa symmetry invariance to fix the following gauge on the parameters of the Cartan form $A$:
\begin{align}
\varepsilon _\a {} ^2 &= \varepsilon _\a {} ^4  = 0 \, , \qquad \bar{\varepsilon}_{2 \da} = \bar{\varepsilon}_{4 \da} = 0 \, , \qquad \chi _1 {} ^\a = \chi _3 {} ^\a = 0 \, , \qquad \bar{\chi}^{\da 1} = \bar{\chi}^{\da 3} = 0 \,. \label{kappagauge}
\end{align}
Written in terms of a $(4 \vert 4)$ supermatrix, this reads as 
\begin{align}
A^{(1)+(3)} = \left( \begin{array} {cccc|cccc}
0 & 0 & 0 & 0 & \bullet & 0 & \bullet & 0 \\
0 & 0 & 0 & 0 & \bullet & 0 & \bullet & 0 \\
0 & 0 & 0 & 0 & 0 & \bullet & 0 & \bullet \\
0 & 0 & 0 & 0 & 0 & \bullet & 0 & \bullet \\ \hline
0 & 0 & \bullet & \bullet & 0 & 0 & 0 & 0 \\
\bullet & \bullet & 0 & 0 & 0 & 0 & 0 & 0 \\
0 & 0 & \bullet & \bullet & 0 & 0 & 0 & 0 \\
\bullet & \bullet & 0 & 0 & 0 & 0 & 0 & 0 \\
\end{array} \right)
\end{align}
In appendix~\ref{app:kappa}, we discuss the possibility of fixing this kappa symmetry gauge for the simplified situation of a straight-line boundary curve to linear order in Gra{\ss}mann variables. While it is possible to fix \eqn{kappagauge}, we find that the fermionic coefficients, which are not set zero, no longer obey the reality constraint $( \varepsilon _\a {} ^A ) ^ \ast = \bar{\varepsilon}_{A \da}$. This is related to working with a Wick rotated superstring action, which implies the appearance of a factor $i$ in the kappa symmetry variations. This subtlety does, however, not affect our further calculation, for which the reality constraint is irrelevant. 

\subsubsection{Equations of Motion}
\label{sec:eom}
We now consider the equations of motion \eqn{boseom}, \eqn{fermeom} order by order in $\tau$. From the leading order $\mathcal{O}(\tau^{-2})$ in the bosonic equation \eqn{boseom} one finds that
\begin{align}
\left( \left( \partial_\tau y \right) r_{\tau \, \mu} \left( P^\mu + K^\mu \right) + \left( \left( \partial_\tau y \right)^2 - r_{\tau}^2 - r_s^2 \right) D \right) _{(0)} = 0 \, , \\
\Rightarrow r_{\tau \, (0)} ^\mu  = 0 \, , \qquad \qquad  (y_{(1)}) ^2 = \left( r_{s \, (0)} \right) ^2 \,.
\end{align}
We identify the leading term of $r_s$ with the supermomentum
\begin{align}
\pi ^\mu = r_{s \, (0)}^\mu = \dx ^\mu + i \tr \big( \dot{\bar{\la}} \s ^\mu \la - \bar{\la} \s ^\mu \dot{\la} \big)
\end{align}
of a superparticle moving along the boundary curve. As $y$ should be positive and the boundary curve space-like we note that $y_{(1)} = \sqrt{\pi^2} = \lvert \pi \rvert$. We now restrict the parametrization of the curve to satisfy
\begin{align}
\lvert \pi \rvert = 1 \quad \Rightarrow \pi \cdot \dot{\pi} = 0 \quad \Rightarrow \quad \dot{\pi}^2 + \pi \cdot \ddot{\pi} = 0 \,.
\end{align}
This is the super analogue of the arc-length condition we employed in section~\ref{sec:bosAdS} and it completely fixes the residual reparametrization invariance in conformal gauge. Considering the fermionic equations \eqn{fermeom} at leading order in $\tau$ leads to the following set of equations:
\begin{align}
\varepsilon_{\tau \, (0)} {} _\a {} ^A + i \pi_{\a \da} \, \bar{\varepsilon} _{\tau \, (0)} {} _B{}^\da K^{BA} &= 0  \label{ferml1} \\
 \bar{\varepsilon}_{\tau \, (0) \, A \da} + i \, \varepsilon_{\tau \, (0)}{}^{\a B} \, \pi _{ \a \da} \, K_{BA} &=0 \label{ferml2}  \\
i \, \bar{\varepsilon}_{s \, (0) A \da} \, \pi ^{\da \a } +  \varepsilon_{s \, (0)} {}^{\a B} \, K_{BA} &=0  \label{ferml3} \\
i \, \pi ^{\da \a} {\varepsilon_{s \, (0)} {} _\a } ^A  + \bar{\varepsilon}_{s \, (0)} {} _B {}^\da \, K^{BA} &= 0 \label{ferml4}
\end{align}
The equations \eqn{ferml1} and \eqn{ferml2} as well as \eqn{ferml3} and \eqn{ferml4} are equivalent to each other as one would expect as they stem from the coefficients of $Q$ and $\Qb$ or $S$ and $\Sb$ in \eqn{fermeom}. Moreover, the real and imaginary parts of these equations are equivalent. Insert for example \eqn{ferml1} into \eqn{ferml2}:
\begin{align}
\Rightarrow \bar{\varepsilon}_{\tau \, (0) \, A \da} = \tilde{\kappa}^2 \, \bar{\varepsilon}_{\tau \, (0) \, C \db} \, \pi ^{\db \a} \, \pi_{\a \da} \, K^{C B} K_{B A} = \tilde{\kappa}^2 \bar{\varepsilon}_{\tau \, (0) \, A \da} 
\end{align}
Here, we have left the parameter $\tilde{\kappa}$ open for a moment in order to show that the equations are less constraining for $\tilde{\kappa}^2 = 1$ when one has kappa symmetry. Consider now equation \eqn{ferml1}. Due to our kappa symmetry gauge \eqn{kappagauge} either $\varepsilon_{\tau \, (0)} {} _\a {}^A$ or $\bar{\varepsilon} _{\tau \, (0)} {} _B {}^\da K^{BA}$ are vanishing for any given value of $A$. Proceeding in the same way for equation \eqn{ferml3} we conclude that
\begin{align}
\varepsilon_{\tau \, (0)} = \varepsilon_{s \, (0)} = 0 \, , \qquad \bar{\varepsilon}_{\tau \, (0)} =  \bar{\varepsilon}_{s \, (0)} = 0 \,.
\end{align}
Note in particular that we do not employ reality conditions such as $\bar{\varepsilon}_{ A \da} = \left( \varepsilon _\a {} ^A \right)^\ast$ as they become problematic for the Wick rotated superstring action. Given that $r_{\tau \, (0)}=0$ and hence $\zeta_{\tau \, (0)}=0$, we conclude that
\begin{align}
\theta_{(1)} = 0 \, , \qquad \btheta_{(1)} = 0 \, , \qquad X_{(1)} = 0 \,. \label{X10}
\end{align}
Setting  $\varepsilon_{s \, (0)} = 0 = \bar \varepsilon_{s \, (0)}$ enforces that
\begin{align}
\vart _{(0)} {} _A {}^\a =  i \,  \dot{\bar{\la}}_{A \da} \, \pi^{\da \a} \, , \qquad \quad \bvart _{(0)} {}^{\da A} = - i \, \pi ^{\da \a} \, \dot{\la} _\a {}^A \,. \label{vart0}
\end{align}
It is thus indeed inconsistent to specify boundary conditions for the $\vart$ variables. 

We now turn to the next-to leading order in the bosonic equation \eqn{boseom}. Due to our findings above, we have $A^{(1) \pm (3)} = \mathcal{O}(\tau ^{1/2})$ and we can hence neglect the fermionic contributions also at the next-to leading order. Moreover we note the following identities for the coefficients of the Cartan form:
\begin{align}
\sigma_i = \partial _i y + \mathcal{O}(\tau^2) \, , \qquad   \la_\tau = \mathcal{O}(\tau) \, , \qquad \la_{s \, \a} {} ^\a - \bar \la _s {} ^\da {} _\da = \mathcal{O}(\tau) \,.
\end{align}
Evaluating \eqn{boseom} then leads to the following equations:
\begin{align}
0&= \gamma ^{i j} \left(  \left( \partial_i y \right)  \left( \partial_j y \right)  - y \, \partial _i \partial _j y - r_i \cdot r_j \right) _{(1)} \, , \label{bosnl1} \\
0&= \gamma ^{i j} \left( 2 \, \partial_i y \,   r_{j \, \a \da} - y \, \partial_i \,  r_{j \, \a \da} + 2i y \, \la \indices{ _{i \,\a} ^\b} r_{j \, \b \da} - 2i y \, r_{j \, \a \db} \, \bar{\la}   \indices{_i ^\db _\da}  \right)_{(1)} \,. \label{bosnl2}
\end{align}
Making use of \eqn{X10} and \eqn{vart0} we find that
\begin{align}
r_\tau ^2 = \mathcal{O}(\tau^2) \, , \qquad r_s^2 = \pi^2 + \mathcal{O}(\tau^2) \, , \qquad \gamma ^{i j} \Big( 2i y  \, \la \indices{ _{i \,\a} ^\b} r_{j \, \b \da} - 2i y \, r_{j \, \a \db} \, \bar{\la}   \indices{_i ^\db _\da} \Big) = \mathcal{O}(\tau^2) \,. 
\end{align}
We can thus conclude that
\begin{align}
y_{(2)} = 0 \, , \qquad \qquad r_{\tau \, (1)}^\mu = \dot{\pi}^\mu \quad \Rightarrow \quad X_{(2)} ^\mu = \dot{\pi}^\mu + i \tr \big( \btheta_{(2)} \s^\mu \la - \bla \s^\mu \theta_{(2)} \big) \,.
\end{align}
Last, we consider the next-to leading order in the fermionic equation \eqn{fermeom}.  We find the following conditions, this time leaving out equivalent conditions:
\begin{align}
 \, \bar{\chi}_{\tau \, (0)} {} ^{A \da} - i \, \pi ^{\da \a} \varepsilon_{s \, (1)} {} _\a {} ^A -  \left(   \bar{\varepsilon}_{s \, (1)} {} _B {} ^\da -i \, \pi ^{\da \a} \chi _{\tau \, (0) B \a} \right) K^{BA} &= 0 \label{fermnl1} \, ,\\
 \bar{\varepsilon}_{\tau \, (1) A \da} - i \, \chi_{s \, (0)} {} _A {} ^\a \, \pi_{\a \da}  -  \left( \bar{\chi}_{s \, (0)} {} _\da {}^B - i \, \varepsilon_{\tau \, (1)} {}^{\a B} \, \pi_{\a \da} \right) K_{BA} &=0 \,. 
\end{align}
Due to our kappa symmetry gauge \eqn{kappagauge}, we can decompose these equations into the conditions
\begin{alignat}{2}
  \bar{\chi}_{\tau \, (0)} {} ^{A \da} + K^{A B} \, \bar{\varepsilon}_{s \, (1)} {} _B {}^\da &= 0 \, , & \quad \qquad \varepsilon_{s \, (1)} {} _\a {} ^A +   K^{A B} \,  \chi _{\tau \, (0) B \a} &=0 \, , \label{fermnl2} \\
\bar{\varepsilon}_{\tau \, (1) A \da}  +  K_{AB} \, \bar{\chi}_{s \, (0)} {} _\da {} ^B &= 0 \, , & \quad \qquad 
 \chi_{s \, (0)} {} _A {}^\a  + K_{A B} \, \varepsilon_{\tau \, (1)} {}^{\a B} &= 0 \,.
\end{alignat}
The latter conditions allow to solve for $\theta_{(2)}$ and $\btheta_{(2)}$:
\begin{equation}
\begin{aligned}
\theta_{(2)} {} _\a {} ^A &= - \dot{\pi}_{\a \da} \, \pi ^{\da \b} \, \dot{\la}  _\b {} ^A + \bar K^{\prime \, A B} \left( 4 \, \big( \dot{\bar{\la}} \pi \dot{\la} \big)_B {} ^C  - i \,  \delta ^C _B \, \partial _s \right) \big( \pi _{\a \db} \, \dot{\bar{\la}}  _C {} ^\db \big) \label{theta2} \\
\btheta _{(2) \, A \da} &= - \, \dbla _{A \db} \, \pi ^{\db \a} \, \dot{\pi}_{\a \da} -  K^\prime _{A B} \left( 4 \, \big( \dbla_C \pi \dot{\la}^B \big) + i \,  \delta ^B _C \, \partial _s \right) \big( \dot{\la}^{\b C} \, \pi_{\b \da} \big)
\end{aligned}
\end{equation}
Here, the matrices $\bar K^{\prime \, A B}$ and $K^\prime _{A B}$ are given by
\begin{align}
K^\prime _{A B} = u _A {} ^C \, u _B {} ^D \, K_{CD} = \left( u K u^T \right)_{AB} \, , \qquad 
\bar K^{\prime \, A B} = \left( \left(u^{-1} \right)^T K u^{-1} \right) ^{AB}
\end{align}
Due to equations \eqn{eqn:su4coset} and \eqn{Kdef}, this can be rewritten as
\begin{align}
K^\prime _{AB} = \left( N^5 \mathbb{I} + N^6 \gamma^5 + i N^a \gamma ^a \right) _A {} ^C \, K_{CB} \, , \qquad 
\bar K^{\prime \, AB} = K^{AC} \left( N^5 \mathbb{I} - N^6 \gamma^5 - i N^a \gamma ^a \right)_C {} ^B \,.
\end{align}
From equation \eqn{fermnl2} we find the following condition for $\vart_{(1)}$:
\begin{align}
\vart_{(1) \, A \a} = i \, K^\prime _{AB} \, \pi_{\a \da} \, \bvart _{(1)} {} ^{\da B} \, , \qquad \bvart _{(1)} {} ^{\da A} = -i \, K^{\prime A B} \, \pi^{\da \a} \, \vart_{(1) \, B \a} \,.
\end{align}
The results of solving the fermionic equations of motion at next-to leading order may conveniently be written in the form
\begin{align}
A_\tau ^{(1)-(3)} = -i \, A_s ^{(1)+(3)} + \mathcal{O}( \tau^{3/2} ) \, , \qquad A_s ^{(1)-(3)} = i \, A_\tau ^{(1)+(3)} + \mathcal{O}( \tau^{3/2} ) \,. \label{fermeqnA}
\end{align}

\subsubsection{Virasoro constraints}
We now turn to the Virasoro constraint
\begin{align}
\left \langle A_\tau ^{(2)}, \, A_\tau ^{(2)} \right \rangle  - \left\langle A_s ^{(2)}, \, A_s ^{(2)} \right\rangle = 0 \,. \label{Vir1}
\end{align}
Recall that $A^{(2)}$ is given by \eqn{A2}
\begin{align*}
A_i ^{(2)} &=  - \frac{r_i ^\mu  + y^2 \,  \kappa_i ^\mu }{2y} \left( P_\mu + K_\mu \right)  - \frac{\s_i}{y} D  - \gamma_i \, C  -  \left(\Lambda \indices{_{i A} ^B} R \indices{^A _B}  + U^{-1} \partial_i U   \right)^{(2)} \,.   
\end{align*} 
From section~\ref{sec:S5} we know that
\begin{align*}
\left\langle P^{(2)} \left( U^{-1} \partial_i U \right), \, P^{(2)} \left( U^{-1} \partial_j U   \right) \right\rangle = \partial _i N^I \, \partial _j N^I \,.
\end{align*}
Noting that $\Lambda_\tau = \mathcal{O}(\tau)$ one finds quickly:
\begin{align}
\left\langle A_\tau ^{(2)}, \, A_\tau ^{(2)}  \right\rangle = \frac{1}{y^2} \left(r_\tau ^2 + \s_\tau ^2 \right) + N_{(1)} ^2  + \mathcal{O}(\tau) \,.
\end{align}
The other term is more involved since $\Lambda _{s \, (0)}$ does not vanish,
\begin{align}
\Lambda \indices{_{s A} ^B} = i \big( u^{-1}  \dot{ \bar{\la}}\, \pi \, \dot{\la} \,  u \big)_A {} ^B   \, + \mathcal{O}(\tau) = : i \big( u^{-1} \Sigma_s u \big)_A {} ^B   \, + \mathcal{O}(\tau)
\end{align}
Making use of the trace identities given in appendix~\ref{app:su224} and the results obtained in section~\ref{sec:S5}, we find:
\begin{equation}
\begin{aligned}
& \left\langle \Lambda \indices{_{s A} ^B}  R \indices{^A _B}  \Lambda \indices{_{s C} ^D}, \, P^{(2)} \left( R \indices{^C _D} \right) \right\rangle = \Lambda \indices{_{s A} ^B} \, \Lambda \indices{_{s C} ^D}  \left( 4 K_{D B} K^{A C} - 4 \delta ^A _D \, \delta ^C _B + 2 \delta ^A _B \, \delta ^C _D \right) \\
& \quad = - 4 \, \bar K^{\prime A C} \, K^\prime _{DB} \, \big( \dot{\bla} \, \pi \, \dot{\la} \big)_A {} ^B \big( \dot{\bla} \, \pi \, \dot{\la}  \big)_C {} ^D + 4 \tr \big( \dot{\bla} \, \pi \dot{\la} \dot{\bla} \, \pi \dot{\la} \big) - 2 \tr \big(\dot{\bla} \, \pi \dot{\la} \big)^2   \\
&\quad = - 4 \tr \big( \bar K^\prime \, \Sigma_s K^{\prime} \Sigma_s ^T \big) + 4 \tr \big( \dot{\bla} \, \pi \dot{\la} \dot{\bla} \, \pi \dot{\la} \big) - 2 \tr \big(\dot{\bla} \, \pi \dot{\la} \big)^2  \, , \\
&\left\langle \Lambda \indices{_{s A} ^B}  R \indices{^A _B}\, , \, P^{(2)} \left( U^{-1} \partial_s U \right) \right\rangle  = 4 \Lambda \indices{_{s A} ^B} \big( A _{S^5} ^{(2)} \big) \indices{_B ^A}   = - 4i \, n^I \dot{n}^J \, \tr \big( \dot{\la} \gamma^{IJ} \dot{\bla} \, \pi \big)\,. 
\end{aligned}
\end{equation}
Thus we find
\begin{align}
\left\langle A_s ^{(2)}, \, A_s ^{(2)}  \right\rangle &= \frac{r_s ^2}{y^2} + 2 \, \pi \cdot \kappa_{s \, (0)} + \dot{n}^2 + 4 \tr \big( \dot{\bla} \, \pi \dot{\la} \dot{\bla} \, \pi \dot{\la} \big) - 2 \tr \big(\dot{\bla} \, \pi \dot{\la} \big)^2 \nn \\ 
& \quad - 4 \tr \big( \bar K^\prime \, \Sigma_s K^{\prime} \Sigma_s ^T \big)  
 - 4i \, n^I \dot{n}^J \, \tr \big( \dot{\la} \gamma^{IJ} \dot{\bla} \, \pi \big) + \mathcal{O}(\tau) \,.
\end{align}
Due to the results obtained in the last section, we can express everything in terms of the boundary data,  
\begin{equation}
\begin{aligned}
r_s^2 &= 1- \tau^2 \left( \dot{\pi}^2 - 2i \tr \big( \btheta _{(2)} \pi \dot{\la} - \dot{\bar{\la}} \pi \theta _{(2)} \big) \right) + \mathcal{O}(\tau^3) \, , \qquad r_\tau ^2 = \tau ^2 \dot{\pi}^2 + \mathcal{O}(\tau^3) \, , \\
\sigma _\tau ^2 &= 1 + \tau^2 \left( 2 y_{(3)} - 4i \tr \big( \btheta _{(2)} \pi \dot{\la} - \dot{\bar{\la}} \pi \theta _{(2)} \big) \right) + \mathcal{O}(\tau^3) \, , \\
\pi \cdot \kappa _ {s \, (0)} &= 6 \tr \big( \dot{\bla} \, \pi \dot{\la} \dot{\bla} \, \pi \dot{\la} \big) -i \tr \big(\dot{\bla}  \pi  \ddot{\la} - \ddot{\bla}  \pi  \dot{\la} \big) +  \mathcal{O}(\tau) \,. \label{pikappa}
\end{aligned}
\end{equation}
Inserting these results into \eqn{Vir1} allows to solve for $y_{(3)}$:
\begin{align}
y_{(3)} & = -\dot{\pi}^2 + 3i \tr \big( \btheta _{(2)} \pi \dot{\la} - \dot{\bar{\la}} \pi \theta _{(2)} \big) + 8 \tr \big( \dot{\bla} \, \pi \dot{\la} \dot{\bla} \, \pi \dot{\la} \big) -i \tr \big(\dot{\bla}  \pi  \ddot{\la} - \ddot{\bla}  \pi  \dot{\la} \big) \nn \\
& \quad -  \tr \big(\dot{\bla} \, \pi \dot{\la} \big)^2  - 2 \tr \big(\bar K^\prime \, \Sigma_s K^{\prime} \Sigma_s ^T \big) -2 i \, n^I \dot{n}^J \, \tr \big( \dot{\la} \gamma^{IJ} \dot{\bla} \, \pi \big)    + \half \left(\dot{n}^2 - N_{(1)}^2 \right) 
\end{align}

\subsubsection{Variational Derivatives of the Minimal Area}
The solutions of the first few orders of the equations of motion allow to extract the divergence of the minimal area. Similarly to the bosonic situation one finds:
\begin{align}
\mathcal{A}_{\mathrm{min}}(C) _{y \geq \varepsilon} = \frac{\mathcal{L}(C)}{\varepsilon} + (\mathrm{finite}) \, , \qquad \mathcal{L}(C) = \int \diff s \, \lvert \pi(s) \rvert \,.
\end{align}
We can thus define a finite functional by 
\begin{align}
\mathcal{A}_\mathrm{ren}(C) = \lim \limits_{\varepsilon \to 0} \left \lbrace \mathcal{A}_{\mathrm{min}}(C) _{y \geq \varepsilon} - \frac{\mathcal{L}(C)}{\varepsilon} \right \rbrace \,.
\end{align}
From the equations of motion we have fixed the expansion of the coordinates $X$ and $\theta$ until second order in the variable $\tau$. As it is the case for the minimal surface in $AdS_5$, the third-order coefficients are not fixed by the equations of motion but can be related to variational derivatives of the minimal surface. Consider therefore the variation of $\mathcal{A}_{\mathrm{min}}(C) _{y \geq \varepsilon}$. Since the parametrization of the minimal area satisfies the equations of motion, the variation only contains boundary terms,  
\begin{align}
\delta \mathcal{A}_{\mathrm{min}}(C) _{y \geq \varepsilon} &= - \int \limits _a ^b \diff s \int \limits _{\tau_0(s)} ^c \diff \tau \left \lbrace \partial _\tau \left\langle g^{-1} \delta g , \, \Lambda ^\tau \right\rangle + \partial _s \left\langle g^{-1} \delta g , \, \Lambda ^s \right\rangle \right \rbrace  \, ,
\end{align}
with $\Lambda^i$ given by \eqn{defLambda}. The boundary $\tau_0(s)$ of the integration domain is defined by demanding $y(\tau_0(s),s) = \varepsilon $. Since the expression $g^{-1} \delta g$ does not contain any derivatives which may be restricted by choosing a special parametrization, we can safely demand that the parametrization satisfy $\lvert \pi \rvert = 1$. Moreover, due to the kappa symmetry invariance of the action, we can restrict ourselves to the kappa symmetry gauge \eqn{kappagauge}. This allows to apply the results of section~\ref{sec:eom}, in particular we have $\tau_0(s) = \varepsilon + \mathcal{O} (\varepsilon^3)$. Thus, using the periodicity of the parametrization in $s$, we find that
\begin{align}
\delta \mathcal{A}_{\mathrm{min}}(C) _{y \geq \varepsilon} =  \int \limits _0 ^L \diff s \left \lbrace \left\langle g^{-1} \delta g , \, \Lambda ^\tau \right\rangle (\tau_0(s),s)  \right \rbrace + \mathcal{O}(\varepsilon) \,.
\end{align}
To compute the above result explicitly, we use the expression \eqn{eqn:MaurerCartanForm} for $g^{-1} \delta g$ and in particular that
\begin{align}
\delta X (\tau , s) = \delta x(s) + \mathcal{O}(\tau^2) \, , \quad \delta \theta (\tau , s) = \delta \lambda (s) + \mathcal{O}(\tau^2)  \, , 
\end{align}
since the first-order coefficients of $X,\theta,\btheta$ vanish identically. Applying the trace-identities given in appendix~\ref{app:su224}, we find
\begin{align}
\delta \mathcal{A}_{\mathrm{min}}(C) _{y \geq \varepsilon} = \frac{\delta \mathcal{L}(C)}{\varepsilon} + ( \mathrm{finite} ) \, ,
\end{align}
and the finite term may be computed as:
\begin{align}
\delta \mathcal{A}_{\mathrm{ren}}(C) = \int \limits _0 ^L \diff s \left \lbrace \delta x_\mu \, b^\mu + \delta \la _\a {} ^A \left(4 \xi _A {} ^\a - i \, \bar{\la}_{A \da} \, b^{\da \a} \right) + \delta \bar{\la}_{A \da} \left( 4 \bar{\xi}^{\da A} - i \, b^{\da \a} \, \la ^A _\a \right) - \delta n^I  N_{(1)} ^I \right \rbrace \label{variation} 
\end{align}
Here, we defined
\begin{align}
b^\mu &= - \half r_{\tau \, (2)} ^\mu - \kappa _{\tau \, (0)} ^\mu + 2i \tr \big( \big( \bvart_{(0)} \vart_{(1)} - \bvart_{(1)} \vart_{(0)} + 2 \, n^I N_{(1)} ^J \,  \bvart_{(0)} \gamma^{IJ} \vart_{(0)} \big) \bar{\sigma}^\mu \big) \, ,  \label{coeffder1} \\
\xi _A {} ^\a &= \vart_{(1)} {} _A {} ^\a +  n^I N_{(1)} ^J \left( \gamma ^{IJ} \vart_{(0)} \right) _A {} ^\a  \, , \qquad \bar{\xi}^{\da A} = - \bvart _{(1)} ^{\da A} +  \, n^I N_{(1)} ^J \left( \bvart_{(0)} \gamma ^{IJ} \right)^{\da A} \,.  \label{coeffder2} 
\end{align}
We read off the functional derivatives of the regulated  minimal area from \eqn{variation}:
\begin{equation}
\begin{alignedat}{2}
b^\mu &=  \frac{\delta \mathcal{A}_{\mathrm{ren}}}{\delta x_\mu (s)} \, , &  \qquad \xi _A {}^\a &= \frac{1}{4} \left( \frac{\delta \mathcal{A}_{\mathrm{ren}}}{\delta \la _\a {} ^A(s)} + i \, \bar{\la}_{A \da} \, \s_\mu ^{\da \a} \, \frac{\delta \mathcal{A}_{\mathrm{ren}}}{\delta x_\mu (s)} \right) \, , \\
N_{(1)} ^I &= - \frac{\delta \mathcal{A}_{\mathrm{ren}}}{\delta n^I (s)} + \left( n^J(s) \frac{\delta \mathcal{A}_{\mathrm{ren}}}{\delta n^J (s)}     \right) n^I (s)   \, , &  \qquad \bar{\xi}^{\da A} &=  \frac{1}{4} \left( \frac{\delta \mathcal{A}_{\mathrm{ren}}}{\delta \bar{\la}_{A \da}(s)} + i\, \s_\mu ^{\da \a} \, \la _\a {} ^A  \, \frac{\delta \mathcal{A}_{\mathrm{ren}}}{\delta x_\mu (s)} \right) \,. \label{varder}
\end{alignedat}
\end{equation}
The relation between the variational derivatives of the minimal area and the coordinates $X_{(3)}, \theta_{(3)}$ and $\vart_{(1)}$ takes a much more complicated form than for the minimal surface in $AdS_5$. The important point for us is however to identify the above coefficients in the Noether current $J_i$, which we do in the next section.

\subsection{The Conserved Charges}

\subsubsection{Local Charges}
In this section we derive the superconformal Ward identities for the super Wilson loop at strong coupling from the condition
\begin{align}
\mathcal{Q}^{(0)} = \int \diff s \, J_\tau  = 0 \, ,  
\end{align} 
which follows from the fact that the curve can be contracted to a point on the minimal surface. Due to current conservation, $\partial _\tau J_\tau + \partial _s J_s = 0$, we have\footnote{Note that the order of $J_i = g \Lambda _i g^{-1}$ can at most be $\mathcal{O}(\tau ^{-2})$. We use the following notation for the coefficients of the Laurent series: $F = F_{(-2)} \tau ^{-2} +  F_{(-1)} \tau ^{-1} +  F_{(0)} + \sum _{n=1} ^\infty  F_{(n)} n^{-1} \tau ^{-n}$.}
\begin{align}
J_s  = J_{s \, (-2)} \,   \tau ^{-2} + J_{s \, (0)} + \mathcal{O}(\tau) \, , \qquad J_\tau = \left(\partial_s J_{s \, (-2)} \right) \tau^{-1} + J_{\tau \, (0)} - \left(\partial_s J_{s \, (0)} \right) \tau + \mathcal{O}(\tau^2) \,.
\end{align} 
We are only interested in the $\tau^0$-component $J_{\tau \, (0)}$, which is given by
\begin{align}
J_{\tau \, (0)} = \left \lbrace g  \Lambda _\tau g^{-1} \right \rbrace _{(0)} = \left \lbrace g \left(A_{\tau} ^2 - \ihalf A_s ^{(1)-(3)} \right) g^{-1} \right \rbrace _{(0)}  = \left \lbrace g \left(A_{\tau} ^2 + \half A_\tau ^{(1)+(3)} \right) g^{-1} \right \rbrace _{(0)} \,.
\end{align}
In the last step, we applied \eqn{fermeqnA} and noted that the $\mathcal{O}(\tau ^{3/2})$-term appearing there does not contribute to the $\tau^0$-order since the conjugation with $g$ lowers the order of a fermionic term at most by $\sqrt{y}$. Doing all the conjugations similarly to the way explained in section~\ref{sec:Cartanform}, we arrive at the following result:
\begin{align}
J_{\tau \, (0)} &= \Big \lbrace \sfrac{1}{\tau} \, e^{X \cdot P + \Omega} \left( - \half \, \dot{\pi}^\mu \, K_\mu - D - \half \left(\vart _{(0)}{} _A {} ^\a  \, S_\a {} ^A + \bvart_{(0)}^{\, \da A} \, \Sb_{A \da} \right) \right)e^{-X \cdot P - \Omega} \Big \rbrace _{(0)} \nn \\
& \quad + \Big \lbrace  \, e^{X \cdot P + \Omega} \left( \half b^\mu \, K_\mu - \xi _A {}^\a \, S_\a {} ^A + \bar{\xi}^{\da A} \, \Sb_{A \da} + n^I N_{(1)} ^J \, \Gamma ^{I J}  \right)e^{-X \cdot P - \Omega} \Big \rbrace _{(0)} 
\end{align}
The first term vanishes since $X = x + \mathcal{O}(\tau^2)$ and $\theta = \lambda + \mathcal{O}(\tau^2)$ and it remains to compute  
\begin{align}
J_{\tau \, (0)} = \, e^{x \cdot P + \left( \la Q + \bla \Qb \right) } \left( \half b^\mu \, K_\mu - \xi _A {}^\a \, S_\a {} ^A + \bar{\xi}^{\da A} \, \Sb_{A \da} + n^I N_{(1)} ^J \, \Gamma ^{I J}  \right)e^{-x \cdot P - \left( \la Q + \bla \Qb \right)} \,. \label{Jtau0}
\end{align}
The coefficients $b$, $\xi$ and $N_{(1)}$ defined in \eqn{coeffder1} and \eqn{coeffder2} are exactly those that were identified with variational derivatives in the last section. We can thus write the resulting expression as
\begin{align}
J_{\tau \, (0)} (s) = j_a (s) \left( \mathcal{A}_\mathrm{ren} \right) \, \hat{T}^a
\end{align}
Here, $\hat{T}^a$ span the dual basis to the generators $T_a$ defined in appendix~\ref{app:su224}, $\big\langle \hat{T}^a, \,  T_b \big\rangle = \delta ^a _b$ and $j_a (s) \left( \mathcal{A}_\mathrm{ren}  \right)$ denotes the action of variational derivative operators on the minimal area $\mathcal{A}_\mathrm{ren}  $. We can read off the action of these operators from the equation
\begin{align}
j_a (s) \left( \mathcal{A}_\mathrm{ren}  \right) = \left\langle J_{\tau \, (0)} (s), \, T_a \right\rangle \,.
\end{align}
We provide the operator densities $j_a (s)$ explicitly in appendix~\ref{app:generators}. They satisfy the commutation relations
\begin{align}
\big[ j_a (s)\, , \,  j_b (s^\prime) \big \rbrace = f \indices{_{ab} ^c} \, j_c (s) \, \delta( s - s^\prime ) \,.
\end{align}
The structure constants $f \indices{_{ab} ^c}$ of these generators are related to the structure constants $\tilde{f} \indices{_{ab} ^c}$ of the generators $T_a$ by 
\begin{align}
\tilde{f} \indices{_{ab} ^c} = - \left( -1 \right) ^{\lvert a \rvert \lvert b \rvert}  f \indices{_{ab} ^c} = f \indices{_{ba} ^c} \,. \label{Strucrelation}
\end{align}
The vanishing of the level-0 charge $\mathcal{Q}^{(0)}$ may be rewritten as the invariance of the super Wilson loop under the level-0 Yangian generators
\begin{align}
\label{Ja0}
J_a^{(0)} = \int \diff s \; j_a (s) \; .
\end{align}

\subsubsection{Multi-local Charges}
In this section we derive higher-level Ward identities from the vanishing of the multi-local charge $\mathcal{Q}^{(1)}=0$. Once more, the interesting part of this equation is contained in the $\tau^0$-component of $\mathcal{Q}^{(1)}$, for which we recall the expression
\begin{align}
\mathcal{Q}^{(1)} &=  \frac{1}{2}  \int \diff s_1 \diff s_2 \, \varepsilon(s_1-s_2) \, \left[ J_\tau(s_1) \, , \, J_\tau (s_2) \right] - \oint \diff s \left( a_s ^{(2)} + \quarter a_s ^{(1)+(3)} \right) \,.
\end{align}
Note that in the above expression, local contributions can also come from the first term due to the appearance of total derivatives in $J_\tau$. It is worthwhile to separate the local and bi-local terms before starting the calculation. This may be achieved by inserting
\begin{align}
J_\tau = \left(\partial_s J_{s \, (-2)} \right) \tau^{-1} + J_{\tau \, (0)} - \left(\partial_s J_{s \, (0)} \right) \tau + \mathcal{O}(\tau^2) \, ,
\end{align}
which follows easily from current conservation. We find that 
\begin{align}
\mathcal{Q}^{(1)} \Big \vert _{\tau^0} &= \frac{1}{2}  \int \diff s_1 \diff s_2 \, \varepsilon(s_1-s_2) \, \Big[ J_{\tau \, (0)}   (s_1) \, , \, J_{\tau \, (0)}   (s_2) \Big] \nn \\
& \quad  - \oint \diff s \Big \lbrace g \big( A_s ^{(2)} + \quarter A_s ^{(1)+(3)} - \tau^2 \big[ \Lambda _s \, , \, \partial_s \Lambda _s - \left[ A_s \, , \, \Lambda _s \right] \big] \big) g^{-1} \Big \rbrace _{(0)} \, , \label{Q1step}
\end{align}
where $J_s = g \Lambda _s g^{-1}$ and $\Lambda _s = A_s^{(2)} + \ihalf A_{\tau} ^{(1)-(3)}$ according to section~\ref{superstring}. Using the results obtained for $J_{\tau \, (0)}$ we may easily compute the non-local term to be 
\begin{align}
\mathcal{Q}^{(1)} _{\mathrm{non-local}} &= \frac{1}{2}  \int \diff s_1 \diff s_2 \, \varepsilon(s_1-s_2) \, \Big[ J_{\tau \, (0)}   (s_1) \, , \, J_{\tau \, (0)}   (s_2) \Big] \nn \\
&= \frac{1}{2} \, \tilde{f} \indices{^{bc} _a}  \int \diff s_1 \diff s_2 \, \varepsilon(s_1-s_2) \, j_b(s_1) \, j_c(s_2) \, \hat{T}^a \nn \\
&= \frac{1}{2} \, f \indices{^{cb} _a}  \int \diff s_1 \diff s_2 \, \varepsilon(s_1-s_2) \, j_b(s_1) \, j_c(s_2) \, \hat{T}^a \,. \label{non-local}
\end{align}
To perform the last step we used that
\begin{align*}
\tilde{f} \indices{^{bc} _a} = \tilde{G}^{bd} \, \tilde{G}^{ce} \tilde{f} \indices{ _{de} ^g} \, \tilde{G}_{ga} =  \left(-1\right)^{\lvert b \rvert + \lvert c \rvert  + \lvert a \rvert} G^{bd} \, G^{ce} f \indices{ _{ed} ^g} \, G_{ga} = \left(-1\right)^{\lvert b \rvert + \lvert c \rvert  + \lvert a \rvert} f \indices{ ^{cb} _a} = f \indices{ ^{cb} _a} \, ,
\end{align*}
since $\left( \lvert b \rvert + \lvert c \rvert  + \lvert a \rvert \right) \in \lbrace 0, 2 \rbrace$ for non-vanishing $f \indices{ ^{cb} _a}$ due to the $\mathbb{Z}_2$ grading of the Lie super algebra. It is interesting to note that for the bi-local part of $\mathcal{Q}^{(1)}$ to lead to the bi-local structure of a level-1 Yangian generator, we need exactly the relation \eqn{Strucrelation} between the structure constants of the differential operators and supermatrix generators.

We now turn to the computation of the local term
\begin{align}
\mathcal{Q}^{(1)} _{\mathrm{local}} = - \oint \diff s \Big \lbrace g \big( A_s ^{(2)} + \quarter A_s ^{(1)+(3)} - \tau^2 \big[ \Lambda _s \, , \, \partial_s \Lambda _s - \left[ A_s \, , \, \Lambda _s \right] \big] \big) g^{-1} \Big \rbrace _{(0)} \,. \label{Qlocal}
\end{align}
While we have pulled out the conjugations with $g$ in order to make use of the cancellations between the two terms, it turns out to be convenient to already discuss the conjugations with $U y^D$ when considering these terms individually, in particular because after the conjugation with $y^D$ the orders in $\tau$ will not get lowered and we may already discard terms that are of order $\mathcal{O}(\tau)$. In the expression \eqn{A2} for $A^{(2)}$, the projections are left implicit for the R-symmetry part. Making use of the explicit form of these projections given in appendix~\ref{app:su224}, we note that    
\begin{align}
\Lambda _{s \, A} {} ^B P^{(2)} \left(R \indices{^A _B} \right) = \sfrac{1}{4} \left( \Lambda _{s \, A} {} ^B  - K^{B C} \Lambda _{s \, C} {} ^D  K_{DA} \right) R \indices{^A _B}
\end{align}
and we find\footnote{The calculation is performed over $\mathfrak{psu}(2,2\vert4)$. If we were doing it over $\mathfrak{su}(2,2\vert4)$, there would be a term proportional to $C$. Note however, that this term would not be conserved.}
\begin{align}
U y^D \left( A_s ^{(2)} + \quarter A_s ^{(1)+(3)} \right) y^{-D} U^{-1} &= \frac{r_{s \, \a \da} + y^2 \kappa _{s \, \a \da}}{4y^2} \, \left( K^{\da \a} + y^2 P^{\da \a} \right) + \quarter \, \rho_A {} ^B  R \indices{^A _B} \nn \\ 
& \quad - \quarter \left( \left(u \chi_s \right)_A {} ^\a S_\a {} ^A + \left( \bar{\chi}_s u^{-1} \right)^{\da A} \Sb_{A \da} \right) + \mathcal{O}(\tau) \,. 
\end{align}
Here, the coefficient of the R-symmetry part is given by
\begin{align}
\rho_A {} ^B =   n^I \dot{n} ^J \, \left(\gamma ^{IJ} \right) _A {} ^B -  i \big( \dot{\bla}  & \pi \dot{\la}  \big)_A {} ^B  + i \bar K^{\prime \, BC}  \big( \dot{\bla} \pi \dot{\la} \big)_C {} ^D  K^\prime _{DA} + \ihalf \delta ^B _A  \tr \big( \dot{\bla}  \pi \dot{\la}  \big)  \,. \label{rhoAB}
\end{align}
We now consider the commutator term in \eqn{Qlocal} and compute
\begin{align}
\left(\ast \right) :=   \big[ U y^D \left( \tau^2 \Lambda _s \right) y^{-D} U^{-1} \, , \, U y^D \left( \partial_s \Lambda _s - \left[ A_s \, , \, \Lambda _s \right] \right) y^{-D} U^{-1} \big] \,.
\end{align}
In this expression we may replace $\Lambda _s = A_s^{(2)} + \ihalf A_{\tau} ^{(1)-(3)} = A_s^{(2)} + \half A_s ^{(1)+(3)} + \mathcal{O}\left( \tau ^{3/2} \right)$ as the unwanted terms are at least of order $\tau$. Consider then the term on the left-hand side. As we shall see shortly, the right-hand side of the commutator is of order $\mathcal{O}(\tau^{-2} )$ and we can hence neglect all terms that are of order $\mathcal{O}(\tau^3 )$ in the computation of the expression on the left-hand side. This leads to finding
\begin{align}
U y^D \left( \tau^2 \Lambda _s \right) y^{-D} U^{-1} & = \frac{\tau^2}{4y^2} \left( r_{s \, \a \da} + y^2 \kappa _{s \, \a \da} \right)    \left( K^{\da \a} + y^2 P^{\da \a} \right) + \sfrac{\tau^2}{4} \, \rho_A {} ^B  R \indices{^A _B} \nn \\
& \quad - \frac{\tau^2}{2} \left( \left(u \chi_s \right)_A {} ^\a S_\a {} ^A + \left( \bar{\chi}_s u^{-1} \right)^{\da A} \Sb_{A \da} \right) + \left( \mathrm{negligible} \right) \,.
\end{align}
Note in particular that only the term proportional to $K^{\da \a}$ is of order $\mathcal{O}(\tau^0)$, while all other terms are of order $\mathcal{O}(\tau^2)$. We need thus only compute the right-hand side of the above commutator up to $\mathcal{O}(\tau^{-2})$ for the generators that commute with $K$ and to $\mathcal{O}(\tau^0)$ for those that don't. With these simplifications we have
\begin{align}
 U y^D \left( \partial_s \Lambda _s - \left[ A_s \, , \, \Lambda _s \right] \right) y^{-D} U^{-1} = \frac{\dot{\pi}_{\a \da}}{4 \tau^2} \left( K^{\da \a} + \tau^2 P^{\da \a} \right) - \frac{r_s^2}{y^2} D + \left( \mathrm{negligible} \right) \,.
\end{align}
Combining these results we find
\begin{align}
\left(\ast \right) &= - \frac{\tau^2 \, r_s ^2}{4 y^4} \left( r_{s \, \a \da} + y^2 \kappa _{s \, \a \da} \right)    \left( K^{\da \a} - y^2 P^{\da \a} \right) + \pi^\mu \dot{\pi}^\nu M_{\mu \nu} \nn \\
& \qquad + \quarter  \left( \left(u \chi_s \right)_A {} ^\a S_\a {} ^A + \left( \bar{\chi}_s u^{-1} \right)^{\da A} \Sb_{A \da} \right)  + \mathcal{O}(\tau) \,.
\end{align}
Accordingly, we have
\begin{align}
& \quad U y^D \left( A_s ^{(2)} + \quarter A_s ^{(1)+(3)} - \tau^2 \big[ \Lambda _s \, , \, \partial_s \Lambda _s - \left[ A_s \, , \, \Lambda _s \right] \big] \right) y^{-D} U^{-1} = \nn \\
&= \frac{ y^2 + \tau ^2 r_s ^2}{4  y^4}  \left( r_s ^{\da \a} + y^2 \kappa_s ^{\da \a} \right) K_{\a \da} + \frac{y^2 - \tau ^2 r_s ^2}{4 y^2}  \, r_s ^{\da \a} \, P_{\a \da} + \quarter \,  \rho_A {} ^B  R \indices{^A _B}    + \pi ^\mu \dot{\pi}^\nu M_{\mu \nu} \nn \\
& \quad +  \half  \left( \left(u \chi_s \right)_A {} ^\a S_\a {} ^A + \left( \bar{\chi}_s u^{-1} \right)^{\da A} \Sb_{A \da} \right)  + \mathcal{O}(\tau) \nn \\
&= \frac{\pi^{\da \a}}{2 \tau^2} \, K_{\a \da} + \quarter \left( r_{s \, (2)}^{\da \a} + 2 \kappa _{s \, (0)} ^{\da \a} + \left( \pi \cdot r_{s \, (2)} - 2 y_{(3)} \right) \pi^{\da \a} \right) \, K_{\a \da} + \quarter \,  \rho_A {} ^B  R \indices{^A _B}    + \pi ^\mu \dot{\pi}^\nu M_{\mu \nu} \nn \\
& \quad + \big( \sfrac{3}{2} \, \dot{\bar{\la}}\pi  \dot{\la} \dot{\bla}  \pi  - \ihalf \, \partial_s \big( \dot{\bla} \pi \big) \big)_A {} ^\a \, S_\a {} ^A + \big( \sfrac{3}{2} \, \pi  \dot{\la} \dot{\bla}  \pi \dot{\la} + \ihalf \, \partial_s \big( \pi \dot{\la} \big) \big)^{\da A} \, \Sb_{A \da}  + \mathcal{O}(\tau) \,.
\end{align}
Rather conveniently, the terms proportional to $P_{\a \da}$ have cancelled out, which simplifies the computation of the conjugations with $e^\eta$. We find that
\begin{align}
& \quad e^{\eta} U y^D \left( A_s ^{(2)} + \quarter A_s ^{(1)+(3)} - \tau^2 \big[ \Lambda _s \, , \, \partial_s \Lambda _s - \left[ A_s \, , \, \Lambda _s \right] \big] \right) y^{-D} U^{-1} e^{-\eta} = \nn \\
&= - \frac{\pi ^\mu}{\tau ^2} K_\mu + \pi ^\mu \dot{\pi}^\nu M_{\mu \nu} - \half N_{(1)} ^2 \, \pi^\mu K_\mu + \half l^\mu K_\mu - \quarter f_A {}^\a \, S_\a {}^A + \quarter \bar{f}^{\da A} \, \Sb_{A \da} + \quarter \,  \rho_A {} ^B  R \indices{^A _B}  \,. \label{etoeta}
\end{align}
Here, the coefficients $l,f$ and $\bar{f}$ are given by (cf.\ equations \eqn{rhoAB}, \eqn{pikappa} and \eqn{theta2})
\begin{equation}
\begin{aligned}
l^\mu &= - r_{s \, (2)} ^\mu - \left( \pi \cdot r_{s \, (2)} - 2 y_{(3)} - N_{(1)}^2 \right) \pi ^\mu + 4i \tr \big( \bvart_{(0)} \, \rho \, \vart_{(0)} \, \bs ^\mu \big) + 3i \tr \big( \dot{\la} \dot{\bla} \big( \dot{\pi} \bs ^\mu \pi -  \pi \bs ^\mu \dot{\pi}  \big) \big) \nn \\
&= - \ddot{\pi}^\mu - \left( \dot{\pi}^2 - \dot{n}^2 \right) \pi ^\mu  + \left( 2 \pi \cdot \kappa_{s \, (0)} - 4i \tr \big( \big( \theta_{(2)} \dot{\bla} - \dot{\la} \btheta_{(2)} + \dot{\la} \rho \dot{\bla} \big) \pi \big)  \right) \pi ^\mu  \nn \\
& \quad - 4i \tr \big(\dot{\la} \rho \dot{\bla} \s ^\mu \big) + 3i \tr \big( \dot{\la} \dot{\bla} \big( \dot{\pi} \bs ^\mu \pi -  \pi \bs ^\mu \dot{\pi}  \big) \big)  \, ,\\
f _A {}^\a  &= \big[ -6  \, \dot{\bar{\la}}\pi  \dot{\la} \dot{\bla}  \pi  + 2i \, \ddot{\bla} \, \pi  + 4i \, \dot{\bla} \, \dot{\pi} + 4i \, \rho \dot{\bla} \pi \big] _A {}^\a   \, ,\\
\bar{f}^{\da A}  &= \big[ 6 \, \pi  \dot{\la} \dot{\bla} \pi \dot{\la} + 2i \, \pi  \ddot{\la} + 4i \, \dot{\pi} \dot{\la} - 4i \, \pi \dot{\la} \rho \big]^{\da A}   \,.
\end{aligned}
\end{equation}
The reader will notice that in \eqn{etoeta} we have written out terms proportional to $K_\mu$ that we could have absorbed into the definition of $l^\mu$. For the term involving $N_{(1)} ^2$, note that this is the only term appearing that involves derivatives of the minimal area, see equation \eqn{varder}. We have thus found the following expression for the local part of $\mathcal{Q}^{(1)}$:
\begin{align}
\mathcal{Q}^{(1)} _{\mathrm{local}} &= - \int \diff s \, e^{x \cdot P + \la Q + \bla \Qb} \Big( \half l^\mu K_\mu - \quarter f_A {} ^\a S_\a {} ^A + \quarter  \bar{f} ^{\da A} \, \Sb_{A \da} + \quarter \,  \rho_A {} ^B  R \indices{^A _B}   \Big) e^{- x \cdot P - \la Q - \bla \Qb} \nn \\
& \quad + \int \diff s \, e^{x \cdot P + \la Q + \bla \Qb} \Big( \half \, N_{(1)} ^2  \pi ^\mu K_\mu \Big) e^{- x \cdot P - \la Q - \bla \Qb} \nn \\
& \quad - \int \diff s  \, \bigg \lbrace e^{X \cdot P + \theta Q + \btheta \Qb} \Big( - \frac{\pi^\mu}{\tau^2} \, K_\mu  + \pi ^\mu \dot{\pi}^\nu M_{\mu \nu} \Big) e^{- X \cdot P - \theta Q - \btheta \Qb} \bigg \rbrace_{(0)} \nn \\
&=: \int \diff s \, j_a ^{(1)} (s) \, \hat{T}^a - \int \diff s \, e^{x \cdot P + \la Q + \bla \Qb} \Big( - \half \, N_{(1)} ^2  \pi ^\mu K_\mu \Big) e^{- x \cdot P - \la Q - \bla \Qb} \,.
\end{align}
The term in the third line evaluated to zero in the bosonic calculation, see \eqn{bosoniczero}. Here we define the densities $j_a ^{(1)\, \prime}$, 
\begin{align}
j_a ^{(1)\, \prime} \, \hat{T}^a &= \bigg \lbrace e^{X \cdot P + \theta Q + \btheta \Qb} \Big( - \frac{\pi^\mu}{\tau^2} \, K_\mu  + \pi ^\mu \dot{\pi}^\nu M_{\mu \nu} \Big) e^{- X \cdot P - \theta Q - \btheta \Qb} \bigg \rbrace_{(0)} \, ,
\end{align}
which we provide in appendix~\ref{app:level1}. The term in the first line resembles the expression found for the level-zero generator in equation \eqn{Jtau0}, 
\begin{align}
J_{\tau \, (0)} &= \, e^{x \cdot P + \la Q + \bla \Qb  } \Big( \half \, \partial^\mu  \, K_\mu - \quarter \left( \partial_A {} ^\a + i \, \bla_{A \da} \, \partial ^{\da \a}  \right) \, S _\a {} ^A \nn \\
& \qquad + \quarter \left( \partial^{\da A} + i \, \partial^{\da \a} \, \la_\a {} ^A  \right) \, \Sb_{A \da}  - \quarter \left( \gamma^{IJ} \right)_A {} ^B \, n^I \partial ^J  \, R ^A {} _B  \Big)e^{-x \cdot P -  \la Q - \bla \Qb } \nn \\
&= j_a (s) \left( \mathcal{A} \right) \, \hat{T}^a = \left( \, j_a ^\mu \, \partial_\mu + j_{a \a} {} ^A \, \partial_A {} ^\a + j_{a \, A \da} \, \partial ^{\da A} + j_{a A} {} ^B \,  \left( \gamma^{IJ} \right)_B {} ^A \, n^I \partial ^J \,    \right) \hat{T}^a \,.
\end{align}
The coefficients of the level-zero densities can be read off the explicit expressions provided in appendix~\ref{app:generators}. We thus have the following expression for the level-1 densities $j_a ^{(1)}$: 
\begin{align}
j_a ^{(1)} &= - j_a ^\mu \, l_\mu - j_{a \a} {} ^A \left(f_A {} ^\a - i \, \bla_{A \da} \, l^{\da \a} \right) - j_{a \, A \da} \left(\bar{f}^{\da A} -i \, l^{\da \a} \la_\a {} ^A \right)   + j_{a A} {} ^B \,  \rho_B {} ^A - j_a ^{(1)\, \prime} \, , 
\end{align}
The combination with the bi-local part of $\mathcal{Q}^{(1)}$ then gives the full level-1 charge
\begin{align}
\mathcal{Q}_a ^{(1)} &= \frac{1}{2} \, f \indices{^{cb} _a}  \int \diff s_1 \diff s_2 \, \varepsilon(s_1-s_2) \, j_b(s_1) (\mathcal{A}_\mathrm{ren}) \, j_c(s_2) (\mathcal{A}_\mathrm{ren}) + \int \diff s \, j_a ^{(1)} (s)  \nn \\
& \qquad + \int \diff s \left( j_a ^ \mu \, \pi _\mu - i \, j_{a \a} {} ^A \, \bla_{A \da} \, \pi^{\da \a} - i \, j_{a \, A \da} \, \pi ^{\da \a} \, \la _\a {} ^A \right) \left( n^I n^J - \delta^{IJ} \right) \frac{\delta \mathcal{A}_\mathrm{ren} }{\delta n^I} \, \frac{\delta \mathcal{A}_\mathrm{ren} }{\delta n^J} \,.
\end{align}
The vanishing of the level-1 charge $\mathcal{Q}^{(1)}$ may be rewritten as the invariance of the super Wilson loop under the level-1 Yangian generators
\begin{align}
\label{Ja1}
J_a ^ {(1)} &= f \indices{^{cb} _a}  \int \diff s_1 \diff s_2 \, \varepsilon(s_1-s_2) \, j_b(s_1) \, j_c(s_2) + \frac{\la}{2 \pi^2} \int \diff s \, j_a ^{(1)} (s)   \nn \\
& \quad   + 2 \int \diff s \left( j_a ^ \mu \, \pi _\mu - i \, j_{a \a} {} ^A \, \bla_{A \da} \, \pi^{\da \a} - i \, j_{a \, A \da} \, \pi ^{\da \a} \, \la _\a {} ^A \right) \left( n^I n^J - \delta^{IJ} \right) \frac{\delta^2 }{\delta n^I \delta n^J} \,. 
\end{align}
Note that in the above equation the $\lambda$ appearing in front of the integral denotes the 't Hooft coupling, whereas the $\lambda$'s appearing under the integrals denote the fermionic piece of the boundary curve. Concretely, applying the above generator to the super Wilson loop gives
\begin{align*}
J_a^{(1)} \left \langle \mathcal{W}(C) \right \rangle = \left( \la \mathcal{Q}_a ^{(1)} + \mathcal{O}\left(\sqrt{\la}\right) \right) \left \langle \mathcal{W}(C) \right \rangle \, .
\end{align*}
To discuss our result, consider the level-1 momentum generator $P^{(1)\, \mu}$, which was also studied in \cite{Muller:2013rta}, 
\begin{align}
P^{(1) \, \mu} &= f \indices{^{cb} _{P^\mu}}  \int \diff s_1 \diff s_2 \, \varepsilon(s_1-s_2) \, j_b(s_1) \, j_c(s_2) +  2 \int \diff s \, \pi ^\mu \left( n^I n^J - \delta^{IJ} \right) \frac{\delta^2 }{\delta n^I \delta n^J}   \nn \\
& \quad + \frac{\la}{2 \pi^2} \int \diff s \, \Big \lbrace \ddot{\pi}^\mu + \left( \dot{\pi}^2 - \dot{n}^2 \right) \pi ^\mu + 4i \tr \big(\dot{\la} \rho \dot{\bla} \s ^\mu \big) - 3i \tr \big( \dot{\la} \dot{\bla} \big( \dot{\pi} \bs ^\mu \pi -  \pi \bs ^\mu \dot{\pi}  \big) \big) \nn \\
& \qquad \qquad \qquad - \tr \big(12 \, \dot{\bla} \pi \dot{\la} \dot{\bla} \pi \dot{\la} + 2i \big(\ddot{\la} \dot{\bar{\la}} -  \dot{\la} \ddot{\bar{\la}} \big) \pi - 4i \big(\theta_{(2)} \dot{\bla} - \dot{\la} \btheta_{(2)} + \dot{\la} \rho \dot{\bla} \big) \pi  \big) \pi ^\mu \Big \rbrace
\end{align}
Compare equations \eqn{theta2} and \eqn{rhoAB} for the variables $\theta_{(2)}$ and $\rho$. The bilocal term has the typical structure of a level-1 Yangian generator and generalizes the bosonic result to the superconformal algebra $\mathfrak{psu}(2,2\vert4)$. In comparison to the $AdS_5$ result \eqn{Level1Old} we find a structurally new contribution in the local term involving two functional derivatives acting on the same point of the loop. This contribution is due to the inclusion of non-trivial boundary curves on $S^5$ and also appears at the purely bosonic order. We also see that the local term reproduces the $AdS_5$ result if we set the fermionic coordinates zero and the sphere vector $n^I$ constant. This shows in particular that the inclusion of fermionic degrees of freedom at strong coupling does not affect the $\la$-dependence of the local term, which was discussed in \cite{Muller:2013rta}. There thus seems to be a non-trivial interpolating function $f(\la)$ in the level-1 Yangian generator, that encodes the symmetry of the super Wilson loop for any value of the coupling constant $\la$, 
\begin{align}
P^{(1) \, \mu} &= f \indices{^{cb} _{P^\mu}}  \int \diff s_1 \diff s_2 \, \varepsilon(s_1-s_2) \, j_b(s_1) \, j_c(s_2) +  2 \int \diff s \, \pi ^\mu \left( n^I n^J - \delta^{IJ} \right) \frac{\delta^2 }{\delta n^I \delta n^J}   \nn \\
& \quad + f(\la) \, \int \diff s \, \Big \lbrace \ddot{\pi}^\mu + \left( \dot{\pi}^2 - \dot{n}^2 \right) \pi ^\mu + 4i \tr \big(\dot{\la} \rho \dot{\bla} \s ^\mu \big) - 3i \tr \big( \dot{\la} \dot{\bla} \big( \dot{\pi} \bs ^\mu \pi -  \pi \bs ^\mu \dot{\pi}  \big) \big) \nn \\
& \qquad \qquad \qquad - \tr \big(12 \, \dot{\bla} \pi \dot{\la} \dot{\bla} \pi \dot{\la} + 2i \big(\ddot{\la} \dot{\bar{\la}} -  \dot{\la} \ddot{\bar{\la}} \big) \pi - 4i \big(\theta_{(2)} \dot{\bla} - \dot{\la} \btheta_{(2)} + \dot{\la} \rho \dot{\bla} \big) \pi  \big) \pi ^\mu \Big \rbrace \,.
\end{align}
Here, we have assumed that the local term has the same contour dependence for any value of $\la$ as the weak coupling calculation for the terms that are of Gra{\ss}mann order zero indicates \cite{Muller:2013rta}. Further evidence for this assumption could be obtained from comparing with the weak coupling result of \cite{Beisert:2015uda}. The generalization to the other level-1 generators should then be obtained from equation \eqn{Ja1}.

\section{Conclusion and Outlook}
In this work we provided a strong coupling description of smooth Wilson loops in $\mathcal{N}=4$ superspace in terms of minimal surfaces of the $AdS_5 \times S^5$ superstring and derived the superconformal and Yangian Ward identities for these super Wilson loops from the classical integrability of the string model. In doing this, we have studied minimal surfaces of the $AdS_5 \times S^5$ superstring and determined their behaviour close to the conformal boundary.

In \cite{Muller:2013rta} the superspace for the super-Wilson loop was adapted to the field content of $\mathcal{N}=4$ SYM by hand. The construction of the superconformal boundary of the string superspace \cite{Ooguri:2000ps} provides a natural candidate for this superspace and we have derived the superconformal symmetry generators that should leave the super Wilson loop invariant. The generators are listed in appendix~\ref{app:generators}. Given the matured understanding of the appropriate superspace and symmetry generators, it would be interesting to revisit the construction of the super Wilson loop as performed in \cite{Muller:2013rta}, which was based only on the requirement of superconformal Ward identities for the super Wilson loop. This provides a different construction principle than the one used in \cite{Ooguri:2000ps,Beisert:2015jxa}, which is based on dimensional reduction from ten-dimensional $\mathcal{N}=1$ superspace and provides the on-shell form of the Wilson loop in $\mathcal{N}=4$ superspace.  

The Yangian symmetry generators we derived show a new structure in the local term, that has not been observed in the weak coupling analysis carried out in \cite{Muller:2013rta}. It will be interesting to compare them to the Yangian symmetry generators for the super Wilson loop at weak coupling \cite{Beisert:2015uda}. As mentioned in the introduction the super Wilson loop we consider may be thought of as the smooth counterpart of the lightlike polygonal non-chiral super Wilson loops of \cite{Beisert:2012xx,CaronHuot:2011ky}. As the cusped Wilson loops stand at the heart of the duality relations to scattering amplitudes or correlators, it would be very interesting to study the relation between smooth and cusped (super) Wilson loops in detail and to find out what implications the Yangian symmetry of the smooth super Wilson loops has for the cusped ones.   

The Yangian symmetry certainly constrains the functional form of the expectation value of the super Wilson loop. In order to make progress towards possible exact results it would be important to understand the structure of Yangian invariants, which should form the building blocks of the exact results. In \cite{Kruczenski:2014bla,Ishizeki:2013hla,Dekel:2015bla} it was shown that the renormalized areas corresponding to a family of boundary curves parametrized by a spectral parameter are the same. The different contours are not related by conformal transformations and it would be interesting to investigate the relation to the hidden symmetries presented in this work.

\subsection*{Acknowledgements}
It is a pleasure to thank Konstantin Zarembo for crucial input as well as for many enlightening discussions.
Moreover, we would like to thank Harald Dorn, Ben Hoare, Thomas Klose, Florian Loebbert, Dennis M\"uller and Jan Plefka for interesting and often very useful discussions as well as Jan Plefka and Konstantin Zarembo for valuable comments on the draft.
Furthermore we would like to thank NORDITA for hospitality in the early phases
of this project.

This research is supported in part by the SFB 647 \textit{``Raum-Zeit-Materie.
Analytische und Geometrische Strukturen''}, the Research Training Group GK 1504
\textit{``Mass, Spectrum, Symmetry''} and the Marie Curie network GATIS of the
European Union’s Seventh Framework Programme FP7/2007-2013/ under REA Grant
Agreement No 317089.

\newpage
\appendix

\section{Spinor Conventions}
\label{app:spinorconv}
The raising and lowering of spinor indices is given by
\begin{align}
\la^\a = \varepsilon ^{\a \b} \la _\b \,, \quad \la _\a = \la ^\b \varepsilon_{\b \a} \,, \quad \bar \la _\da = \varepsilon_{\da \db} \bar \la ^\db \, , \quad \bar \la ^\da = \bar \la _\db \varepsilon ^{\db \da}
\end{align}
The epsilon tensor is defined by 
\begin{align}
\varepsilon^{1 2} = \varepsilon_{1 2} = 1 \, , \quad \varepsilon^{\dot{1} \dot{2}} = \varepsilon_{\dot{1} \dot{2}} = -1 \quad \Rightarrow \varepsilon ^{\a \b} \varepsilon _{\g \b} = \delta ^\a _\g \, , \quad \varepsilon ^{\da \db} \varepsilon _{\dg \db} = \delta ^\da _\dg
\end{align}
We introduce the following convention for sigma matrices
\begin{equation}
\begin{alignedat}{2}
\left( \sigma ^\mu \right) ^{\da \a} &= \left( \mathbb{I}_2 , \vec{\s} \right) ^{\da \a} \, , & \qquad \left( \bs ^\mu \right) _{\a \da} &= \left( \mathbb{I}_2, -\vec{\s} \right) _{\a \da} \\
\left( \sigma ^{\mu \nu} \right) \indices{_\a ^\b} &= \ihalf \left( \bs ^\mu \s ^\nu - \bs ^\nu \s ^\mu \right) \indices{_\a ^\b} \, , & \qquad \left( \bs ^{\mu \nu} \right)\indices{^\da _\db} &= \ihalf \left( \s ^\mu \bs ^\nu - \s ^\nu \bs ^\mu \right) \indices{^\da _\db}
\end{alignedat}
\end{equation}
With the mostly plus metric $\eta = \diag(-,+,+,+)$, these matrices satisfy the following identities:
\begin{equation}
\begin{alignedat}{2}
\bs ^\mu _{\a \dg} \, \s ^{\nu \, \dg \b} + \bs ^\nu _{\a \dg} \, \s ^{\mu \, \dg \b} &= - 2\, \eta^{\mu \nu} \, \delta ^\b _\a \, , \qquad \qquad &  \s ^{\mu \, \da \a} \, \bs _{\mu \, \b \db} &=- 2 \, \delta ^\a _\b \, \delta ^\da _\db \, ,\\
\s ^{\mu \, \da \g} \, \bs ^\nu _{\g \db} +  \s ^{\nu \, \da \g} \, \bs ^\mu _{\g \db}  &= - 2\, \eta^{\mu \nu} \, \delta ^\da _\db \, , & \bs ^\mu _{\a \da} \, \s ^{\nu \, \da \a} &= - 2 \, \eta^{\mu \nu} \,.
\end{alignedat}
\end{equation}
We also note the following trace-identities:
\begin{equation}
\begin{aligned}
\half \, \Tr(\overline\sigma^{\mu}\, \sigma^{\nu}\, \overline\sigma^{\rho}\,\sigma^{\kappa}) &=
\eta^{\mu\nu}\, \eta^{\rho\kappa} + \eta^{\nu\rho}\, \eta^{\mu\kappa} -
\eta^{\mu\rho}\, \eta^{\nu\kappa} -i\,\epsilon^{\mu\nu\rho\kappa} \, , \\
\half \, \Tr(\sigma^{\mu}\, \overline\sigma^{\nu}\, \sigma^{\rho}\,\overline\sigma^{\kappa}) &=
\eta^{\mu\nu}\, \eta^{\rho\kappa} + \eta^{\nu\rho}\, \eta^{\mu\kappa} -
\eta^{\mu\rho}\, \eta^{\nu\kappa} +i\,\epsilon^{\mu\nu\rho\kappa} \,.
\end{aligned}
\end{equation}
We assign bispinors to four-vectors and two-tensors by 
\begin{equation}
\begin{aligned}
x^{\da \a} &= \s ^{\mu \, \da \a} \, x_\mu \, , \qquad x_{\a \da} = \bs ^\mu _{\a \da} \, x_\mu \, , \qquad x^\mu = - \half \sigma^{\mu \, \da \a} \, x_{\a \da} \, , \\
B_\a {}^\b &= B_{\mu \nu} \, \left( \sigma ^{\mu \nu} \right) \indices{_\a ^\b} \, , \hspace{31mm} \bar{B}\indices{^\da _\db} = B_{\mu \nu} \,  \left( \bs ^{\mu \nu} \right)\indices{^\da _\db} \,.
\end{aligned}
\end{equation}
For these we have the following identities:
\begin{align}
x^{\da \a} \, x_{\a \db} = -x^2 \, \delta ^\da _\db \, , \quad x_{\a \da} \, x^{\da \b} = -x^2 \delta ^\b _\a , \qquad x_{\a \da} \, y^{\da \a} = - 2 \, x y 
\end{align}
In our paper the fermionic coordinates and related quantities have canonical index positions, which are given by:
\begin{alignat}{3}
&{\theta_\a}^A \, , \btheta_{A \da} & \quad &\text{for variables conjugate to} \quad {Q_A}^\a \, , \, \Qb^{\da A} \, , \nn \\
&{\vart_A}^\a \, , \bvart^{\da A} & \quad &\text{for variables conjugate to} \quad {S_\a}^A \, , \, \Sb_{A \da} \, \,. \nn
\end{alignat}
Whenever a spinor index is raised or lowered into a different position, we spell out the indices explicitly. In a matrix product, the indices of bispinors are positioned accordingly. We provide the following examples for clarity:
\begin{align*}
\tr \big( \btheta \bvart + \vart \theta \big) = \btheta_{A \da} \bvart^{\da A} + \vart_A {} ^\a \theta_\a {} ^A \, , \qquad \big[ \dot{\bar{\la}}\pi \dot{\la} \dot{\bla}  \pi \big]_A {} ^\a =  \dot{\bar{\la}}_{A \db} \, \pi ^{\db \b} \, \dot{\la}_\b {}^B \, \dot{\bla}_{B \dg} \, \pi^{\dg \a} \,.
\end{align*}

\section{Transformation Behaviour of the Bosonic Local Term}
\label{app:transf}

In this appendix we prove that the local term 
\begin{align}
J_{a, \, \mathrm{local}} ^{(1)} = \frac{\la}{2 \pi ^2} \int \limits _0 ^L \diff s \, \xi^\mu _a (x) \left( \dx_\mu \, \ddot{x}^2 + \dddot{x}_\mu \right) \label{local}
\end{align}
of the bosonic Yangian symmetry generators derived in \cite{Muller:2013rta} indeed transforms as
\begin{align}
\left[ J^{(0)} _a \, , \,  J_{b , \,  \mathrm{local}} ^{(1)} \right] = f \indices{_{ab}^c} \, J_{c  , \, \mathrm{local}} ^{(1)} \,.  \label{level1transf}
\end{align}
Here, $\xi^\mu _a (x)$ denote conformal Killing vectors which satisfy the identities
\begin{equation}
\begin{alignedat}{2}
\xi^\rho _a  \,  \partial _\rho \, \xi^\mu _b  - \xi^\rho _b  \,  \partial _\rho \, \xi^\mu _a  &= f _{a b} {} ^c  \xi^\mu _c  \,  , &  \qquad  \partial ^\mu \, \xi^\nu _a + \partial ^\nu \, \xi^\mu _a &= \half \left( \partial_\kappa \xi ^\kappa _a \right) \eta^{\mu \nu} \, ,  \\
\quarter \left( \eta^{\mu \lambda} \, \partial^\nu + \eta ^{\nu \lambda} \, \partial^\mu - \eta ^{\mu \nu} \, \partial^\lambda \right) \left(\partial_\kappa \, \xi ^\kappa _a \right) &= \partial ^\mu \partial^\nu \, \xi ^\lambda _a \,  ,&  \qquad \partial_\mu \partial_\nu \partial_\lambda \, \xi ^\kappa _a &= 0 \,. \label{Killing}
\end{alignedat}
\end{equation}
Moreover, in \eqn{local} the parametrization is fixed to satisfy $\dx^2 = 1$, which implies that $\dx \cdot \ddot{x} = 0$. We indicate the use of an arc-length parametrization by stating the boundaries $0$ and $L$ of the integration domain. In order to derive the transformation behaviour \eqn{level1transf}, we need to rewrite \eqn{local} as a reparametrization invariant curve integral, since $\delta \lvert \dx \rvert \neq 0$ also if we have fixed a parametrization for which $\lvert \dx \rvert = 1$. One may easily convince oneself that 
\begin{align}
J_{a , \, \mathrm{local}} ^{(1)} = \frac{\la}{2 \pi ^2} \int  \diff s \, \xi^\mu _a (x) \, \left( \dx_\mu \left( \frac{1}{\lvert \dx \rvert} \partial_s \left( \frac{\dx_\mu}{\lvert \dx \rvert } \right) \right)^2  +      \partial_s \left( \frac{1}{\lvert \dx \rvert }  \partial_s \left( \frac{\dx_\mu}{\lvert \dx \rvert } \right) \right) \right)
\end{align}
is reparametrization invariant and reproduces \eqn{local} for an arc-length parametrization. We are thus able to compute the variation of $J_{b , \,  \mathrm{local}} ^{(1)}$:
\begin{align}
\delta J_{b , \,  \mathrm{local}} ^{(1)} &= \frac{\la}{2 \pi ^2} \int \limits _0 ^L \diff s \Big \lbrace \left( \partial _\rho \, \xi ^\mu _b \right) \left( \dx_\mu \, \ddot{x}^2 + \dddot{x}_\mu \right) - \partial_s \left[ \xi ^\mu _b \left( \eta_{\mu \rho} \, \ddot{x}^2 - 4 \, \dx_\mu \dx_\rho \, \ddot{x}^2 \right)  \right] \nn \\ 
& \qquad \qquad + 2 \partial_s ^2 \left( \xi ^\mu _b \, \dx_\mu \dx_\rho \right) - \partial_s \left[ \left( \partial_s \, \xi ^\mu _b \right) \ddot{x}_\mu \dx_\rho + \left( \partial_s ^2 \, \xi ^\mu _b \right) \left( \eta_{\mu \rho} - \dx_\mu \dx_\rho \right)   \right] \Big \rbrace \delta x^\rho(s)
\end{align}
Note that we have reverted back to an arc-length parametrization after calculating the variation. Using the above result, one finds:
\begin{align}
\left[ J^{(0)} _a \, , \,  J_{b , \,  \mathrm{local}} ^{(1)} \right] &= \int \limits _0 ^L \diff s \,  \xi ^\rho _a (x) \, \dfrac{\delta J_{b , \,  \mathrm{local}} ^{(1)} }{\delta x^\rho(s)} \nn \\
&=  \frac{\la}{2 \pi ^2} \int \limits _0 ^L \diff s \Big \lbrace \xi ^\rho _a \left( \partial _\rho \, \xi ^\mu _b \right) \left( \dx_\mu \, \ddot{x}^2 + \dddot{x}_\mu \right) + \left( \partial_s \, \xi ^\rho _a \right) \xi_{\rho b} \, \ddot{x}^2 + \left( \partial_s ^3 \, \xi ^\rho _a \right) \xi_{\rho b}  \nn \\
& + \left( \partial_s \, \xi ^\rho _a \right) \big[ -4 \, \xi ^\mu _b \dx_\mu \dx_\rho \, \ddot{x}^2 - 2  \, \partial_s  \left( \xi ^\mu _b \, \dx_\mu \dx_\rho \right) + \left( \partial_s \, \xi ^\mu _b \right) \ddot{x}_\mu \dx_\rho - \left( \partial_s ^2 \, \xi ^\mu _b \right) \, \dx_\mu \dx_\rho  \big] \Big \rbrace \label{step1}
\end{align}
Due to the use of an arc-length parametrization and the identities \eqn{Killing} one finds that
\begin{align}
\partial _s ^3 \, \xi ^\rho _a = \left( \partial ^\lambda \, \xi ^\rho _a \right) \, \dddot{x}_\lambda + 3 \partial_s \left( \partial ^\lambda \, \xi ^\rho _a \right) \ddot{x}_\lambda = -\left( \partial ^\rho \, \xi ^\lambda _a \right) \, \dddot{x}_\lambda + \half \left(\partial_\kappa \, \xi ^\kappa _a \right) \dddot{x}^\rho + 3 \partial_s \left( \partial ^\lambda \, \xi ^\rho _a \right) \ddot{x}_\lambda
\end{align}
We can thus rewrite the first line of \eqn{step1} as 
\begin{align}
\int \limits _0 ^L \diff s \, \big \lbrace \left( \xi ^\rho _a  \partial _\rho \, \xi ^\mu _b - \xi ^\rho _b  \partial _\rho \, \xi ^\mu _a  \right) \left( \dx_\mu \, \ddot{x}^2 + \dddot{x}_\mu \right) + \half \, \xi ^\rho _b \big[ \left( \partial_\kappa \, \xi ^\kappa _a \right) \left( \dx_\rho \, \ddot{x}^2 + \dddot{x}_\rho \right) + 6 \, \partial_s \left( \partial ^\lambda \, \xi ^\rho _a \right) \ddot{x}_\lambda  \big]    \big \rbrace \,.
\end{align}
Accordingly, we find
\begin{align}
\left[ J^{(0)} _a \, , \,  J_{b , \,  \mathrm{local}} ^{(1)} \right] = f \indices{_{ab}^c} \, J_{c  , \, \mathrm{local}} ^{(1)} + R_{ab}
\end{align}
and still need to show that
\begin{align}
R_{ab} &=  \frac{\la}{4 \pi ^2} \int \limits _0 ^L \diff s \, \Big \lbrace \left( \partial_s \, \xi ^\rho _a \right) \big[ -4 \, \xi ^\mu _b \dx_\mu \dx_\rho \, \ddot{x}^2 - 2  \, \partial_s  \left( \xi ^\mu _b \, \dx_\mu \dx_\rho \right) + \left( \partial_s \, \xi ^\mu _b \right) \ddot{x}_\mu \dx_\rho - \left( \partial_s ^2 \, \xi ^\mu _b \right) \, \dx_\mu \dx_\rho  \big] \nn \\
& \qquad \qquad \quad + \xi ^\rho _b \big[ \left( \partial_\kappa \, \xi ^\kappa _a \right) \left( \dx_\rho \, \ddot{x}^2 + \dddot{x}_\rho \right) + 6 \, \partial_s \left( \partial ^\lambda \, \xi  _{\rho a} \right) \ddot{x}_\lambda  \big] \Big \rbrace = 0 \,. \label{step2}
\end{align}
This may be achieved by applying the conformal Killing vector identities \eqn{Killing}. Consider for example the first term:
\begin{align}
\int \limits _0 ^L \diff s \left( \partial_s \, \xi ^\rho _a \right) \left( -4 \, \xi ^\mu _b \dx_\mu \dx_\rho \, \ddot{x}^2 \right) &= \int \limits _0 ^L \diff s \left( \half \, \dx ^\rho \left( \partial_\kappa \, \xi ^\kappa _a \right) - \left( \partial^\rho \, \xi ^\lambda _a \right) \dx_\lambda \right)  \left( -4 \, \xi ^\mu _b \dx_\mu \dx_\rho \, \ddot{x}^2 \right) \nn \\
&= - 2 \int \limits _0 ^L \diff s \left( \partial_\kappa \, \xi ^\kappa _a \right) \xi ^\mu _b \dx_\mu \, \ddot{x}^2 - \int \limits _0 ^L \diff s \left( \partial_s \, \xi ^\la _a \right) \left( -4 \, \xi ^\mu _b \dx_\mu \dx_\la \, \ddot{x}^2 \right) \nn \\
\Rightarrow \int \limits _0 ^L \diff s \left( \partial_s \, \xi ^\rho _a \right) \left( -4 \, \xi ^\mu _b \dx_\mu \dx_\rho \, \ddot{x}^2 \right) &= - \int \limits _0 ^L \diff s \left( \partial_\kappa \, \xi ^\kappa _a \right) \xi ^\mu _b \dx_\mu \, \ddot{x}^2
\end{align}
Analogously one finds that
\begin{align}
\int \limits _0 ^L \diff s \left( \partial_s \, \xi ^\rho _a \right) \left( - \left( \partial_s ^2 \, \xi ^\mu _b \right) \, \dx_\mu \dx_\rho \right) &= - \frac{1}{4} \int \limits _0 ^L \diff s \left( \partial_\kappa \, \xi ^\kappa _a \right) \left( \partial_s ^2 \, \xi ^\mu _b \right) \, \dx_\mu \, , \nn \\
\int \limits _0 ^L \diff s \left( \partial_s \, \xi ^\rho _a \right) \left( \partial_s \, \xi ^\mu _b \right) \ddot{x}_\mu \dx_\rho &= \frac{1}{4} \int \limits _0 ^L \diff s \left( \partial_\kappa \, \xi ^\kappa _a \right) \left( \partial_s \, \xi ^\mu _b \right) \ddot{x}_\mu \,.
\end{align}
Moreover, we have
\begin{align}
3 \int \limits _0 ^L \diff s \, \partial_s \left( \partial ^\la \, \xi ^\rho _a \right) \, \xi_{\rho b} \, \ddot{x}_\la &= \frac{3}{4}  \int \limits _0 ^L \diff s \left( \eta ^{\la \rho} \partial ^\mu \left( \partial_\kappa \, \xi ^\kappa _a \right) + \eta ^{\mu \rho} \partial ^\la \left( \partial_\kappa \, \xi ^\kappa _a \right) \right) \dx_\mu \ddot{x} _\la \, \xi_{\rho b} \nn \\
&= \frac{3}{4}  \int \limits _0 ^L \diff s \left(  \partial _s \left( \partial_\kappa \, \xi ^\kappa _a \right) \xi^\mu _b \, \ddot{x}_\mu  +  \partial ^\la \left( \partial_\kappa \, \xi ^\kappa _a \right) \ddot{x}_\la \xi^\mu _b \, \dx_\mu  \right)  \nn \\
&= \frac{3}{4}  \int \limits _0 ^L \diff s  \left( \partial_\kappa \, \xi ^\kappa _a \right) \left( \left(\partial_s ^2 \xi^\mu _b \right) \, \dx_\mu  + \left(\partial_s  \xi^\mu _b \right) \, \ddot{x}_\mu  \right) \, , \\
2 \int \limits _0 ^L \diff s \left( \partial_s ^2 \xi ^\rho _a \right) \ddot{x}_\rho \, \xi^\mu _b \dx_\mu &= \frac{1}{2} \int \limits _0 ^L \diff s \left( \partial_\kappa \, \xi ^\kappa _a \right)  \left( \xi^\mu _b \, \dx_\mu \ddot{x}^2 - \partial _s ^2 \left( \xi ^\mu _b \dx_\mu \right) \right) \,.
\end{align}
Inserting these results into \eqn{step2}, one indeed finds $R_{ab} = 0$ which concludes the proof.

\section{The fundamental Representation of \texorpdfstring{$\mathfrak{su}(2,2\vert4)$}{su(2,2;4)}}
\label{app:su224}
In this appendix, we introduce the fundamental representations of $\mathfrak{su}(2,2 \vert 4)$, which is used throughout this paper. We begin with the fundamental representation of the $R$-symmetry part $\mathfrak{su}(4)$. 

\subsection*{The fundamental Representation of \texorpdfstring{$\mathfrak{su}(4)$}{su(4)}}
\label{app:su4}
Following \cite{Arutyunov:2009ga}, we fix an explicit representation of Dirac matrices:
\begin{equation}
\begin{alignedat}{3}
\gamma ^1 &= \begin{pmatrix} 0 & -i \sigma^2 \\ i \sigma^2 & 0  \end{pmatrix} \, , & \qquad \gamma ^2 &= \begin{pmatrix} 0 & i \sigma^1 \\ -i \sigma^1 & 0  \end{pmatrix} \, ,& \qquad \gamma ^3 &= \begin{pmatrix} 0 & \mathbb{I}_2 \\ \mathbb{I}_2 & 0  \end{pmatrix} \, , \\
\gamma ^4 &= \begin{pmatrix} 0 & -i \sigma^3 \\ i \sigma^3 & 0  \end{pmatrix} \, , & \qquad \gamma ^5 &= \begin{pmatrix} \mathbb{I}_2 & 0 \\ 0 & \mathbb{I}_2  \end{pmatrix} 
\end{alignedat}
\end{equation}
These matrices satisfy the Clifford algebra for $SO(5)$,
\begin{align}
\left \lbrace \gamma ^a \, , \, \gamma ^b \right \rbrace = 2 \, \delta^{a b} \, \mathbb{I}_4  \, , \qquad a,b \in \left \lbrace 1 \, , \ldots , 5 \right \rbrace \,. 
\end{align}
Based on these matrices we may construct a set of matrices $\gamma^{I J} = - \gamma^{J I}$, which form a basis of $\mathfrak{su}(4) \simeq \mathfrak{so}(6)$:
\begin{equation}
\begin{alignedat}{3}
\gamma^{a b} &= \quarter \, \left[ \gamma ^a \, , \, \gamma ^b \right] \, ,& \qquad \gamma^{a 5} &= \ihalf \gamma^a \, ,& \qquad   a,b &\in \left \lbrace 1 \, , \ldots , 4 \right \rbrace \\
\gamma^{a 6} &= \quarter \, \left[ \gamma ^a \, , \, \gamma ^5 \right] \, , & \qquad \gamma^{5 6} &= - \ihalf \gamma^5 \,.
\end{alignedat} 
\end{equation}
They satisfy the commutation relations
\begin{align}
\left[ \gamma^{I J} \, , \, \gamma^{K L} \right] &= \delta ^{I L} \, \gamma^{J K} + \delta ^{J K} \, \gamma^{I L} - \delta ^{I K} \,  \gamma^{J L} - \delta ^{J L} \, \gamma^{I K} \,.
\end{align}
The matrices $\left \lbrace \gamma^{a b} \, , \gamma ^{a 6} \, \vert a,b \in   \left \lbrace 1 \, , \ldots , 4 \right \rbrace \right \rbrace$ span the sub-algebra $\mathfrak{so}(5) \subset \mathfrak{so}(6)$. We also introduce the matrix
\begin{align}
K = - \, \gamma ^2 \, \gamma ^4 = \begin{pmatrix} -i \sigma ^2 & 0 \\ 0 & -i \sigma^2 \end{pmatrix} \, , \qquad K \, \left( \gamma ^a \right)^T = \gamma ^a \, K \,. \label{Kdef} 
\end{align}
This matrix can be used to define a $\mathbb{Z}_2$-grading on $\mathfrak{su}(4)$ according to the definitions
\begin{align}
\mathfrak{su}(4) ^{(0)} := \left \lbrace m \in  \mathfrak{su}(4) \, \vert \,  m^t = - K m K^{-1} \right \rbrace , \, \, \mathfrak{su}(4) ^{(2)} := \left \lbrace m \in  \mathfrak{su}(4) \, \vert \, m^t = K m K^{-1} \right \rbrace \,.
\end{align}
Using the relation $(\gamma ^a)^t = K \gamma^a K^{-1} \, , \, a \in \left \lbrace 1 \, , \ldots , 5 \right \rbrace$ one finds that 
\begin{equation}
\begin{aligned}
\mathfrak{su}(4) ^{(0)} &= \mathrm{span} \left \lbrace \gamma^{a b} \, , \gamma ^{a 6} \, \vert a,b \in   \left \lbrace 1 \, , \ldots , 4 \right \rbrace \right \rbrace \simeq \mathfrak{so}(5) \, , \\
\mathfrak{su}(4) ^{(2)} &= \mathrm{span} \left \lbrace \gamma^{a 5} \, , \gamma ^{5 6} \, \vert a \in   \left \lbrace 1 \, , \ldots , 4 \right \rbrace \right \rbrace \,.
\end{aligned}
\end{equation}
Thus the grading gives rise to the decomposition $\mathfrak{so}(6) = \mathfrak{so}(5) \oplus \mathfrak{f}$, which may be employed to construct the coset space $SO(6)/SO(5)\simeq S^5$. 

\subsection*{Supermatrix Representation of \texorpdfstring{$\mathfrak{su}(2,2\vert4)$}{su(2,2;4)}}
The superalgebra $\mathfrak{su}(2,2\vert 4)$ can be defined as the set of $(4\vert 4)$ supermatrices satisfying $\str(B) =0$ and the following reality condition:
\begin{align}
B = \begin{pmatrix}
m & \theta \\ \eta & n
\end{pmatrix} = \begin{pmatrix}
-H\, m^\dag H^{-1} & -H\, \eta^\dag   \\ -\theta^\dag H^{-1} & - n ^\dag
\end{pmatrix} = - \begin{pmatrix}
H & 0 \\ 0 & \mathbb{I}_4 
\end{pmatrix} B^\dag \begin{pmatrix}
H^{-1} & 0 \\ 0 & \mathbb{I}_4 
\end{pmatrix} \,.
\end{align}
Here, the matrix $H$ is given by
\begin{align}
H = \begin{pmatrix}
0 & \mathbb{I}_2  \\ \mathbb{I}_2 & 0  
\end{pmatrix} \,.
\end{align}
This choice of the matrix $H$ is better adapted to the choice of generators that are typically used on the field theory side and it differs from \cite{Arutyunov:2009ga}. The different choices for the matrix $H$ are related by a unitary transformation. To endow $\mathfrak{su}(2,2\vert4)$ with a $\mathbb{Z}_4$-grading consider the automorphism
\begin{align}
B \mapsto \Omega(B) = - \mathcal{K} \, B^{\mathrm{st}} \, \mathcal{K}^{-1} \, , \qquad \mathcal{K} = \begin{pmatrix}
K & 0 \\ 0 & K
\end{pmatrix} \,.
\end{align}
This automorphism gives a decomposition of the superalgebra $\mathfrak{sl}(4 \vert 4)$, according to the eigenspaces of $\Omega$: 
\begin{equation}
\begin{gathered}
\mathfrak{sl}(4 \vert 4) = \mathfrak{sl}(4 \vert 4)^{(0)} \oplus \mathfrak{sl}(4 \vert 4)^{(2)} \oplus \mathfrak{sl}(4 \vert 4)^{(1)} \oplus \mathfrak{sl}(4 \vert 4)^{(3)} \, , \\
	\mathfrak{sl}(4 \vert 4)^{(k)} \coloneqq \left \lbrace B \in \mathfrak{sl}(4 \vert 4) \, \vert \, \Omega(B) = i^k B \right \rbrace \,.
\end{gathered}
\end{equation}
Any element of $\mathfrak{sl}(4\vert4)$ may be projected onto an eigenspace by the prescription
\begin{align}
P^{(k)}(B) =  B^{(k)} = \quarter \left( B + i^{3k} \Omega(B) +  i^{2k} \Omega^2(B) +  i^{k} \Omega^3(B) \right) \, , \qquad \Omega (B^{(k)}) = i^k \, B^{(k)} \,.
\end{align}
While the automorphism $\Omega$ may not be restricted to $\mathfrak{su}(2,2\vert4)$, the projectors $P^{(k)}$ can, i.e.\
\begin{align}
B \in \mathfrak{su}(2,2\vert4) \Rightarrow B^{(k)} \in  \mathfrak{su}(2,2\vert4) \,.
\end{align}
This property does not depend on our particular choice of reality constraint. Thus one may define a grading on $\mathfrak{su}(2,2\vert4)$ by:
\begin{align*}
\mathfrak{su}(2,2\vert4) &= \mathfrak{g}^{(0)} \oplus \mathfrak{g}^{(2)} \oplus \mathfrak{g}^{(1)} \oplus \mathfrak{g}^{(3)} \, , \quad \text{where} \quad \mathfrak{g}^{(k)} := \left\lbrace  P^{(k)}(B) \, \vert \, B \in \mathfrak{su}(2,2\vert4) \right \rbrace \, , \\
\left[ \mathfrak{g}^{(k)} \, , \, \mathfrak{g}^{(l)} \right] &\subset \mathfrak{g}^{(k+l) \, \mathrm{mod} 4 }  \,.
\end{align*}
Following \cite{Drummond:2009fd}, we choose an explicit basis of the superalgebra $\mathfrak{su}(2,2\vert 4)$: 
\begin{align}
\left( \begin{array}{cc|c} 0 & P_\mu & Q _A {} ^\a \\
K_\mu & 0 & \bar{S}_{A \da} \\ \hline
S _\a {} ^A & \bar{Q}^{\da A} & R \indices{^A _B}
\end{array} \right) = 
\left( \begin{array}{cc|c} 0 & i \bs_\mu & 2 \, E \indices{^\a _A} \\
i \s_\mu & 0 & 2 \, E_{\da A} \\ \hline
- 2 \, E \indices{^A _\a} & -2 \, E^{A \da} & 4 \, E \indices{^A _B} - \delta ^A _B \, \mathbb{I}_4
\end{array} \right) 
\end{align}
This equation is to be read as
\begin{align}
P_\mu = \left( \begin{array}{cc|c} 0 & i \bs_\mu & 0 \\
0 & 0 & 0 \\ \hline
0 & 0 & 0
\end{array} \right)
\end{align}
and similarly for the other generators. The notation $E \indices{^A _B}$ denotes a matrix with entry 1 in the position $(A,B)$ and all other entries vanishing. The remaining generators of  $\mathfrak{su}(2,2\vert4)$ are given by
\begin{align}
M_{\mu \nu}  = -\frac{i}{2} \left( \begin{array}{c|c}
		 \begin{array}{cc}
		\s_{\mu \nu} & 0 \\ 0 & \bs_{\mu \nu} 
\end{array}	& 0 \\
		\hline
		0 & 0 
	\end{array} \right) \, , &&
D  = \frac{1}{2} \left( \begin{array}{c|c}
		 \begin{array}{cc}
		\mathbb{I}_2  & 0 \\ 0 & -\mathbb{I}_2 
\end{array}	& 0 \\
		\hline
		0 & 0 
	\end{array} \right) \, , &&
	C = \half \,  \mathbb{I} _8  \,.
\end{align}
Note that the fermionic generators do not satisfy the reality constraint, but the linear combinations $\theta _\a {} ^A Q_A {}^\a + \btheta _{A \da} \bar{Q}^{\da A}$ and $\vart _A {} ^\a S_\a  {}^A  + \bvart ^{\da A} \Sb_{A \da}$ do, provided that
\begin{align}
\btheta _{A \da} = \left( \theta _\a {} ^A  \right) ^\ast \, , \qquad \bvart ^{\da A} = \left( \vart_A {}^\a \right) ^\ast \, ,
\end{align}
such that they form the components of a ten-dimensional Majorana-Weyl spinor. Also the R-symmetry generators $R\indices{^A _B}$ do not satisfy the reality condition. Instead, the $\mathfrak{su}(4)$ sub-algebra is spanned by the matrices
\begin{align}
\Gamma ^{IJ} = \begin{pmatrix} 0 & 0 \\  0 & \gamma ^{I J} \end{pmatrix} \, ,
\end{align}
which are related to the $R\indices{^A _B}$ by $\Gamma^{IJ} = \quarter \, \left( \gamma^{IJ} \right){ } \indices{_A ^B} R \indices{^A _B} $. The generators $R\indices{^A _B}$ are however more convenient to write down commutation relations. We have the following for our specific choice of basis:
\begin{equation}
\begin{alignedat}{3}
 \Big[ D \, , \, P_\mu \Big] &= + P_\mu   \, & \quad \Big[ P_\mu \, , \, K_\nu \Big] &= + 2 \eta _{\mu \nu} \, D - 2 M_{\mu \nu}   \, & \quad  \Big[ D \, , \, K_\mu \Big] &= - K_\mu \\
\Big \lbrace Q _A {}^\a \, , \, \bar{Q}^{\da B} \Big \rbrace &= -2i \, \delta ^B _A P^{\da \a} \, & \quad \Big \lbrace S _\a {} ^A \, , \, \bar{S}_{B \da} \Big \rbrace &= -2i \, \delta ^A _B K_{\a \da} \, & \quad  \Big[ D \, , \, Q_A {} ^\a \Big] &= + \half Q_A {} ^\a \\
\Big[ K_{\a \da} \, , \, Q_A {}^\b \Big] &= -2i \, \delta^\b _\a \bar{S}_{A \da} \, & \quad \Big[ K_{\a \da} \, , \, \bar{Q}^{\db A} \Big] &= + 2i \, \delta^\db _\da S _\a {}^A \, & \quad  \Big[ D \, , \, S_\a ^A \Big] &= -\half  S_\a ^A \\
\Big[ P^{\da \a} \, , \, S_\b {} ^A \Big] &= + 2i \, \delta^\a _\b \bar{Q}^{\da A} \, & \quad \Big[ P^{\da \a} \, , \, \bar{S}_{A \db} \Big] &= - 2i \, \delta^\da _\db Q_A {}^\a
\end{alignedat}
\end{equation}
The commutators with the generators $M$ and $R$ only depend on the set of indices and their position:
\begin{equation}
\begin{alignedat}{2}
\Big[ M \indices{_\a ^\b} \, , \, J_\g \Big] &= 2i \, \delta ^\b _\g J_\a - i \delta ^\b _\a J_\g \, & \qquad \qquad  \Big[ M \indices{_\a ^\b} \, , \, J^\g \Big] &= - 2i \, \delta ^\g _\a J^\b + i \delta ^\b _\a J^\g \\
\Big[ \overline{M} \, \indices{^\da _\db} \, , \, J^\dg \Big] &= 2i \, \delta ^\dg _\db J^\da - i \delta ^\da _\db J^\dg \, & \qquad \qquad  \Big[ \overline{M} \, \indices{^\da _\db} \, , \, J_\dg \Big] &= - 2i \, \delta ^\da _\dg J_\db + i \delta ^\da _\db J_\dg \\
\Big[ R \indices{^A _B} \, , \, J^C \Big] &= 4 \, \delta ^C _B J^A - \delta ^A _B J^C \, & \qquad \qquad \Big[ R \indices{^A _B} \, , \, J_C \Big] &= - 4 \, \delta ^A _C J_B + \delta ^A _B J_C
\end{alignedat}
\end{equation}
The remaining non-vanishing commutators are given by
\begin{equation}
\begin{aligned}
\Big \lbrace Q _A {} ^\a \, , \, S_\b {}^B \Big \rbrace = -2i \, \delta ^B _A \, M \indices{_\b ^\a} - \delta ^\a _\b \, R \indices{^B _A} - 2 \, \delta ^B _A \, \delta ^\a _\b \left(D + C \right) \, ,\\
\Big \lbrace \bar{Q}^{\da A} \, , \, \bar{S}_{B \db} \Big \rbrace = -2i \, \delta ^A _B \, \overline{M} \, \indices{^\da _\db} - \delta ^\da _\db \, R \indices{^A _B} + 2 \, \delta ^A _B \, \delta ^\da _\db \left(D - C \right) \,.
\end{aligned}
\end{equation}
We collectively denote the generators defined above by $T_a$ and their structure constants by $\tilde{f}_{ab} {} ^c$, 
\begin{align}
\Big[ T_a \, , \, T_b \Big \rbrace = \tilde{f}_{ab} {} ^c \, T_c \,.
\end{align}
A set of generators of $\mathfrak{psu}(2,2 \vert 4)$ may be obtained by projecting out the central element $C$, i.e.\ identifying $T_a \sim T_a + \a \, C$, which is consistent since $C$ generates an ideal, $\left[ C , T_a \right] = 0$. We denote the structure constants of the corresponding basis of $\mathfrak{psu}(2,2 \vert 4)$ also by $\tilde{f}_{ab} {} ^c$ as they are the same with just a different range of indices. 

Let us now work out the projections of the supermatrix generators onto the graded components. For a general supermatrix
\begin{align}
B =  \begin{pmatrix}
m & \theta \\ 
\eta & n
\end{pmatrix} 
\end{align}
these projections can be given explicitly \cite{Arutyunov:2009ga},
\begin{alignat}{2}
B^{(0)} = \frac{1}{2} \, \begin{pmatrix}
m - K m^t K^{-1} & 0 \\
0 & n - K n^t K^{-1}
\end{pmatrix} \, , & \qquad  B^{(2)} = \frac{1}{2} \, \begin{pmatrix}
m + K m^t K^{-1} & 0 \\
0 & n + K n^t K^{-1}
\end{pmatrix} \, , \nn \\
B^{(1)} = \frac{1}{2} \, \begin{pmatrix}
0 & \theta - i K \eta^t K^{-1} \\
\eta + i K \theta^t K^{-1}  & 0
\end{pmatrix} \, , & \qquad  B^{(3)} = \frac{1}{2} \, \begin{pmatrix}
0 & \theta + i K \eta^t K^{-1} \\ 
\eta - i K \theta^t K^{-1}  & 0
\end{pmatrix} \,. \nn 
\end{alignat}
Making use of these identities we find: 
\begin{align}
\mathfrak{g}^{(0)} = \mathrm{span} \left \lbrace  M_{\mu \nu} , \half \left(P_\mu - K_\mu \right), \Gamma ^{i j} , \Gamma ^{i 6} \right \rbrace  \, , && \mathfrak{g}^{(2)} &= \mathrm{span} \left \lbrace  C , D , \half \left(P_\mu + K_\mu \right) , \Gamma ^{i 5} , \Gamma^{5 6} \right \rbrace
\end{align}
For the fermionic generators we introduce the notation $A^{(1) \pm (3)} = A^{(1)} \pm A^{(3)}$ and note that $ ( Q , S , \bar{Q} , \bar{S} ) ^{(1)+(3)} =  ( Q , S , \bar{Q} , \bar{S} )$ and
\begin{equation}
\begin{alignedat}{2}
\left( Q^\a _A \right) ^{(1)-(3)} &= i \, K_{A B} \, \epsilon^{\a \b} \, S_\b ^B \, , & \qquad \left( \bar{Q}^{A \da} \right) ^{(1)-(3)} &= - i \, K^{A B}  \, \bar{S}_{B \db} \, \epsilon^{\db \da} \, , \\
\left( S_\a ^A \right) ^{(1)-(3)} &= i \, K^{A B}  \, Q^\b _B \, \epsilon_{\b \a} \, ,& \qquad \left( \bar{S}_{A \da} \right) ^{(1)-(3)} &= - i \, K_{A B} \, \epsilon_{\da \db}  \, \bar{Q}^{B \db} \,.
\end{alignedat}
\end{equation}
A metric $\tilde{G}_{a b} = \left \langle T_a  ,  T_b \right \rangle = \str (T_a T_b)$ on the algebra is given by:
\begin{equation}
\begin{alignedat}{3}
\langle P^{\da \a} , K_{\b \db} \rangle &= - 4 \, \delta ^\a _\b \, \delta ^\da _\db & \quad  \langle D , D \rangle &= 1 & \quad \langle M \indices{_\a ^\b} , M \indices{_\g ^\eps} \rangle &= - 4 \, \delta ^\b _\g \, \delta ^\eps _\a + 2\, \delta ^\b _\a \, \delta ^\eps _\g  \\
\langle \bar{Q}^{\da A} , \bar{S}_{B \db} \rangle &=  4 \, \delta ^A _B \, \delta ^\da _\db  & \quad  \langle Q _A {} ^\a , S_\b {} ^B \rangle &= -4 \, \delta ^B _A \, \delta ^\a _\b  & \quad  \langle \bar{M} \indices{^\da _\db} , \bar{M} \indices{^\dg _{\dot{\eps}}} \rangle &= - 4 \, \delta ^\dg _\db \, \delta ^\da _{\dot{\eps}} + 2\, \delta ^\da _\db \, \delta ^\dg _{\dot{\eps}}   \\
\langle \bar{S}_{B \db} ,  \bar{Q}^{\da A} \rangle &= - 4 \, \delta ^A _B \, \delta ^\da _\db  & \quad  \langle S_\b {}^B ,  Q _A {} ^\a  \rangle &= 4 \, \delta ^B _A \, \delta ^\a _\b  & \quad  \langle R \indices{^A _B} , R \indices{^C _D} \rangle &= - 16 \, \delta ^A _D \, \delta ^C _B + 4 \, \delta ^A _B \, \delta ^C _D  \label{groupmetric}
\end{alignedat}
\end{equation}
All other entries are vanishing. We note moreover that
\begin{equation}
\begin{alignedat}{2}
\langle \Gamma ^{IJ} , \Gamma^{KL} \rangle &= 16 \left( \delta^{IL} \delta^{JK} - \delta^{IK} \delta^{JL} \right) \, , & \vspace{-10mm}
\langle R \indices{^A _B} , \Gamma^{I J} \rangle &= -4 \, \left( \gamma^{I J} \right) \indices{_B ^A} \, , \\
\Rightarrow \; \Gamma^{IJ} &= \quarter \, \left( \gamma^{IJ} \right) \indices{_A ^B} R \indices{^A _B}  \, , \\
\langle P^{(2)}  R \indices{^A _B} ,  P^{(2)}  R \indices{^C _D} \rangle &=  \langle  R \indices{^A _B} , P^{(2)}  R \indices{^C _D} \rangle = - 4 K_{B D}  & K^{A C} - 4 \delta ^A _D  \delta ^C_B + 2 \delta ^A _B & \delta ^C _D \,.
\end{alignedat}
\end{equation}
If we restrict to $\mathfrak{psu}(2,2 \vert 4 )$, the metric $\tilde{G}_{a b}$ becomes non-degenerate and we denote its inverse by $\tilde{G}^{a b}$,  $\tilde{G}^{a b} \tilde{G}_{b c} = \delta ^a _c$. Note also that $\tilde{G}_{a b}$ satisfies the symmetry property $\tilde{G}_{a b} = \left(-1 \right) ^{\lvert a \rvert} \tilde{G}_{b a}$, where $\lvert a \rvert = \mathrm{deg}(T_a)$ denotes the Gra{\ss}mann degree of a (homogeneous) basis element, $\lvert a \rvert = 0 \, (1)$ for an even (odd) generator.

\section{Densities of the Yangian Generators}
\subsection{Level 0}
\label{app:generators}
In this appendix we provide the differential generators $j_a (s)$ obtained from
\begin{align}
j_a (s) \left( \mathcal{A} \right) = \left\langle J_{\tau \, (0)}\, , \, T_a \right\rangle \, ,
\end{align}
which we write out explicitly in the form $p^\mu (s) \left( \mathcal{A} \right) = \left\langle J_{\tau \, (0)} (s)\, , \, P^\mu \right\rangle$ and similarly for all other generators. Here, we use the short-hand notation
\begin{align}
\partial^\mu = \frac{\delta }{\delta x_\mu (s)} \, , \quad \partial^{\da \a} = \sigma _\mu ^{\da \a} \, \partial ^\mu  \, , \quad \partial ^\a _A = \frac{\delta }{\delta \la _\a ^A(s)} \, , \quad \bar{\partial} ^{A \da} =  \frac{\delta }{\delta \bar{\la}_{A \da}(s)} \, , \quad \partial ^I = \frac{\delta }{\delta n^I (s)} \,.
\end{align}
Then, we have:
\begin{equation}
\begin{alignedat}{2}
p^{\mu} &= \partial^{\mu}  \, & \qquad  d &= \half \left( \la \, \partial _\la + \bar{\la} \, \bar{\partial} _{\la} \right) + x \cdot \partial _x \\
q^\a _A &= - \partial ^\a _A + i \bar{\la}_{A \da} \, \partial ^{\da \a} \, & \qquad  \bar{q}^{A \da} &= - \bar{\partial}^{A \da} + i \la ^A _\a \, \partial ^{\da \a} \label{levelzero1}  \\
m \indices {_\a ^\b} &= n \indices {_\a ^\b} - \half \, \delta ^\b _\a \, n \indices {_\g ^\g} \, & \qquad  \bar{m} \indices{ ^\da _\db} &= \bar{n} \indices{ ^\da _\db} - \half \delta ^\da _\db \, \bar{n} \indices{ ^\dg _\dg} \\
n \indices {_\a ^\b} &= -2i \, \la ^A _\a \, \partial ^\b _A + i \, x_{\a \da} \, \partial ^{\da \b} & \qquad  \bar{n} \indices{ ^\da _\db} &= 2i \, \bar{\la}_{A \db} \, \bar{\partial}^{A \da} - i \, x_{\a \db} \, \partial ^{\da \a} \\
 r \indices{^A _B} &= 4 \big( \big( \gamma ^{IJ} \big) _B {} ^A  \, n^I \partial ^J  + \bla _{B \da} \, \bar{\partial} ^{A \da}  -  & \la ^A _\a \, \partial ^\a _B  \big) -   \delta ^A _B  \big(  \bla \, \bar{\partial}_\la  &- \la \,  \partial _\la \big)   
\end{alignedat}
\end{equation}
Finally, we note the generators of special superconformal transformations: 
\begin{equation}
\begin{aligned}
s^A _\a &= i \, x^{-} _{\a \da} \, \bar{\partial}^{A \da} + x^{+} _{\a \da} \, \la ^A _\b \, \partial ^{\da \b} - 4 \, \la ^B _\a \, \la ^A _\b \, \partial^\b _B + 4 \la ^B _\a \, \big( \gamma ^{IJ} \big) _B {} ^A \, n^I \partial ^J
\\
\bar{s}_{A \da} &= -i \, x^{+} _{\a \da} \, \partial ^\a _A - x^{-} _{\a \da}  \, \bla_{A \db} \, \partial ^{\db \a} - 4 \, \bla _{A \db} \, \bla _{B \da} \bar{\partial}^{B \db} + 4 \big( \gamma ^{IJ} \big) _A {} ^B \, \bla _{B \da} \, n^I \partial ^J
\\
k_{\a \da} &= i \, x^+ _{\a \db} \, \bar{n}^\db {} _\da - i x^- _{\b \da} \, n _\a {} ^\b - x^+_{\a \db} \, x^- _{\b \da} \, \partial ^{\db \b} - 8i \left( \la \, \gamma ^{IJ} \, \bla \right)_{\a \da} \, n^I \partial ^J   \label{levelzero2} 
\end{aligned}
\end{equation}
Here, we introduced the chiral and anti-chiral coordinates
\begin{align}
x^+ _{\a \da} = x_{\a \da} + 2i \, \la ^A _\a \, \bla _{A \da} \, , \qquad \qquad x^- _{\a \da} = x_{\a \da} - 2i \, \la ^A _\a \, \bla _{A \da} \,.
\end{align}  
These generators satisfy the following commutation relations:
\begin{equation}
\begin{alignedat}{3}
 \Big[ d \, , \, p_\mu \Big] &= - p_\mu   \, & \qquad \Big[ p_\mu \, , \, k_\nu \Big] &= - 2 \eta _{\mu \nu} \, d + 2 m_{\mu \nu}   \, & \qquad  \Big[ d \, , \, k_\mu \Big] &= + k_\mu \\
\Big \lbrace q^\a _A \, , \, \bar{q}^{B \da} \Big \rbrace &= -2i \, \delta ^B _A p^{\da \a} \, & \qquad \Big \lbrace s^A _\a \, , \, \bar{s}_{B \da} \Big \rbrace &= -2i \, \delta ^A _B k_{\a \da} \, & \qquad  \Big[ d \, , \, q^\a _A \Big] &= - \half  q^\a _A \\
\Big[ k_{\a \da} \, , \, q^\b _A \Big] &= + 2i \, \delta^\b _\a \bar{s}_{A \da} \, & \qquad \Big[ k_{\a \da} \, , \, \bar{q}^{A \db} \Big] &= - 2i \, \delta^\db _\da s_\a ^A \, & \qquad  \Big[ d \, , \, s_\a ^A \Big] &= +\half  s_\a ^A \\
\Big[ p^{\da \a} \, , \, s_\b ^A \Big] &= - 2i \, \delta^\a _\b \bar{q}^{A \da} \, & \qquad \Big[ p^{\da \a} \, , \, \bar{s}_{A \db} \Big] &= + 2i \, \delta^\da _\db q^\a _A
\end{alignedat}
\end{equation}
The commutators with the generators $m$ and $r$ only depend on the set of indices and their position:
\begin{equation}
\begin{alignedat}{2}
\Big[ m \indices{_\a ^\b} \, , \, J_\g \Big] &= - 2i \, \delta ^\b _\g J_\a + i \delta ^\b _\a J_\g \, & \qquad \qquad  \Big[ m \indices{_\a ^\b} \, , \, J^\g \Big] &= + 2i \, \delta ^\g _\a J^\b - i \delta ^\b _\a J^\g \\
\Big[ \overline{m} \, \indices{^\da _\db} \, , \, J^\dg \Big] &= - 2i \, \delta ^\dg _\db J^\da + i \delta ^\da _\db J^\dg \, & \qquad \qquad  \Big[ \overline{m} \, \indices{^\da _\db} \, , \, J_\dg \Big] &= + 2i \, \delta ^\da _\dg J_\db - i \delta ^\da _\db J_\dg \\
\Big[ r \indices{^A _B} \, , \, J^C \Big] &= - 4 \, \delta ^C _B J^A + \delta ^A _B J^C \, & \qquad \qquad \Big[ r \indices{^A _B} \, , \, J_C \Big] &= + 4 \, \delta ^A _C J_B - \delta ^A _B J_C
\end{alignedat}
\end{equation}
The remaining non-vanishing commutators are given by
\begin{equation}
\begin{aligned}
\Big \lbrace q^\a _A \, , \, s_\b ^B \Big \rbrace = -2i \, \delta ^B _A \, m \indices{_\b ^\a} - \delta ^\a _\b \, r \indices{^B _A} - 2 \, \delta ^B _A \, \delta ^\a _\b \, d  \\
\Big \lbrace \bar{q}^{A \da} \, , \, \bar{s}_{B \db} \Big \rbrace = -2i \, \delta ^A _B \, \overline{m} \, \indices{^\da _\db} - \delta ^\da _\db \, r \indices{^A _B} + 2 \, \delta ^A _B \, \delta ^\da _\db \, d 
\end{aligned}
\end{equation}
Note in particular, that the differential operators form a representation of $\mathfrak{psu}(2,2 \vert 4)$ as the central charge is vanishing identically, which can be seen from the commutators of $q$ and $s$ given above. A comparison to the commutation relations of the generators given in appendix~\ref{app:su224} shows that all except the odd-odd commutators have a different sign,
\begin{equation}
\begin{aligned}
\Big[ T_a \, , \, T_b \Big \rbrace &= \tilde{f} \indices{_{ab} ^c} \, T_c \, , \qquad \\
\Big[ j_a (s) \, , \, j_b (s^\prime) \Big \rbrace &= f \indices{_{ab} ^c}  \, \delta(s - s^\prime) \, j_c (s) \, , \qquad \tilde{f} \indices{_{ab} ^c} = - (-1)^{ \lvert a \rvert \lvert b \rvert}  f \indices{_{ab} ^c} = f \indices{_{ba} ^c} \,.
\end{aligned}
\end{equation}
In order to see how the metric \eqn{groupmetric} translates to the basis of differential operators we may consider a representation of the $j_a$ in terms of the $\mathfrak{psu}(2,2 \vert 4)$-generators we introduced in~\ref{app:su224}. The assignment is given by 
\begin{equation}
\begin{aligned}
R(j_a) = \widetilde{T_a}  = \begin{cases}
- T_a, & \text{if $\Delta _a \in \left \lbrace 0, \, \half \right \rbrace$} \\
T_a , & \text{if  $\Delta _a \in \left \lbrace -1, \, -\half, \, 1 \right \rbrace$}
\end{cases} \, , \qquad \big[ \widetilde{T_a} , \widetilde{T_b} \big \rbrace = f \indices{_{ab} ^c} \, \widetilde{T_c} 
\end{aligned}
\end{equation}
where $\Delta_a$ denotes the weight of $T_a, \big[ D , T_a \big] = \Delta_a \, T_a$. A straightforward calculation shows that the components of $G$ in terms of the generators $\widetilde{T_a}$ are given by
\begin{align}
\left \langle \widetilde{T_a} , \widetilde{T_b} \right \rangle =   G_{ab} = \left(-1 \right)^{\lvert a \rvert}  \tilde{G}_{ab} = \tilde{G}_{ba} \, , \qquad G^{ab} = \left(-1 \right)^{\lvert a \rvert}  \tilde{G}^{ab} = \tilde{G}^{ba} \,.
\end{align}

\subsection{Level 1}
\label{app:level1}
In this appendix, we provide the parts of the level-1 densities, which are defined by
\begin{align}
j_a ^{(1)\, \prime} \, \hat{T}^a &=  \bigg \lbrace e^{X \cdot P + \theta Q + \btheta \Qb} \Big( - \frac{\pi^\mu}{\tau^2} \, K_\mu  + \pi ^\mu \dot{\pi}^\nu M_{\mu \nu} \Big) e^{- X \cdot P - \theta Q - \btheta \Qb} \bigg \rbrace_{(0)}
\end{align}
A direct calculation gives
\begin{equation}
\begin{alignedat}{2}
\left( p^{(1)\, \prime} \right)^{\mu} &= 0  \, & \qquad \qquad  d^{(1)\, \prime} &= i \tr \big( \bla \pi \theta_{(2)} - \btheta_{(2)} \pi \la \big) \\
\left( q^{(1)\, \prime} \right)^\a _A &= -2i\,  \btheta_{(2)\, A \da} \, \pi^{\da \a} \, & \qquad \qquad \left( \bar{q}^{(1)\, \prime} \right)^{A \da} &= -2i \, \pi^{\da \a} \theta _{(2)} {} _\a {} ^A 
\end{alignedat}
\end{equation}
For the remaining generators we have
\begin{equation}
\begin{aligned}
\left( m^{(1)\, \prime} \right) \indices {_\a ^\b} &= - 4 \big( \la \btheta_{(2)}  \, \pi \big) _\a {} ^\b + 2 \, \delta ^\b _\a \, \tr \big( \la \btheta_{(2)}  \, \pi \big) \\
\left( \bar{m}^{(1)\, \prime} \right) \indices{ ^\da _\db} &= 4 \, \big( \pi \,  \theta_{(2)} \bla \big) ^\da {} _\db - 2 \, \delta ^\da _\db \, \tr \big( \pi \,  \theta_{(2)} \bla \big) \\
\left( r^{(1)\, \prime} \right) \indices{^A _B} &= 8i \big( \bla \pi \theta_{(2)} + \btheta_{(2)} \pi \la  \big)_B {} ^A   - 2i \, \delta ^A _B \, \tr \big( \bla \pi \theta_{(2)} +  \btheta_{(2)}  \pi \la \big)  \, \\
\left( s^{(1)\, \prime} \right)_\a {} ^A &= \left( - 4 \left( \dot{\pi} - 2i \, \theta_{(2)} \bla \right)  \, \pi \, \la - 2 \, x^- \, \pi \theta_{(2)} \right)_\a {} ^A \\
\left( \bar{s}^{(1)\, \prime} \right)_{A \da} &= \left( 4 \, \bla \, \pi \left( \dot{\pi} + 2i \, \la \btheta_{(2)} \right)  + 2 \, \btheta_{(2)} \, \pi \, x^+ \right) _{A \da} \\
\left(k^{(1)\, \prime} \right) _{\a \da} &= 4i \left(x^+ \, \pi \, \theta_{(2)} \bla - \la \btheta_{(2)} \, \pi \, x^- \right)_{\a \da}
\end{aligned}
\end{equation}

\section{Kappa symmetry gauge fixing}
\label{app:kappa}
In order to provide some evidence for the possibility to fix the kappa symmetry gauge \eqn{kappagauge}, we consider the simple case of a straight-line boundary curve. The boundary conditions are given by
\begin{align}
X(0,s) = (0 , 0 , 0 , s) \, , \qquad y(0,s) = 0  \, , \qquad  N(0,s) = const. \,.
\end{align}
We do not need specific boundary condition for the $\theta$-variables as we will be working to linear order in Gra{\ss}mann variables, such that the $\theta$-variables do not appear. For the same reason, we are only interested in the parts of the bosonic variables which are of Gra{\ss}mann order zero. These are subject to the equations of motion for a bosonic string on $AdS_5$. It is easy to see that they are solved by (we fix conformal gauge)
\begin{align}
X(\tau, s) = (0 , 0 , 0 , s) \, , \qquad y(\tau,s) = \tau \,. \label{bosonicsol}
\end{align}
We recall that the kappa symmetry parameter is given by \eqn{kappatransform}
\begin{align*}
\kappa^{(1)} &= A_{i , -} ^{(2)} \, \mathcal{K} _+ ^{(1),i} + \mathcal{K} _+ ^{(1),i} \, A_{i , -} ^{(2)} \, ,  \qquad \qquad
\kappa^{(3)} = A_{i , +} ^{(2)} \, \mathcal{K}_- ^{(3),i} + \mathcal{K} _- ^{(3),i} \, A_{i , +} ^{(2)} \, , \quad \text{where}\\
V_{\pm} ^i &= P_{\pm} ^{i j} \, V_j = \half \left( \gamma^{ij} \pm i \, \eps^{i j} \right) V_j \,.
\end{align*}
In conformal gauge we may simplify this to
\begin{align}
\kappa = \kappa ^{(1)+(3)} = \left \lbrace A_s ^{(2)} \, , \, \mathcal{K} ^{(1)+(3)} \right \rbrace + i \left \lbrace A_s ^{(2)} \, , \, \mathcal{K} ^{(1)-(3)} \right \rbrace \label{kappaspec}
\end{align}
As we are only working to linear order in Gra{\ss}mann variables, we only need to determine those parts of $A_i ^{(2)}$, which are of Gra{\ss}mann order zero. They follow easily from \eqn{bosonicsol} and we have   
\begin{align}
A_s^{(2)} = \frac{\dot{X}_{\a \da} }{4y} \left( P^{\da \a} +  K^{\da \a} \right) \, , \qquad  A_\tau^{(2)} = - \frac{1}{y} D \,.
\end{align}
The anti-commutators of supermatrices appearing in \eqn{kappaspec} can be related to commutators due to the particular form of the supermatrix generators given in appendix~\ref{app:su224}. We have the following identities for $\mathcal{P} \in \left \lbrace P_\mu , K_\mu, D \right \rbrace$:
\begin{align}
\left \lbrace \mathcal{P} \, , \, Q \right \rbrace = \left[ \mathcal{P} \, , \, Q \right] \, , && 
\left \lbrace \mathcal{P} \, , \, \Sb \right \rbrace = \left[ \mathcal{P} \, , \, \Sb \right]\, , &&
\left \lbrace \mathcal{P} \, , \, S \right \rbrace = - \left[ \mathcal{P} \, , \, S \right] \, , &&
\left \lbrace \mathcal{P} \, , \, \Qb \right \rbrace = - \left[ \mathcal{P} \, , \, \Qb \right] \,.
\end{align}
This allows to compute the variation parameter $\kappa$. Parametrizing the arbitrary fermionic matrix $\mathcal{K}$ as
\begin{align}
\mathcal{K} = a_\a {} ^A Q_A {} ^\a - \bar{a}_{A \da} \Qb^{\da A} + b_A {} ^\a S_\a {} ^A - \bar{b}^{\da A} \Sb_{A \da} \, 
\end{align}
we obtain
\begin{align}
\kappa &= \frac{1}{2y} \Big[ \left( i \dx_{\a \da} \,  \bar{b}^{\da A} + K^{AB} \, b_{B \a} \right) Q _A {} ^\a + \left(-i b  _A {} ^\a \, \dx_{\a \da} -K_{AB} \, \bar{b}  _\da {}^B \right) \Qb^{\da A} \nn \\
& \qquad + \left( i  \bar{a}_{A \da} \, \dx^{\da \a} + K_{AB} \, a^{\a B} \right)  S _\a {} ^A + \left(-i \dx ^{\da \a} \,  a_\a {} ^A - K^{AB} \, \bar{a}^\da {} _B  \right) \Sb_{A \da} \Big] \nn \\
&= c_\a {} ^A Q_A {} ^\a - \bar{c}_{A \da} \,  \Qb^{\da A} + d_A {} ^\a S_\a {} ^A - \bar{d}^{\, \da A} \Sb_{A \da} \, .
\end{align}
The parameters $c,\bar{c},d, \bar{d}$ are related among each other by 
\begin{align}
\bar{c}_{A \da} = i \, \eps_{\da \db} \, \dx ^{\db \da} \,  c_\a {} ^B  \, K_{BA} \, , \qquad \bar{d}^{\, \da A} = -i   \, K^{AB} \, d_B {} ^\a \, \dx_{\a \db} \, \eps^{\db \da} \,.
\end{align}
This means that fixing $c_1$ and $\bar{c_1}$ determines $c_2$ and $\bar{c}_2$, fixing $c_3$ and $\bar{c_3}$ determines $c_4$ and $\bar{c}_4$ and likewise for $d, \bar{d}$. In particular, we observe that the kappa transformation has half the degrees of freedom of a generic fermionic element, as it should be. We also see that we cannot enforce $\kappa \in \mathfrak{su}(2,2 \vert 4 )$ by constraining $\mathcal{K}$. One may e.g.\ enforce $(c_\a {} ^1)^\ast = \bar{c}_{1 \da}$ but that leads to $(c_\a {} ^2)^\ast = - \bar{c}_{2 \da}$. We go on to calculate the kappa variation of the fermionic part of $A$, which is given by
\begin{align}
A^{(1)+(3)} \mapsto A^\prime {}^{(1)+(3)} = A^{(1)+(3)} -\diff \kappa + \left[ A^{(0)+(2)} \, , \, \kappa \right]
\end{align}
For the terms that are of Gra{\ss}mann order zero we have
\begin{align}
A_{\tau}^{(0)+(2)} = - \frac{1}{y} \, D \, , \qquad A_s^{(0)+(2)} = \frac{\dx_{\a \da}}{2y} \, P^{\a \da} \, ,
\end{align}
and we find the following transformations for the parameters $\varepsilon$ and $\chi$:
\begin{equation}
\begin{alignedat}{2}
\chi ^\prime _\tau {} _A {} ^\a &= \chi _\tau {} _A {} ^\a - \left(\partial_\tau + \sfrac{1}{2y} \right) d_A {} ^\a\, ,& \qquad \bar \chi^\prime _\tau {} ^{\da A}  &= \bar \chi _\tau {} ^{\da A} - \left(\partial_\tau + \sfrac{1}{2y} \right) \bar{d} ^{\, \da A}   \, , \\
\chi ^\prime _s {} _A {} ^\a &= \chi _s {} _A {} ^\a - \partial_s \, d_A {} ^\a \, ,& \qquad \bar \chi^\prime _s {} ^{\da A}  &= \bar \chi _s{} ^{\da A} - \partial_s \, \bar{d} ^{\, \da A} \, , \\
\varepsilon^\prime _\tau {} _\a {} ^A &= \varepsilon _\tau {} _\a {} ^A - \left(\partial_\tau -\sfrac{1}{2y} \right) c_\a {} ^A \, ,& \qquad \bar \varepsilon^\prime _\tau {} _{A \da} &= \bar \varepsilon _\tau {} _{A \da}- \left(\partial_\tau -\sfrac{1}{2y} \right) \bar{c} _{A \da} \, , \\
\varepsilon^\prime _s {} _\a {} ^A &= \varepsilon _s {} _\a {} ^A - \partial_s \, c_\a {} ^A - \sfrac{i}{2y} \, \dx_{\a \da} \, \bar{d}^{\, \da A}    \, ,& \qquad \bar \varepsilon^\prime _s {} _{A \da} &= \bar \varepsilon _s {} _{A \da} - \partial_s \, \bar{c} _{A \da}  + \sfrac{i}{2y} \, d_A {} ^\a \, \dx_{\a \da}   \, , 
\end{alignedat}
\end{equation}
Due to the form of the relations between the parameters $c , \bar{c}, d , \bar{d}$ explained above, it is thus clear that we can indeed reach the kappa symmetry gauge
\begin{align}
\varepsilon _\a {} ^2 &= \varepsilon _\a {} ^4  = 0 \, , \qquad \bar{\varepsilon}_{2 \da} = \bar{\varepsilon}_{4 \da} = 0 \, , \qquad \chi ^\a _1 = \chi^\a _3 = 0 \, , \qquad \bar{\chi}^{1 \da} = \bar{\chi}^{3 \da} = 0 \,.
\end{align}

\newpage

\bibliographystyle{nb}
\bibliography{botany}

\end{document}